\documentclass[12pt]{article}
\usepackage{amsmath,amssymb}
\usepackage{braket}
\usepackage{color}
\usepackage{cancel}
\usepackage{ulem}
\usepackage{epsf,epsfig}

\def\dAlembert{\kern1pt\vbox{\hrule height 1.2pt\hbox{\vrule width 1.2pt\hskip 3pt
   \vbox{\vskip 6pt}\hskip 3pt\vrule width 0.6pt}\hrule height 0.6pt}\kern1pt}

\begin{document}


\begin{center}
{\bf Quantum backreaction and stability of topological wormholes}\\
\end{center}
\bigskip
\begin{center}
{\bf Haris Mehulic and Tomislav Prokopec}
\end{center}

\begin{center}
{\rm ITP, EMMEPH and Spinoza Institute, Utrecht University, 
Princetonplein 5, 3584 CC Utrecht, The Netherlands}
\end{center}

\bigskip
\begin{center}
{\bf Abstract}
\end{center}

We investigate the quantum stability of a timelike topological wormhole with a simple geometry $M_2 \times S^2$, supported classically by anisotropic fluid. We compute the one-loop quantum backreaction generated by the vacuum fluctuations of a minimally coupled, massive scalar field propagating on the wormhole background. Using dimensional regularization we renormalize the one-loop energy-momentum tensor and identify the necessary gravitational counterterms. We then solve the semiclassical Einstein equations
to linear order in $\hbar$ for both time-dependent and static metric {\it Ans\"{a}tze}.
Depending on the choice of finite counterterms,
the quantum effects can induce either negative or positive angular pressure, which will tend to destabilize or stablize the wormhole, respectively.
 We also show that a classically traversable wormhole will remain traversable 
when the quantum backreaction is taken into account.
It would be of interest to investigate whether these 
conclusions remain true when the equations of semiclassical gravity are 
self-consistently solved.


\section{Introduction}

Wormholes~\cite{Einstein:1935tc,Godel:1949ga,MisnerThorneZurek:2009} are among the most intriguing solutions of general relativity, providing hypothetical shortcuts between distant regions of spacetime. Classical analyses~\cite{Morris:1988cz} have long shown that traversable wormholes typically require stress--energy tensors violating null energy condition, motivating the study of exotic matter sources, modified gravity theories, and semiclassical effects as possible mechanisms for realizing such geometries. Nevertheless, it is of interest to explore configurations that evade such requirements at the classical level.

A particularly simple example is a wormhole with spacetime geometry \( M_2 \times S^2 \), where \( M_2 \) denotes two--dimensional Minkowski space and the curvature is entirely localized on the spherical section 
$S^2 $. The spacetime is otherwise flat, and the Einstein equations admit a solution supported by a homogeneous anisotropic stress--energy tensor, effectively corresponding to a two--dimensional cosmological constant along \( M_2 \). Related constructions have appeared in scalar--tensor theories, where spherically symmetric solutions supported by scalar stresses exhibit nontrivial global structure and wormhole--like properties without requiring localized material sources~\cite{Bronnikov:1973fh,Kardashev:2006nj,Bronnikov:2013coa,Bronnikov:2021uta}.

Classical consistency, however, does not guarantee quantum stability. Even in classically well-behaved spacetimes, quantum vacuum fluctuations of matter fields generically produce nontrivial backreaction effects~\cite{Khabibullin:2005ad,Sushkov:2008zz}. In semiclassical gravity~\cite{Birrell:1982ix} these effects are encoded in the renormalized expectation value of the energy--momentum tensor, which acts as a source in Einstein’s equations. Assessing whether such quantum corrections preserve or destabilize a given classical geometry is therefore essential, particularly for wormholes whose viability depends sensitively on their stress--energy content.

Early investigations of this question demonstrated that Casimir--type vacuum energies can influence wormhole geometries. In a seminal work, Khabibullin, Khusnutdinov, and Sushkov~\cite{Khabibullin:2005ad,Sushkov:2008zz} studied the Casimir effect in wormhole spacetimes and showed that vacuum polarization can generate negative energy densities, although typically at microscopic scales and often requiring additional structure to sustain traversability. Related studies of Casimir energies in curved and compact backgrounds, including analyses of the thermal Casimir effect in the Einstein universe with spherical boundaries, further illustrate how global geometry and boundary conditions shape quantum stresses when the background spacetime is treated as fixed \cite{Butcher:2014lea,Mota:2022qpf,Garattini:2024uso,Garattini:2024jkr}. Casimir effects have been revisited in the context of semiclassical wormhole constructions, where vacuum energy is combined with additional classical or semiclassical sources to obtain self--consistent solutions of the semiclassical 
Einstein equations~\cite{Hochberg:1996ee} and to explore conditions for sustained traversability~\cite{Avalos:2025hfw,Agrawal:2024dam}.

In this work, we address the closely related question of whether quantum vacuum fluctuations stabilize or destabilize a classically admissible wormhole geometry. We also study how traversability is 
affected by the inclusion of the quantum backreaction.
For analytical tractability, we consider a wormhole described by the product spacetime \( M_2 \times S^2 \), with the two--sphere $S^2$ of constant radius \( a \). At the classical level, the geometry is supported by a homogeneous anisotropic stress--energy tensor
and by localized negative energy shells at both mouths rather than by localized matter at a throat. The wormhole length \( L_0 \) is assumed to be much larger than its radius, allowing for a simplified treatment. The classical stress--energy tensor exhibits a thin shell at the interface between the wormhole region and the surrounding Minkowski space, which would be smoothed out in more realistic models.

We quantize a minimally coupled massive scalar field on this background and compute the renormalized one--loop expectation value of its energy--momentum tensor using dimensional regularization and appropriate renormalization conditions. We choose the finite cosmological constant counterterm 
such that the observed the cosmological constant vanishes both in the Minkowski spaces outside the wormhole and inside the wormhole. 
The finite Newton constant counterterm is chosen such that the observed Newton constant has identical value inside and outside the wormhole.

Treating gravity semiclassically, we then study the backreaction of the one-loop quantum stress--energy on the classical geometry.
We find that, depending on the choice of finite higher derivative counterterms, quantum effects can induce either a negative or a positive angular pressure. 
When the angular pressure is negative (positive), the quantum backreaction 
will tend to destablize (stabilize) the wormhole. In both cases a traversable wormholes will remain traversable.

\section{Classical wormhole}
\label{Classical wormhole}

\begin{figure}[t]
	\centering
\vskip -2.5cm
	\includegraphics[width=0.45\textwidth]{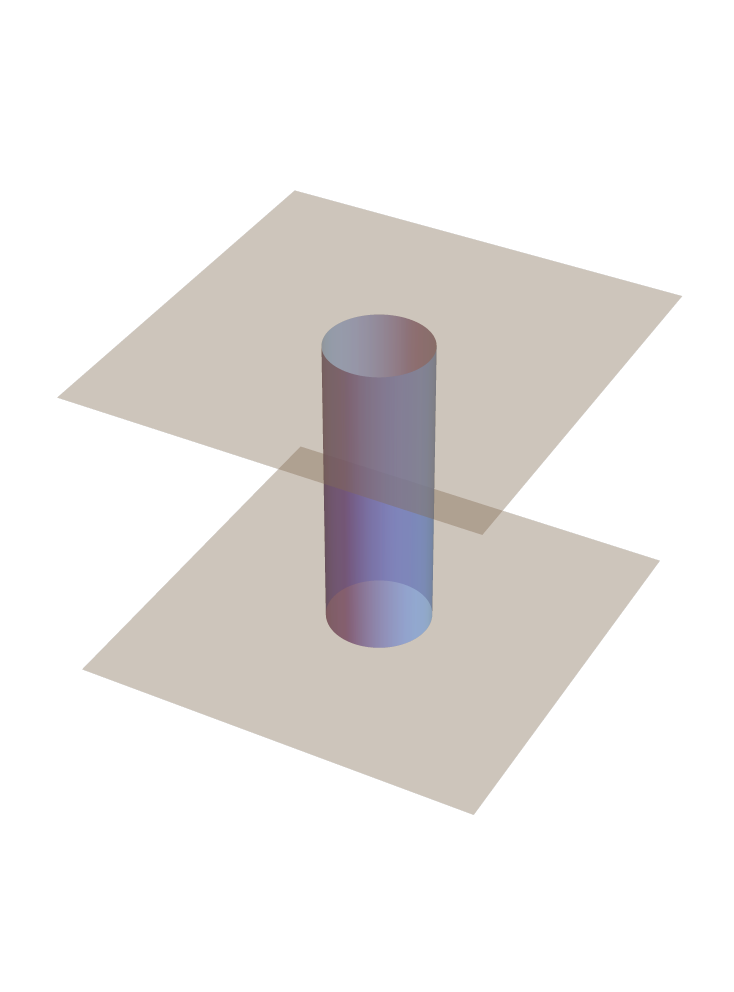}
\vskip -1cm
    \caption{
    The wormhole geometry described by the metric in Eq.~(\ref{wormhole metric}) shown as $\mathbb{R}\times S^1$ for illustrative purpose. The cylinder length is $L_0$ and its radius is $a$. The points of contact between the flat Minkowski spaces and the wormhole 
    is the boundary of the cylinder, shown as the two circles of radius $a$  at $z=\pm L_0/2$.}
	\label{figure: wormhole}
\end{figure}

In this work we consider a topological wormhole, whose geometry 
is given by $\mathbb{R}\times \mathbb{R}\times S^2$,~\footnote{ Note that the wormhole in 
Eq.~(\ref{wormhole metric})
has a flat five-dimensional Minkowski space, $M_{5}=M_{2}\times\mathbb{R}^{3}$, as the embedding space.
}
and four-dimensional metric of the wormhole is,
\begin{equation}
{\rm d}s^2 = -{\rm d}t^2 
+ {\rm d} z^2
+ a^2 {\rm d}\Omega_2^2
\,,\qquad \big({\rm d}\Omega_2^2 = {\rm d}\vartheta^2
                 + \sin^2(\vartheta){\rm d}\varphi^2\big)
\,,\qquad
\label{wormhole metric}
\end{equation}
where $a$ is the wormhole radius, $\vartheta\in[0,\pi)$ and $\varphi\in[0,2\pi)$
are the usual spherical coordinates on $S^2$,
$z\in(-\infty,\infty)$ 
is the longitudinal
coordinate along the wormhole and $t\in(-\infty,\infty)$ is time.
In this paper we use units in which the speed of light $c=1$.
The wormhole in Eq.~(\ref{wormhole metric}) is illustrated in Figure~\ref{figure: wormhole}.

The spherical section of the wormhole is curved, so
nonvanishing Chrisoffel symbols are, 
\begin{equation}
    \Gamma^\vartheta_{\varphi\varphi} 
    = -\sin(\vartheta)\cos(\vartheta)
\,,\qquad
 \Gamma^\varphi_{\vartheta\varphi} 
    =\Gamma^\varphi_{\varphi\vartheta} 
    = \frac{\cos(\vartheta)}{\sin(\vartheta)}
\,.\qquad
\label{Chrisoffel symbols}
\end{equation}
The wormhole in Eq.~(\ref{wormhole metric}) is attached to 
flat Minkowski spaces above and below as shown 
in figure~\ref{figure: wormhole}. It is instructive to analyze
how the attachment works in some detail.
The corresponding flat metrics are, 
\begin{equation}
{\rm d}s^2 = -\, {\rm d} t^2 + {\rm d}r_\pm^2 + r_\pm^2\big({\rm d}\vartheta^2 +\sin^2(\vartheta){\rm d}\varphi^2\big)
\,,\qquad (r_\pm\geq a)
\,,\qquad
\label{metrics: flat space}
\end{equation}
where at the points of contact between the wormhole and
the flat spaces $r_\pm = a \pm z - L_0/2$, where $L_0$ is the wormhole length, and $z=0$ at the center of the wormhole. 
Nonvanishing Christoffel symbols 
for the metrics in Eqs.~(\ref{wormhole metric}) and~(\ref{metrics: flat space}) are, 
\begin{eqnarray}
\Gamma^{r_\pm}_{\vartheta\vartheta} 
&\!\!=\!\!& - \sum_{\pm}r_\pm\!\times\!\Theta (r_\pm\!-\!a)
\,,\qquad 
\Gamma^{r_\pm}_{\varphi\varphi}
= - \sum_{\pm}r_\pm\sin^2(\vartheta)\!\times\!\Theta (r_\pm\!-\!a)
\,,
\nonumber\\
\Gamma^\vartheta_{{r_\pm}\vartheta} &\!\!=\!\!& 
             \sum_{\pm}\frac{1}{r_\pm}\!\times\!\Theta (r_\pm\!-\!a)
\,,\qquad 
\Gamma^\vartheta_{\varphi\varphi} 
            = -\sin(\vartheta)\cos(\vartheta)
\,,\quad 
\nonumber\\
\Gamma^\varphi_{\vartheta\varphi} &\!\!=\!\!&  \Gamma^\varphi_{\varphi\vartheta}
    =\frac{\cos(\vartheta)}{ \sin(\vartheta)}
\,,\;\,
\label{classical Christoffel symbols}
\end{eqnarray}
where the Heaviside theta functions $\Theta(x)$ indicate that the Christoffel symbols in Eq.~(\ref{classical Christoffel symbols}) exhibit a jump at $r_\pm=a$ ($z=\pm L_0/2$). 

Nonvanishing components of the Riemann curvature tensor are then,
\begin{eqnarray}
R_{r_\pm\vartheta r_\pm\vartheta} 
&\!\!=\!\!& R_{\vartheta r_\pm\vartheta r_\pm} 
= - R_{\vartheta  r_\pm r_\pm\vartheta} 
= - R_{r_\pm\vartheta \vartheta r_\pm} 
    = - a\sum_\pm\delta(r_\pm\!-\!a)
\,,\qquad 
\label{Riemann tensor: singular 1}\\
R_{r_\pm\varphi r_\pm\varphi} 
&\!\!=\!\!& R_{\varphi r_\pm\varphi r_\pm} 
= - R_{\varphi  r_\pm r_\pm\varphi} 
= - R_{r_\pm\varphi \varphi r_\pm} 
    = - a\sin^2(\vartheta)\sum_\pm\delta(r_\pm\!-\!a)
\,,\qquad 
\label{Riemann tensor: singular 2}\\
R_{\vartheta\varphi\vartheta\varphi}
&\!\!=\!\!&  R_{\varphi\vartheta\varphi\vartheta} 
= - R_{\vartheta\varphi\varphi\vartheta} 
= - R_{\varphi\vartheta \vartheta\varphi} 
    = a^2\sin^2(\vartheta)\Theta\Big(\frac{L_0}{2}-|z|\Big)
\,.\qquad 
\label{Riemann tensor: regular}
\end{eqnarray}
These relations imply for nonvanishing components of the Ricci tensor and Ricci scalar,
\begin{eqnarray}
R_{r_\pm r_\pm} &\!\!=\!\!& -\frac{2}{a}\sum_\pm\delta(r_\pm\!-\!a)
\,,\qquad
\nonumber\\
R_{\vartheta\vartheta} &\!\!=\!\!& - a\sum_\pm\delta(r_\pm\!-\!a)
   + \Theta\Big(\frac{L_0}{2}-|z|\Big)
\,,\qquad
\nonumber\\
R_{\varphi\varphi} &\!\!=\!\!& 
  -a\sin^2(\vartheta)\sum_\pm\delta(r_\pm\!-\!a)
  + \sin^2(\vartheta)\Theta\Big(\frac{L_0}{2}-|z|\Big)
\,,\qquad 
\nonumber\\
R &\!\!=\!\!& -\frac{4}{a}\sum_\pm\delta(r_\pm\!-\!a)
+ \frac{2}{a^2}\Theta\Big(\frac{L_0}{2}-|z|\Big)
\,,\qquad
\label{Ricci tensor and Ricci scalar}
\end{eqnarray}
such that nonvanishing components of the Einstein curvature tensor,
$G^{\mu\nu}=R^{\mu\nu}-\frac12 g^{\mu\nu} R$, are
\begin{eqnarray}
 G_{tt} &\!\!=\!\!& -\frac{2}{a}\sum_\pm\delta(r_\pm\!-\!a)
 +\frac{1}{a^2}\Theta\Big(\frac{L_0}{2}-|z|\Big)
\,,\qquad
\nonumber\\
  G_{zz} &\!\!=\!\!& - \frac{1}{a^2}\Theta\Big(\frac{L_0}{2}-|z|\Big)
\,,\qquad
\nonumber\\
  G_{\vartheta\vartheta} &\!\!=\!\!& 
  a\sum_\pm\delta(r_\pm\!-\!a)
\,,\qquad
\nonumber\\
  G_{\varphi\varphi} &\!\!=\!\!& 
  a\sin^2(\vartheta)\sum_\pm\delta(r_\pm\!-\!a)
\,.\qquad
\label{Einstein tensor}
\end{eqnarray}
The Einstein equation, 
\begin{equation}
    G_{\mu\nu} = 8\pi G T_{\mu\nu}
\,,\qquad
\label{Einstein equation}
\end{equation}
then implies that the classical energy-momentum tensor that supports
our topological wormhole is given by, 
\begin{eqnarray}
    T_{t}^{\;t} &\!\!=\!\!&
    \frac{1}{4\pi G a}\sum_\pm\delta(r_\pm- a)
 -\frac{1}{8\pi G a^2}\!\times\!\Theta\Big(\frac{L_0}{2}-|z|\Big)
    \,,\qquad
\nonumber\\
 T_{z}^{\;z} &\!\!=\!\!&
  - \frac{1}{8\pi G a^2}\!\times\!\Theta\Big(\frac{L_0}{2}-|z|\Big)
\nonumber\\
 T_{\vartheta}^{\;\vartheta}&\!\!=\!\!& T_{\varphi}^{\;\varphi} 
   =  \frac{1}{8\pi G a}\sum_\pm\delta(r_\pm- a)
\,,\qquad
\label{energy-momentum tensor: topological wormhole}
\end{eqnarray}
which contains {\it regular} contributions inside the wormhole and {\it singular} contributions at the boundary between the wormhole and the flat spacetimes.
Inside the wormhole 
the energy momentum tensor in Eq.~(\ref{energy-momentum tensor: topological wormhole}) corresponds to an anisotropic fluid with the classical energy-momentum tensor of an anisotropic fluid, 
 \begin{equation}
 T_{\mu}^{\;\nu} = {\rm diag}\big(\!-\rho,p_z,p_\perp,p_\perp\big)
   \,,\qquad
\label{energy-momentum tensor: topological wormhole 2}
\end{equation}
with $p_z=-\rho =- 1/(8\pi Ga^2)$ and $p_\perp = 0$, which is large when converted to SI units,
$\rho\approx 5\times 10^{42}\big(1{\rm m}/a\big)^2~\big[{\rm kg/(ms^2)}\big]$,
which corresponds to a mass density 
$\rho/c^2\approx 5\times 10^{25}\big(1{\rm m}/a\big)^2~{\rm \big[kg/m^3\big]}$,
where $a$ is here expressed in meters. This means that for a wormhole with $a=10~{\rm km}$, mass density is about $5\times 10^{17}~{\rm kg/m^3}$, which is comparable to that in a neutron star.
This means that the amount of energy needed to build a wormhole of radius $a$ (in meters) is
$E\approx 2\times 10^{43}(a/1{\rm m})~[{\rm J}]$, which is comparable with the Earth's rest energy,
$E_\oplus = m_\oplus c^2 \simeq 5.36\times 10^{41}~{\rm J}$.
As expected, the energy needed to build a wormhole is very large~\cite{Morris:1988cz}, simply because it takes a large amount of energy to curve space.~\footnote{The Universe is known to be spatially flat to a high precision~\cite{Planck:2018vyg}. If the observed Universe were positively or negatively curved, one would need an equally large amount of energy to curve it. Even though this energy is large, one does not use it 
as an argument against a spatially curved universe.}
Therefore, the wormhole is supported by a fluid with a negative pressure along the $z$ direction, and a vanishing pressure in the angular directions, {\it i.e.} it behaves as a two-dimensional cosmological constant.
 Note that, for null vectors $\ell_1^\mu=(1,1,0,0)$
 and $\ell_2^\mu=(1,0,a^{-1},0)$ ($\ell_1^2=0$, $\ell_2^2=0$), we have,
 \begin{eqnarray}
    T_{\mu\nu}\ell_1^\mu\ell_1^\nu 
 = T_{00} + T_{zz} = \rho + p_z = 0
 \,,\qquad
 \label{NEC 1}\\
     T_{\mu\nu}\ell_2^\mu\ell_2^\nu 
 = T_{00}+\frac{1}{a^2}T_{\vartheta\vartheta} = \rho + p_\perp
  = \frac{1}{8\pi G a^2} > 0
 \,,\qquad
 \label{NEC 2}
 \end{eqnarray}
 implying that null energy condition (NEC) is {\it marginally satisfied} inside the wormhole for the regular contribution in 
 Eq.~(\ref{energy-momentum tensor: topological wormhole 2}).
On the other hand, the singular contribution in 
Eq.~(\ref{energy-momentum tensor: topological wormhole}) corresponds to a surface energy density tensor $\sigma_{\mu\nu}$ of matter with,
 \begin{equation}
\sigma_{\mu}^{\;\nu} = {\rm diag}\big(\!-\sigma_t,\sigma_z,\sigma_\perp,\sigma_\perp\big)
\,,\quad\! \sigma_t = - \frac{1}{4\pi G a}
\,,\quad\! \sigma_z = 0
\,,\quad\! \sigma_\perp = \frac{1}{8\pi G a}
\,,
\label{surface density}
 \end{equation}
 which is also needed to support the wormhole.~\footnote{ 
 Including the negative surface energy in Eq.~(\ref{surface density}) 
 will violate NEC.
 }
 The negative value of $\sigma_t$ makes this wormhole somewhat problematic.
 In realistic wormholes the singular shells of
 energy and pressure
 in Eq.~(\ref{energy-momentum tensor: topological wormhole})  
 would be smoothed out. Here we keep the geometry singular in order to make our analytic calculations tractable. One way of incorporating the singular shells into an energy condition is by introducing
 an averaged energy-momentum tensor of the wormhole,
 \begin{equation}
    \langle T_{\mu\nu}\rangle_V 
    \equiv \frac{1}{L_0}\int_{-\frac{L_0}{2} - \delta}^{\frac{L_0}{2} + \delta}
    {\rm d}z \, T_{\mu\nu}
\,,\qquad
\label{averaged energy-momentum tensor}
\end{equation}
with $\delta>0$ an infinitesimal quantity, resulting in,
\begin{eqnarray}
    \langle T_{t}^{\;t}\rangle_V   &\!\!=\!\!&
     - \langle \rho\rangle_V  
   = - \frac{1}{8\pi G a^2}\Big(1-\frac{4a}{L_0}\Big)
    \,,\qquad
\nonumber\\
 \langle T_{z}^{\;z}\rangle_V   &\!\!=\!\!&  \langle p_z\rangle_V  
   = - \frac{1}{8\pi G a^2}
   \,,
\nonumber\\
 \langle T_{\vartheta}^{\;\vartheta}\rangle_V   &\!\!=\!\!& 
 \langle T_{\varphi}^{\;\varphi} \rangle_V  = \langle p_\perp\rangle_V  
   =  \frac{1}{4\pi G a L_0}
\,.\qquad
\label{energy-momentum tensor: topological wormhole: averaged}
\end{eqnarray}
Using these results in Eqs.~(\ref{NEC 1})--(\ref{NEC 2})
one obtains for average null energy condition (ANEC),
 \begin{eqnarray}
    \langle T_{\mu\nu}\rangle_V  \ell_1^\mu\ell_1^\nu 
&\!\!=\!\!& \langle T_{00}\rangle_{V}  + \langle T_{zz}\rangle_V  
   = \langle\rho\rangle_V  + \langle p_z\rangle_V  
    = -  \frac{1}{2\pi G a L_0} <0
 \,,\qquad
 \label{ANEC 1}\\
    \langle T_{\mu\nu}\rangle_V \ell_2^\mu\ell_2^\nu 
&\!\!=\!\!& \langle T_{00}\rangle_V  +\frac{1}{a^2}
 \langle T_{\vartheta\vartheta}\rangle_V  = \langle\rho\rangle_V  
    + \langle p_\perp\rangle_V  
  = \frac{1}{8\pi G a^2}\Big(1-\frac{2a}{L_0}\Big)
 \,,\qquad\;
 \label{ANEC 2}
 \end{eqnarray}
 implying that ANEC is {\it violated} inside the wormhole,
 which is in agreement with the literature~\cite{Morris:1988cz,Friedman:1993ty,Visser:1995cc,Hochberg:1997wp,Hochberg:1998ii,Krasnikov:2010vw}. 
This means that the singular shells
in Eq.~(\ref{energy-momentum tensor: topological wormhole}) 
contribute towards violation of ANEC.
In order to make our calculations analytically tractable, in this work we assume that the wormhole is supported by the classical energy-momentum tensor given in 
 Eq.~(\ref{energy-momentum tensor: topological wormhole}).
However, recent literature~\cite{Maldacena:2013xja,Maldacena:2017axo,Maldacena:2018lmt,Garcia-Garcia:2019poj,Zhou:2020wgh,Maldacena:2018gjk,Maldacena:2020sxe}
 has established that 
 even standard model matter can be used to form traversable wormholes
 which may, but need not, be stable.

Curiously, the wormhole in Eq.~(\ref{wormhole metric}) neither violates the weak energy condition (WEC) 
nor the average weak energy condition (AWEC),
 \begin{eqnarray}
    T_{\mu\nu} t ^\mu t ^\nu 
&\!\!=\!\!&  - T_{0}^{\;0} = \rho  = \frac{1}{8\pi G a^2} > 0
 \,,\qquad
 \label{WEC}\\
  \langle T_{\mu\nu}\rangle_V  t^\mu t^\nu 
&\!\!=\!\!& - \langle T_{0}^{\;0}\rangle_{V} 
    = \langle \rho \rangle_{V} 
   =  \frac{1}{8\pi G a^2}\Big(1-\frac{4a}{L_0}\Big) > 0
 \,,\quad  \big(L_0> 4a\big)
  \,,\qquad \;
 \label{AWEC}
 \end{eqnarray}
where $t^\mu=(1,0,0,0)$ ($t^2=-1$) is a timelike vector. To get the last inequality in Eq.~(\ref{AWEC})
we assumed that the wormhole is sufficiently long, $L_0> 4a$.

\medskip\noindent
{\bf A note on topology.}
The wormhole spacetime in Eq.~(\ref{wormhole metric}) 
considered in this work 
is topological in the sense that the second homotopy group of the the wormhole manifold is nontrivial, 
$\pi_2(M_2\times S^2) =\mathbb{Z}$, and in this sense it accords with the definition promoted in Ref.~\cite{Visser:1995cc}. Note that higher homotopy groups are also nontrivial. For example, 
$\pi_3(M_2\times S^2) =\mathbb{Z}$ and $\pi_4(M_2\times S^2) =\pi_5(M_2\times S^2) =\mathbb{Z}_2$. The fact that the wormhole in Eq.~(\ref{wormhole metric}) is continuously connected to two flat Minkowski spaces as illustrated in figure~\ref{figure: wormhole} does not change the wormhole topology. On the other hand, when the two wormhole ends are attached to different spatial sections of the same Minkowski space, then spacetime topology changes to $\mathbb{R}\times S^1\times S^2$. In this case, in addition to $\pi_2$, also fundamental homotopy group $\pi_1$ of the manifold becomes nontrivial, $\pi_1\big(\mathbb{R}\times S^1\times S^2\big)=\mathbb{Z}$.
It is reasonable to assume that topology is preserved by the classical evolution. However, topology 
can be changed by the quantum evolution~\cite{Morris:1988cz,Maldacena:2018gjk}. 
Since here we are primarily interested in the classical and quantum dynamics of the wormhole geometry, and topological differences between $M_2\times S^2$ and $\mathbb{R}\times S^1\times S^2$
do not affect the wormhole dynamics, the fate of topology will not be our primary concern in this work.

{

\section{Quantum backreaction}
\label{Quantum backreaction}

In this section we compute the one-loop quantum backreaction induced by a massive scalar field on
the classical geometry in Eq.~(\ref{wormhole metric}). We begin by introducing the theory, and then proceed towards computation of the quantum backreaction.

\subsection{Bare theory}
\label{Bare theory}

We consider a noninteracting, massive real scalar field, whose bare matter action is given by,
\begin{equation}
    S_{\tt b}[g_{\mu\nu},\phi] 
    = \int {\rm d}^Dx \sqrt{-g} 
    \left[ -\frac12 
     g^{\mu\nu} (\partial_\mu\phi_{\tt b})(\partial_\nu\phi_{\tt b})
     -\frac12 m_{\tt b}^2\phi^2_{\tt b}\right]
\,,\qquad
\label{bare matter action}
\end{equation}
where $g_{\mu\nu}$ and $g^{\mu\nu}$ is the metric tensor and 
its inverse, $g={\rm det}[g_{\mu\nu}]$, $\phi_{\tt b}$ is the bare field 
and $m_{\tt b}$ is the bare mass. Upon defining:
\begin{equation}
\phi_{\tt b} = \sqrt{Z_{\tt b}}\phi
 \,, \qquad
 Z_{\tt b}  = 1 + \delta Z
 \,, \qquad Z_{\tt b}m_{\tt b}^2 = m^2 + \delta m^2
 \,,\qquad
\label{bare matter action: couplings}
\end{equation}
where $\phi$ is classical (tree-level) scalar field, $Z_{\tt b}$
is field strength renormalization, 
$m^2$ is tree-level mass, and
$\delta m^2$ and $\delta Z$ are used to renormalize 
loop divergences of the theory. With these remarks in mind, 
the bare action in Eq.~(\ref{bare matter action}) splits into two parts, the classical action 
$S_\phi$ and the counterterm action $S_\phi^{\tt ct}$,
\begin{eqnarray}
    S_{\tt b}[g_{\mu\nu},\phi] &\!=\!&
     S_\phi[g_{\mu\nu},\phi] + S_\phi^{\tt ct}[g_{\mu\nu},\phi]
\,,\qquad
\nonumber\\
   S_\phi[g_{\mu\nu},\phi]  &\!=\!& \int {\rm d}^Dx \sqrt{-g} 
    \left[ -\frac12 g^{\mu\nu} (\partial_\mu\phi)(\partial_\nu\phi)-\frac12 m^2\phi^2\right]
\,,\qquad
\label{classical matter action}\\
   S_\phi^{\tt ct}[g_{\mu\nu},\phi]  &\!=\!&
   \int {\rm d}^D x\sqrt{-g}
    \left[ -\frac12\delta Z g^{\mu\nu} (\partial_\mu\phi)(\partial_\nu\phi)-\frac12 \delta m^2\phi^2\right]
\,.\qquad
\label{counterterm matter action}
\end{eqnarray}
In what follows we canonically quantize $\phi$ to study how its quantum fluctuations backreact onto the spacetime.
We shall use the semiclassical gravity approximation, according to which gravity is classical, and matter is quantum. For the purposes of this work, we shall use Einstein's gravity,
whose bare action is given by,
\begin{equation}
    S_{\tt g}[g_{\mu\nu}] 
    = \frac{1}{\kappa^2_{\tt b}}\int {\rm d}^Dx \sqrt{-g} 
    \big[R - 2\Lambda_{\tt b}\big]
\,,\qquad
\label{bare gravitatinal action}
\end{equation}
where $R=R(g_{\mu\nu})$ is the Ricci curvature scalar
and $\kappa_{\tt b}^2 = 16\pi G_{\tt b}$ 
and $\Lambda_{\tt b}$ are the bare gravitational coupling and the cosmological constant, respectively.
The bare gravitational couplings can be split into their classical parts and quantum parts (needed for renormalization),
\begin{equation}
 \frac{1}{\kappa^2_{\tt b}} 
 = \frac{1}{\kappa^2} + \delta \left(\frac{1}{\kappa^2}\right)
 \,, \qquad 
 \frac{\Lambda_{\tt b}}{\kappa^2_{\tt b}} 
 = \frac{\Lambda}{\kappa^2} 
          + \delta\left( \frac{\Lambda}{\kappa^2}\right)
 \,,\qquad
\label{bare gravitational action: couplings}
\end{equation}
where $\kappa^2 = 16\pi G$ and $\Lambda$ are the tree level gravitational coupling constant and cosmological constant ($G$ is the tree-level Newton constant), respectively,
and $\delta \left(\frac{1}{\kappa^2}\right)$ 
and $\delta\left( \frac{\Lambda}{\kappa^2}\right)$ are the corresponding 
Ricci scalar and cosmological constant counterterm couplings. 
When Eq.~(\ref{bare gravitational action: couplings}) is inserted into Eq.~(\ref{bare gravitatinal action}),
the bare gravitational action $S_{\tt g}[g_{\mu\nu}]$ splits into the tree-level and counterterm action,
$S_{\tt g}[g_{\mu\nu}] = S_{\tt HE}[g_{\mu\nu}] +S_{\tt HE}^{\rm ct}[g_{\mu\nu}] $, where
\begin{eqnarray}
    S_{\tt HE}[g_{\mu\nu}] 
    &\!\!=\!\!&  \frac{1}{\kappa^2}\int {\rm d}^Dx \sqrt{-g} 
    \big[R - 2\Lambda\big]
\,,\qquad
\label{tree kevel gravitational action}\\
    S_{\tt HE}^{\rm ct}[g_{\mu\nu}] 
   &\!\!=\!\!&  \int {\rm d}^Dx \sqrt{-g} 
    \left[\delta\bigg(\frac{1}{\kappa^2}\bigg)R - 2\delta\bigg(\frac{\Lambda}{\kappa^2}\bigg)\right]
\,.\qquad
\label{counterterm gravitational action}
\end{eqnarray}
In the renromalization procedure we shall also use the following higher dimensional counterterm gravitational
action, 
\begin{equation}
    S_{\rm Riem^2}[g_{\mu\nu}] = \delta\alpha_{\tt Riem^2} \int {\rm d}^Dx \sqrt{-g} 
      R_{\alpha\beta\gamma\delta}R^{\alpha\beta\gamma\delta}  
\,,\qquad
\label{Riemann squared counterterm action}
\end{equation}
where $R_{\alpha\beta\gamma\delta}$ denotes the Riemann curvature tensor and $\delta\alpha_{\tt Riem^2}$
is the corresponding coupling constant.
For simplicity, 
we shall assume that the observed cosmological constant vanishes in the Minkowski spaces that surround the wormhole.


\subsection{Canonical quantization}
\label{Canonical quantization}

We are interested in canonical quantization in
the background of a classical topological wormhole,
whose $D$-dimensional geometry~\footnote{We consider a classical topological wormhole in general $D$ spacetime dimensions, needed for dimensional regularization which we shall use to renormalize the divergences of the one-loop energy momentum tensor. After renormalization one can set $D=4$.} is given by 
$\mathbb{R}_t \times \mathbb{R}^{D-3} \times S^2$, 
and whose classical metric is, 
\begin{equation}
{\rm d}s^2 = -{\rm d}t^2 + {\rm d}\mathbf{z}\!\cdot\!{\rm d}\mathbf{z}
+ a^2 {\rm d}\Omega_2^2
\,,\qquad
\label{flat metric}
\end{equation}
where $a$ is radius of the wormhole, and
${\rm d}\Omega_2^2={\rm d}\vartheta^2
           + \sin^2(\vartheta) {\rm d}\varphi^2$
is the metric on the two-dimensional sphere. 

The classical matter action~(\ref{classical matter action}) implies a canonical momentum,
$\pi(x) = \delta S_\phi/\delta \partial_0\phi(x)
 = - \sqrt{-g}g^{0\nu}(\partial_\nu\phi)$, which in a flat space background simplifies to, $\pi(x) = \dot \phi(x)$,
 where $\dot X \equiv \frac{{\rm d}}{{\rm d}t}X$.
 Canonical quantization is exacted by, 
\begin{equation}
   \bigl[\hat\phi(t,\mathbf{x}),\hat\pi(t,\mathbf{x}')\bigr] 
     = i\hbar\delta^{D-1}(\mathbf{x}\!-\!\mathbf{x}')
\,.\qquad
\label{canonical quantization}
\end{equation}
Varying Eq.~(\ref{classical matter action}) with respect to $\phi(x)$
yields a Klein-Gordon equation for the quantum field
$\hat\phi$,
\begin{equation}
(\Box_x - m^2)\hat\phi(x) = 0
\,,\qquad 
\label{Klein-Gordon equation: quantum field}
\end{equation}
where 
$\Box = \frac{1}{\sqrt{-g}}\partial_\mu\sqrt{-g}g^{\mu\nu}\partial_\nu$ is the d'Alembertian (wave) operator
for curved spacetimes, which for the metric
in Eq.~(\ref{flat metric}) simplifies to,
\begin{equation}
\Box_x = -\partial_t^2 + \Delta_{D-3} 
   + \frac{1}{a^2}\Delta_{S^2}
\,,\qquad
\end{equation}
Where here 
\begin{equation}
   \Delta_{S^2} = \frac{1}{\sin(\vartheta)}\frac{\partial}{\partial\vartheta}\sin(\vartheta)\frac{\partial}{\partial\vartheta}
+\frac{1}{\sin^2(\vartheta)}\partial_\varphi^2 
\,,\qquad
\label{wave operator on sphere}
\end{equation}
is the wave operator on the sphere $S^2$ and 
$\Delta_{D-3}$ is Laplace operator on the $(D\!-\!3)$-dimensional flat Euclidean space, which in Cartesian coordinates reads,
\begin{eqnarray}
\Delta_{D\!-\!3} 
 &\!\!=\!\!& \delta^{ab}_{D-3}\frac{\partial}{\partial x^a}
    \frac{\partial}{\partial x^b}
\,,\qquad
\label{wave operator on (D-4)-sphere}
\end{eqnarray}
where $\delta^{ab}_{D\!-\!3}$ is a flat metric on the
$(D\!-\!3)$-dimensional sub-space.

Taking account of the background symmetries of the metric~(\ref{flat metric}), 
the field operator can be decomposed in terms of mode functions 
$u_{\ell m}(t,\mathbf{k})$ as,
\begin{eqnarray}
\hat\phi(t,\mathbf{z},\vartheta,\varphi) 
&\!\!=\!\!& \sum_{\ell m}\int \frac{{\rm d}^{D\!-\!3}k}{(2\pi)^{D-3}}
\Big({\rm e}^{i \mathbf{k}\cdot\mathbf{z}}
   u_{\ell m}(t,\mathbf{k}) \hat{a}_{\ell m}(\mathbf{k})
   Y_{\ell m}(\vartheta,\varphi)
\nonumber\\
&&\hskip 3.cm
+\,{\rm e}^{- i \mathbf{k}\cdot\mathbf{z}}
     u^*_{\ell m}(t,\mathbf{k}) 
              \hat{a}^\dagger_{\ell m}(\mathbf{k}) 
              Y^*_{\ell m}(\vartheta,\varphi)
 \Big)
 \,,\qquad\;
\label{field decomposition}
\end{eqnarray}
where $\hat{a}^\dagger_{\ell m}(\mathbf{k})$
and $\hat{a}_{\ell m}(\mathbf{k})$ 
and the creation and annihilation operators which 
satisfy, 
\begin{eqnarray}
  \big[\hat{a}_{\ell m}(\mathbf{k}),
  \hat{a}^\dagger_{\ell' m'}(\mathbf{k}')\big] 
  &\!\!=\!\!& (2\pi)^{D-3}
  \delta^{D-3}(\mathbf{k}\!-\!\mathbf{k}')
  \delta_{\ell \ell'}\delta_{mm'}
\nonumber\\
  \big[\hat{a}_{\ell m}(\mathbf{k}),
  \hat{a}_{\ell' m'}(\mathbf{k}')\big] 
  &\!\!=\!\!& 0 
  = \big[\hat{a}^\dagger_{\ell m}(\mathbf{k}),
         \hat{a}^\dagger_{\ell' m'}(\mathbf{k}')\big] 
\,,\qquad
\label{commutation relation a and a+}
\end{eqnarray}
such that
$\hat{a}_{\ell m}(\mathbf{k})|\Omega\rangle = 0$
destroys the vacuum $|\Omega\rangle$ and 
$\hat{a}^\dagger_{\ell m}(\mathbf{k})$ creates a particle
with quantum numbers $\{\mathbf{k},\ell,m\}$, where 
$\ell\in\{0,1,2,\cdots\}$, $-\ell\leq m\leq \ell$
and $\mathbf{k}\in \mathbb{R}^{D-3}$. 
Furthermore, $Y_{\ell m}(\vartheta,\varphi)$ are the 
usual spherical harmonics, and
$ u_{\ell m}(t,\mathbf{k})$ and 
$u^*_{\ell m}(t,\mathbf{k})$ are the mode functions
of the problem 
that obey, 
\begin{equation}
 \left(\frac{{\rm d}^2}{{\rm d}t^2}+\omega_\ell^2\right)
    u_{\ell m}(t,\mathbf{k}) = 0
\,,\qquad
\Big(\omega_\ell^2 = \|\mathbf{k}\|^2 + m^2 
  +\frac{\ell(\ell+1)}{a^2}\Big)
\,.\qquad
\label{mode function eom}
\end{equation}
One can then show that canonically normalized mode functions
are,
\begin{equation}
  u_{\ell m}(t,\mathbf{k}) 
  = \frac{1}{a}\sqrt{\frac{\hbar}{2\omega_\ell}}
   {\rm e}^{-i\omega_\ell t}
\,,\qquad
  u^*_{\ell m}(t,\mathbf{k}) 
  = \frac{1}{a}\sqrt{\frac{\hbar}{2\omega_\ell}}
   {\rm e}^{i\omega_\ell t}
\,.\qquad
\label{canonically normalized mode functions}
\end{equation}
It is useful to recall that spherical harmonics
are eigenfunctions of the Laplace operator $ \Delta_{S^2}$,
\begin{equation}
  \Delta_{S^2} Y_{\ell m}(\vartheta,\varphi) 
     = -\ell(\ell+1) Y_{\ell m}(\vartheta,\varphi) 
\,,\qquad
\end{equation}
they are orthonormal polynomials on the sphere,
\begin{equation}
\int_0^\pi {\rm d}\vartheta \sin(\vartheta)
\int_0^{2\pi} {\rm d}\varphi\,
Y_{\ell m}(\vartheta,\varphi)
\,Y^*_{\ell' m'}(\vartheta,\varphi)
  = \delta_{\ell\ell'}\delta_{mm'}
\,,\qquad
\label{spherical harmonics: orthonormal}
\end{equation}
and they satisfy a completeness relation,
\begin{equation}
\sum_{\ell,m}  Y_{\ell m}(\vartheta,\varphi)
Y^*_{\ell m}(\vartheta',\varphi')
  = \delta\big(\cos(\vartheta)-\cos(\vartheta')\big)
     \delta(\varphi-\varphi')
  = \frac{\delta(\vartheta-\vartheta')}{\sin(\vartheta)}    
     \delta(\varphi-\varphi')
\,,\qquad
\label{spherical harmonics: completeness relations}
\end{equation}
the coincidence limit of which is, 
\begin{eqnarray}
\sum_{m}  Y_{\ell m}(\vartheta,\varphi)
Y^*_{\ell m}(\vartheta,\varphi)
  &\!\!=\!\!&
  \frac{2\ell+1}{4\pi}
\nonumber\\
&& \hskip -4cm 
\;\Longrightarrow \;
\sum_{\ell,m}  Y_{\ell m}(\vartheta,\varphi)
Y^*_{\ell m}(\vartheta,\varphi)  
= \sum_{\ell}  \frac{2\ell+1}{4\pi}
= \frac{2\zeta(-1)+\zeta(0)+1}{4\pi}
\,,\qquad
\label{spherical harmonics: completeness relation 2}
\end{eqnarray}
such that the sum over $\ell$ diverges.
The last equality expresses the divergent sums in terms of 
the Riemann zeta functions, defined as 
$\zeta(z)=\sum_{n=1}^\infty\frac{1}{n^z}$.

\subsection{One-loop energy-momentum tensor}
\label{One-loop energy-momentum tensor}

The classical energy-momentum tensor associated with 
the scalar field $\phi$ is defined as, 
\begin{equation}
    T^{\mu\nu}(x) = \frac{2}{\sqrt{-g}}
     \frac{\delta S_\phi}{\delta g_{\mu\nu}(x)}
     = (\nabla^\mu \phi)(\nabla^\nu \phi) 
     + g^{\mu\nu}
\left(\!- \frac{1}{2} g^{\alpha\beta} 
 (\partial_\alpha \phi) (\partial_\beta \phi) 
- \frac{1}{2} m^2 \phi^2 \right)
\,,\qquad
\label{classical energy-momentum tensor}
\end{equation}
where $S_\phi$ is the classical action in 
Eq.~(\ref{classical matter action}), $\nabla^\mu \phi=g^{\mu\nu}\partial_\nu \phi$,
and $\phi$ satisfies
the classical equation of motion
({\it cf.} Eq.~(\ref{Klein-Gordon equation: quantum field})),
\begin{equation}
(\Box_x - m^2)\phi(x) = 0
\,,\qquad 
\label{Klein-Gordon equation: classical field}
\end{equation}
where $\phi(x)=\langle\hat\phi(x)\rangle$,
$\langle\hat\phi(x)\rangle 
= {\rm Tr}\big[\hat\rho(t)\hat\phi(x)\big]$ 
and $\hat\rho(t)$ is the density operator.
For the purpose of the paper, the density operator 
corresponds to the pure vacuum state $|\Omega\rangle$, and 
therefore it can be written as,
\begin{equation}
   \hat\rho(t)=|\Omega\rangle\langle\Omega|
\,.\qquad
\label{density operator: pure vacuum state}
\end{equation}

In this paper we are not interested in studying classical 
gravitational backreaction generated by the energy-momentum tensor
in Eq.~(\ref{classical energy-momentum tensor}), as that depends on
the classical field $\phi(x)$, which satisfies 
Eq.~(\ref{Klein-Gordon equation: classical field}), but otherwise it is 
not constrained. Instead, we consider here the quantum backreaction induced by the one-loop contribution 
to the energy-momentum tensor, which is defined by,
\begin{eqnarray}
    \big\langle\Delta \hat{T}^{\mu\nu}(x)\big\rangle
     &\!\!=\!\!& \big\langle T^* \big[\nabla^\mu\hat\phi(x)
               \nabla^\nu\hat\phi(x)\big]\big\rangle
\nonumber\\
  &&\hskip -0.cm   
  +\, g_{\mu\nu}
\left(\!- \frac{1}{2} g^{\alpha\beta} 
 \big\langle T^* \big[\partial_\alpha\hat\phi(x)
               \partial_\beta\hat\phi(x)\big]\big\rangle
\!- \frac{1}{2} m^2 \big\langle T^* \big[\hat\phi^2(x)\big] \big\rangle  \right)
\label{one loop energy-momentum tensor: T*}\\
     &\!\!=\!\!& \left[\nabla^\mu {\nabla'}^\nu i\Delta_\phi(x;x')
       \right]_{x'\rightarrow x} 
\nonumber\\
  &&\hskip -0.cm   
  +\, g^{\mu\nu}
\left(\!- \frac{1}{2} g^{\alpha\beta} 
 \left[\partial_\alpha\partial_\beta' i\Delta_\phi(x;x')
 \right]_{x'\rightarrow x}
- \frac{1}{2} m^2 i\Delta_\phi(x;x)  \right)
\label{one loop energy-momentum tensor: T*2}
\,,\qquad
\end{eqnarray}
where $i\Delta_\phi(x;x')$ is the Feynman (time-ordered) propagator,
defined by,
\begin{equation}
 i\Delta_\phi(x;x') 
   = \Theta(t\!-\!t')\langle \hat\phi(x)\hat\phi(x') \rangle
   +\Theta(t'\!-\!t)\langle \hat\phi(x')\hat\phi(x) \rangle
\,,\qquad
\label{Feynman propagator}
\end{equation}
which obeys,
\begin{equation}
 \sqrt{-g}\big(\Box - m^2\big)i\Delta_\phi(x;x') = i\hbar\delta^D(x\!-\!x')
\,.\qquad
\label{Feynman propagator: eom}
\end{equation}
To get the second equality~(\ref{one loop energy-momentum tensor: T*2}) we used a $T^*$ ordering,
according to which the Heaviside $\Theta$ functions in the time ordering operator $T^*$ commute with the vertex derivatives. One can show that the difference between the more-usual T-ordering prescription and 
the $T^*$-ordering used here are ultralocal terms containing Dirac $\delta$ functions 
evaluated at spacetime coincidence and/or derivatives acting on $\delta$ functions evaluated at spacetime coincidence, which all vanish in dimensional regularization used here. Our considerations in the Appendix  imply that, for the purposes of this work, the difference between the $T^*$- and $T$-ordering is immaterial.

Here we are interested in studying the 
backreaction generated by the one-loop energy momentum tensor of the quantum scalar field $\hat\phi(x)$, which are universal
in the sense that they are completely specified once the background metric is fixed.~\footnote{This is only true if the quantum fluctuations are evaluated in a vacuum which respects the symmetries of the background metric, which is what we assume here. This is however not always possible. For example, in
de Sitter space quantum fluctuations of a massless scalar field violate 
de Sitter symmetry even at the tree level~\cite{Allen:1987tz,Onemli:2002hr}, 
which is then manifested in secular contributions 
to the propagator which grow in time. Quite unexpectedly, an analogous violation of de Sitter symmetry
occurs in the gauge sector of the photon field in de Sitter space in general covariant 
gauges~\cite{Glavan:2022dwb,Glavan:2022nrd}.}
The self-consistent semiclassical metric
can then be obtained from the semiclassical Einstein equation,
\begin{equation}
    G^{\mu\nu} = \frac{\kappa^2}{2}\left( T^{\mu\nu}+
    \big\langle\Delta \hat{T}^{\mu\nu}(x)\big\rangle 
      +{T}_{\tt ct}^{\mu\nu}(x) \right)
\label{semiclassical Einstein equation}
\,,\qquad
\end{equation}
where $G^{\mu\nu}$ denotes the Einstein curvature tensor, 
$T^{\mu\nu}$ is the classical energy-momentum 
tensor  in Eq.~(\ref{energy-momentum tensor: topological wormhole})
needed to support the wormhole, its quantum contribution is given 
in Eq.~(\ref{one loop energy-momentum tensor: T*2}), and the corresponding counterterm contribution
is defined as,
\begin{equation}
  {T}_{\tt ct}^{\mu\nu}(x)=\frac{2}{\sqrt{-g}}
  \frac{\delta S_{\tt ct}}{\delta g_{\mu\nu}(x)} 
\,,\qquad
\label{counterterm action}
\end{equation}
where $S_{\tt ct}$ is a sum of the counterterm actions 
in Eqs.~(\ref{counterterm matter action}),
(\ref{counterterm gravitational action}) and~(\ref{Riemann squared counterterm action}).
In this work we assume that classical scalar field $\phi(x)$ vanishes, and therefore 
we shall not need the matter counterterm action Eq.~(\ref{counterterm matter action})
to renormalize the one-loop energy-momentum tensor in Eq.~(\ref{one loop energy-momentum tensor: T*2}).

Solving self-consistently the system of equations of 
semiclassical gravity in Eqs.~(\ref{Feynman propagator: eom})
and~(\ref{semiclassical Einstein equation})
is a hard problem and it cannot be solved in general.
Instead, 
we shall resort to perturbative methods and construct a perturbative solution, according to which
the right-hand-side of Eq.~(\ref{semiclassical Einstein equation}) 
can be viewed as the source for the left-hand-side. In this scheme,
the metric, the Einstein tensor, the propagator, and the energy-momentum tensors can all be
expanded in powers of $\hbar$ as, 
\begin{eqnarray}
    g_{\mu\nu} &\!\!=\!\!& g_{\mu\nu}^{(0)} + g_{\mu\nu}^{(1)}
        + g_{\mu\nu}^{(2)} + \mathcal{O}\big(\hbar^3\big)
\,,\qquad
\label{perturbative expansion: metric}\\
    G^{\mu\nu} &\!\!=\!\!& {G^{\mu\nu}}^{(0)}  + {G^{\mu\nu}}^{(1)} 
        + {G^{\mu\nu}}^{(2)}  + \mathcal{O}\big(\hbar^3\big)
\,,\qquad
\label{perturbative expansion: Einstein tensor}\\
   i\Delta_\phi(x;x')  &\!\!=\!\!& 
     {i\Delta_\phi}^{(1)}(x;x') + 
     {i\Delta_\phi}^{(2)}(x;x')
     + \mathcal{O}\big(\hbar^3\big)
\,,\qquad
\label{perturbative expansion: propagator}\\
    T^{\mu\nu}_{\rm ct} &\!\!=\!\!& {T^{\mu\nu}_{\rm ct}}^{(1)} 
     + {T^{\mu\nu}_{\rm ct}}^{(2)} + \mathcal{O}\big(\hbar^3\big)
\,,\qquad
\label{perturbative expansion: Tmn ct}\\
    \big\langle\Delta \hat T^{\mu\nu}\big\rangle 
     &\!\!=\!\!& {\big\langle\Delta \hat T^{\mu\nu}\big\rangle}^{(1)} 
     + {\big\langle\Delta \hat T^{\mu\nu}\big\rangle}^{(2)} + \mathcal{O}\big(\hbar^3\big)
\,,\qquad
\label{perturbative expansion: Tmn quantum}
\end{eqnarray}
where $ g_{\mu\nu}^{(0)}$ is the classical wormhole metric
in Eq.~(\ref{flat metric}).
Since the source $\big\langle\Delta \hat{T}^{\mu\nu}(x)\big\rangle$
in Eq.~(\ref{semiclassical Einstein equation}) is the one-loop energy-momentum tensor, which is $\mathcal{O}\big(\hbar\big)$, the expansion in
Eq.~(\ref{perturbative expansion: metric}) is indeed an expansion in powers of $\hbar$. The equations of 
motion~(\ref{Feynman propagator: eom})--(\ref{semiclassical Einstein equation}) can be expanded as,
\begin{equation}
    {G^{\mu\nu}}^{(0)} = \frac{\kappa^2}{2}  {T}^{\mu\nu}
\,,\qquad
    {G^{\mu\nu}}^{(1)} = \frac{\kappa^2}{2}\left(
    {\big\langle\Delta \hat{T}^{\mu\nu}\big\rangle}^{(1)}  
      + {{T}_{\tt ct}^{\mu\nu}}^{(1)}\right)
\label{semiclassical Einstein equation: order hbar}
\,,\qquad
\end{equation}
and ({\it cf.} Eq.~(\ref{Feynman propagator: eom})),
\begin{eqnarray}
\sqrt{-g_{(0)}}
  \big(\Box_{(0)} - m^2\big){i\Delta_\phi}^{(1)}(x;x') 
   &\!\!=\!\!& i\hbar\delta^D(x\!-\!x')
\,,\quad 
\nonumber\\
\Box_{(0)}&\!\!=\!\!& \frac{1}{\sqrt{-g_{(0)}}}
  \partial_\mu\sqrt{-g_{(0)}} g_{(0)}^{\mu\nu} \partial_\nu 
\,,\qquad
\label{Feynman propagator: eom: order hbar}
\end{eqnarray}
and to leading order in $\hbar$ 
Eq.~(\ref{one loop energy-momentum tensor: T*2}) reads, 
\begin{eqnarray}
    \big\langle\Delta \hat{T}^{\mu\nu}(x)\big\rangle^{(1)}
     &\!\!=\!\!& \left[\partial_{(0)}^\mu {\partial'}^\nu_{(0)} i\Delta_\phi^{(1)}(x;x')
       \right]_{x'\rightarrow x} 
\nonumber\\
&&\hskip -0.5cm   
  +\, g_{(0)}^{\mu\nu}
\left(\!- \frac{1}{2} g_{(0)}^{\alpha\beta} 
 \left[\partial_\alpha\partial_\beta' i\Delta_\phi^{(1)}(x;x')
 \right]_{x'\rightarrow x}
\!- \frac{1}{2} m^2 i\Delta_\phi^{(1)}(x;x)  \right)
\,.\qquad
\label{energy-momentum: order hbar}
\end{eqnarray}
An interested reader can easily write down the relevant equations
at $\mathcal{O}(\hbar^2)$. In this work however, we limit ourselves
to studying the leading order solution $\mathcal{O}(\hbar)$ 
of the quantum backreaction problem, for which 
Eqs.~(\ref{semiclassical Einstein equation: order hbar})--(\ref{energy-momentum: order hbar})
suffice.

\medskip
In the Appendix we compute the one-loop energy-momentum
tensor in Eq.~(\ref{one loop energy-momentum tensor: T*2}),
and determine the counterterm action $S_{\tt ct}$ needed to 
renormalize it. It turns out that the Hilbert-Einstein counterterm action in 
Eq.~(\ref{counterterm gravitational action}) is sufficient to remove the divergences
in the one-loop energy momentum tensor and to renormalize it. However, our analysis shows
that one can also use a finite higher derivative counterterm action in 
Eq.~(\ref{Riemann squared counterterm action})
to further simplify the quantum backreaction.

With this in mind, we can write down nonvanishing components of 
Eq.~(\ref{semiclassical Einstein equation: order hbar}) as
(see Eqs.~(\ref{Einstein tensor}),  (\ref{energy-momentum tensor: topological wormhole})  
and~(\ref{T tt 11})--(\ref{T thetatheta 11})),
\begin{eqnarray}
    {G^{\,t}_t}^{(1)} &\!\!=\!\!& {G^{\,z}_z}^{(1)} 
=  \frac{\kappa^2}{2}\Big[ T_t^{\;t} \!+\!
    {\big\langle\Delta \hat{T}^{\,t}_t\big\rangle}_{\tt ren}^{(1)}  
\Big]
\nonumber\\
    &\!\!=\!\!&-\frac{1}{a^2} - \frac{\hbar \kappa^2}{128\pi^2}
\bigg\{\bigg(m^4\!-\!\frac{2m^2}{3a^2}\!+\!\frac{2}{15a^4}\bigg)
  \bigg[\ln\bigg(\frac{m^2 { - \frac{1}{4a^2}}}{4\pi\mu^2}\bigg)
             \!+\!\gamma_E-\frac32\bigg]
\nonumber\\
&&\hskip -0.cm
-\,\frac{1}{12a^2}\Big(m^2\!-\!\frac{31}{40a^2}\Big)
    { 
    \!-\!\frac{31}{4032a^6}\frac{1}{m^2\!-\!\frac{1}{4a^2}}
    }
\!+\!\mathcal{O}\Big(a^{-8}\,,{\rm e}^{-2\pi {am}}\Big)
\bigg\}
\nonumber\\
&&\hskip -0.cm
   +\,\frac{2\kappa^2}{a^4} \alpha_{\tt Riem^2} 
\!+\!\frac{\kappa^2}{\bar\kappa^2}\frac{1}{a^2}  \!-\!\Lambda
\,,\qquad
\label{sc Einstein equation: order hbar: tt}\\
    {G^{\,\vartheta}_\vartheta}^{(1)} &\!\!=\!\!&{G^{\,\varphi}_\varphi}^{(1)}
= \frac{\kappa^2}{2}
    {\big\langle\Delta \hat{T}^{\,\vartheta}_\vartheta\big\rangle}_{\tt ren}^{(1)}  
\nonumber\\
&\!\!=\!\!& -\frac{\hbar \kappa^2}{128\pi^2}
\bigg\{\bigg(m^4\!-\!\frac{2}{15a^4}\bigg)
  \bigg[\ln\bigg(\frac{m^2{ - \frac{1}{4a^2}}}{4\pi\mu^2}\bigg)
             \!+\!\gamma_E-\frac32\bigg]
\nonumber\\
&&\hskip -0.cm
+\,\frac{1}{4a^2}\Big(m^2\!-\!\frac{27}{40a^2}\Big)
    {  
           \!+\!\frac{457}{20160a^6}\frac{1}{m^2\!-\!\frac{1}{4a^2}}
    }
\!+\!\mathcal{O}\Big(a^{-8}\,,{\rm e}^{-2\pi {am}}\Big)
\bigg\}
\nonumber\\
&&\hskip -0.cm
            +\,\frac{\kappa^2}{a^4} \alpha_{\tt Riem^2}  
 \!-\! \Lambda
\,,\qquad
\label{sc Einstein equation: order hbar: thetatheta}
\end{eqnarray}
where $\bar\kappa^{-2}$, $\Lambda$ and $\alpha_{\tt Riem^2}$ are finite counterterm coefficients,  
and we dropped the $\mathcal{O}\big(1/(a^8)\big)$ terms.
The $\mu$-dependence in Eqs.~(\ref{sc Einstein equation: order hbar: tt})--(\ref{sc Einstein equation: order hbar: thetatheta}) was introduced by the counterterms needed to renormalize the theory. Indeed, making a different choice of the scale $\mu$, 
the quantum backreaction changes as, 
\begin{eqnarray}
 \frac{\kappa^2}{2}{\big\langle\Delta \hat{T}^{\,t}_t\big\rangle}_{\tt ren}^{(1)}(\mu')  
&\!\!=\!\!& \frac{\kappa^2}{2}{\big\langle\Delta \hat{T}^{\,t}_t\big\rangle}_{\tt ren}^{(1)}(\mu) 
+ \frac{\hbar \kappa^2}{64\pi^2}
\bigg(\!m^4\!-\!\frac{2m^2}{3a^2}\!+\!\frac{2}{15a^4}\!\bigg)
 \ln\bigg(\frac{{\mu'}}{\mu}\bigg)
\,,\qquad
\label{sc Einstein equation: order hbar: tt 2}\\
  \frac{\kappa^2}{2}
    {\big\langle\Delta \hat{T}^{\,\vartheta}_\vartheta\big\rangle}_{\tt ren}^{(1)}(\mu')   
&\!\!=\!\!&  \frac{\kappa^2}{2}
    {\big\langle\Delta \hat{T}^{\,\vartheta}_\vartheta\big\rangle}_{\tt ren}^{(1)}(\mu)   
+ \frac{\hbar \kappa^2}{64\pi^2}
\bigg(m^4\!-\!\frac{2}{15a^4}\bigg) \ln\bigg(\frac{{\mu'}}{\mu}\bigg)
\,.\qquad
\label{sc Einstein equation: order hbar: thetatheta 2}
\end{eqnarray}
This scale dependence of the quantum backreaction 
is also present in Minkowski space, when the cosmological constant backreaction is 
considered~\cite{Koksma:2011cq}.

A natural choice of the finite cosmological constant in 
Eqs.~(\ref{sc Einstein equation: order hbar: tt})--(\ref{sc Einstein equation: order hbar: thetatheta}) is that for which the cosmological constant in Minkowski space vanishes,
\begin{eqnarray}
 \Lambda &\!\!=\!\!&
                          - \frac{\hbar \kappa^2m^4}{128\pi^2}
  \bigg[\ln\bigg(\frac{m^2}{4\pi\mu^2}\bigg)
             \!+\!\gamma_E-\frac32\bigg]
\,,\qquad
\label{Lambda: vanishing backreaction in Mink}
\end{eqnarray}
with the price of introducing a fine tuning. 
Eqs.~(\ref{sc Einstein equation: order hbar: tt 2})--(\ref{sc Einstein equation: order hbar: thetatheta 2}) suggest that the fine tuning needs to be repeated for every choice of scale $\mu$.
This is however an artifact of the perturbative approximation~\cite{Lucat:2018slu}. 
Indeed, when the geometric cosmological constant is tuned 
to cancel the energy associated with minimum of the RG-improved effective potential of the 
theory, then fine tuning ought to be done once,
 thus resolving the problem of dependence of 
the quantum backreaction on choice of the renormalization scale $\mu$,
see Ref.~\cite{Lucat:2018slu} for more details.    

The considerations of Ref.~\cite{Lucat:2018slu} can be extended to the problem at hand
as follows.  The $\mu$-dependent fine tuning of 
$\Lambda$  in Eq.~(\ref{Lambda: vanishing backreaction in Mink})
which removes the perturbative one-loop quantum backreaction at some scale $\mu$, 
can be generalized to the RG-improved effective action of the theory 
such that the total cosmological constant remains zero in the surrounding 
Minkowski space. However, even when this is done, the wormhole will harbour
a nonvanishing quantum energy-momentum tensor,
\begin{eqnarray}
    {G^{\,t}_t}^{(1)} &\!\!=\!\!& {G^{\,z}_z}^{(1)} 
= \frac{\kappa^2}{2}\Big[ T_t^{\;t} \!+\!
    {\big\langle\Delta \hat{T}^{\,t}_t\big\rangle}_{\tt ren}^{(1)}  
\Big]
\nonumber\\
    &\!\!=\!\!&  \frac{1}{a^2}\Big(\frac{\kappa^2}{\bar\kappa^2}-1 \Big) 
\!+\!\frac{\hbar \kappa^2m^2}{192\pi^2a^2}
\bigg\{\bigg(1\!-\!\frac{1}{5(am)^2}\bigg)
  \bigg[\ln\bigg(\frac{m^2}{4\pi\mu^2}\bigg)
             \!+\!\gamma_E\bigg]
\nonumber\\
&&\hskip -0.2cm
-\,\Big(1\!-\!\frac{4}{105(am)^4}\Big)
\!+\!\mathcal{O}\Big(a^{-8}\,,{\rm e}^{-2\pi {am}}\Big)
\bigg\}+2\frac{\kappa^2}{a^4}\alpha_{\tt Riem^2}
\,,\qquad
\label{sc Einstein equation: order hbar: tt 3}\\
    {G^{\,\vartheta}_\vartheta}^{(1)} &\!\!=\!\!&{G^{\,\varphi}_\varphi}^{(1)}
= \frac{\kappa^2}{2}
    {\big\langle\Delta \hat{T}^{\,\vartheta}_\vartheta\big\rangle}_{\tt ren}^{(1)}  
\nonumber\\
&\!\!=\!\!& \frac{\hbar \kappa^2}{960\pi^2a^4}
\bigg\{\ln\bigg(\frac{m^2}{4\pi\mu^2}\bigg)
             \!+\!\gamma_E \!-\!\frac{8}{21(am)^2}\Big)
\!+\!\mathcal{O}\Big(a^{-8}\,,{\rm e}^{-2\pi {am}}\Big)
\bigg\} 
\nonumber\\
&& \hskip 0cm
+\,\frac{\kappa^2}{a^4}\alpha_{\tt Riem^2}
\,.\qquad
\label{sc Einstein equation: order hbar: thetatheta 3}
\end{eqnarray}
There are two more counterterms in Eqs.~(\ref{sc Einstein equation: order hbar: tt 3}) and~(\ref{sc Einstein equation: order hbar: thetatheta 3}) that remain unfixed, namely those of the Ricci scalar
and Riemann tensor squared.
One way of fixing the Ricci scalar
counterterm  is to fix the finite part of it to zero. In this case 
the term $\kappa^2/(a^2\bar\kappa^2)$ in Eq.~(\ref{sc Einstein equation: order hbar: tt 3}) 
vanishes, leaving us with the classical contribution $-1/a^2$ plus the quantum contribution
$\propto \hbar$. 
Another choice,
the one we will be using in this paper, is to fix $\bar\kappa^2$ to remove as much
of the quantum source in Eq.~(\ref{sc Einstein equation: order hbar: tt 3}) as possible,
but such that $\bar\kappa^2$ remains independent on the wormhole size $a$,

\begin{equation}
\frac{1}{\bar\kappa^2}= 
   -\frac{\hbar m^2}{192\pi^2}
\bigg[\ln\bigg(\frac{m^2}{4\pi\mu^2}\bigg)
             \!+\!\gamma_E-1\bigg]
\,.\qquad
\label{sc Einstein equation: order hbar: tt 3G}
\end{equation}
This choice represents a second fine tuning. Just like for the cosmological constant tuning in 
Eq.~(\ref{Lambda: vanishing backreaction in Mink}), the choice of the Newton constant counterterm
in Eq.~(\ref{sc Einstein equation: order hbar: tt 3G}) guarantees continuity of the Newton constant
at the boundary between the wormhole and the asymptotic Minkowski spaces shown in 
figure~\ref{figure: wormhole}, and from that point of view 
is not a fine tuning, but a good way of fixing the counterterms such that 
the tree-level Newton constant in Eq.~(\ref{tree kevel gravitational action}) 
corresponds to the observed one on large scales both in the Minkowski spaces and inside the wormhole.
An analogous choice of the Newton constant counterterm was used in Ref.~\cite{Jimu:2024xqm}
to prove a {\it gravitational decoupling theorem} between the ultraviolet and infrared massive scalar fluctuations
in Minkowski space,
constituting the gravitational version of the Appelquist-Carazzone decoupling theorem, 
originally proved for gauge fields in Ref.~\cite{Appelquist:1974tg}.

Finally, one can also fix the finite Riemann 
counterterm in Eqs.~(\ref{sc Einstein equation: order hbar: tt 3})--(\ref{sc Einstein equation: order hbar: thetatheta 3})
to remove as much as possible of the quantum energy-momentum tensor in 
Eq.~(\ref{sc Einstein equation: order hbar: thetatheta 3}),
\begin{equation}
\alpha_{\tt Riem^2} = - \frac{\hbar}{960\pi^2}
\bigg[\ln\bigg(\frac{m^2}{4\pi\mu^2}\bigg)
             \!+\!\gamma_E\bigg]
\,,\qquad
\label{alpha Riem2: fine tuning}
\end{equation}
representing yet another fine tuning. 
This fine tuning can be justified by physical considerations as follows. 
Upon inspecting the dependence on the radius $a$ of the quantum backreaction in 
Eqs.~(\ref{sc Einstein equation: order hbar: tt 3})--(\ref{sc Einstein equation: order hbar: thetatheta 3})
one sees that the choice in Eq.~(\ref{alpha Riem2: fine tuning}) corresponds to
choosing a vanishing coupling constant in the contribution to the effective action 
of the form given in Eq.~(\ref{Riemann squared counterterm action}).  
Adding a finite contribution in Eq.~(\ref{alpha Riem2: fine tuning}),
$\alpha_{\tt Riem^2} \rightarrow - [\hbar/(960\pi^2)]
\big[\ln\big(m^2/(4\pi\mu^2)\big)+\gamma_E\big]+\alpha$ would imply that 
the measured coupling constant in the higher derivative action in 
Eq.~(\ref{Riemann squared counterterm action}) is $\alpha$ both inside and outside the wormhole.
Since observable modifications of gravity induced by higher derivative terms such as the one in 
Eq.~(\ref{alpha Riem2: fine tuning}) are tiny in the infrared, our constraints 
on higher derivative couplings are very weak, see {\it e.g.} Ref.~\cite{Jimu:2024xqm}.~\footnote{
An exception is Starobinsky's inflation~\cite{Starobinsky:1980te}, 
in which the value of the $R^2$ coupling is constrained by
the normalisation of scalar cosmological perturbations to be $\sim 10^9$.
} 
Sensitivity to higher derivative terms inside the wormhole is much higher however,
and could be, in principle, used to measure the higher derivative coupling $\alpha$.
With these choices of the counterterms
the quantum sources in Eqs.~(\ref{sc Einstein equation: order hbar: tt 3}) simplify to,
\begin{eqnarray}
    {G^{\,t}_t}^{(1)} &\!\!=\!\!& {G^{\,z}_z}^{(1)} 
= \frac{\kappa^2}{2}\Big[ T_t^{\;t} \!+\!
    {\big\langle\Delta \hat{T}^{\,t}_t\big\rangle}_{\tt ren}^{(1)}  
\Big]
\nonumber\\
    &\!\!=\!\!& -\frac{1}{a^2}- \frac{\hbar \kappa^2}{320\pi^2a^4}
  \bigg[\ln\bigg(\frac{m^2}{4\pi\mu^2}\bigg) \!+\!\gamma_E\!-\!\frac{4}{63(am)^2}
  \bigg]
\nonumber\\
&& \hskip 0cm
    +\,\mathcal{O}\Big(a^{-8}\,,{\rm e}^{-2\pi {am}}\Big)
\,,\qquad
\label{sc Einstein equation: order hbar: tt 3G2}\\
    {G^{\,\vartheta}_\vartheta}^{(1)} &\!\!=\!\!&{G^{\,\varphi}_\varphi}^{(1)}
= \frac{\kappa^2}{2}
    {\big\langle\Delta \hat{T}^{\,\vartheta}_\vartheta\big\rangle}_{\tt ren}^{(1)}  
= - \frac{\hbar \kappa^2}{2520\pi^2a^6m^2}
\!+\!\mathcal{O}\Big(a^{-8}\,,{\rm e}^{-2\pi {am}}\Big)
\,.\qquad
\label{sc Einstein equation: order hbar: thetatheta 5}
\end{eqnarray}
Note that the remaining angular contribution is tiny in the limit of large wormholes ($am\gg 1$).
In what follows we shall investigate perturbative 
quantum backreaction induced by the general quantum one-loop 
energy-momentum tensor in Eqs.~(\ref{sc Einstein equation: order hbar: tt 3})
and~(\ref{sc Einstein equation: order hbar: thetatheta 3}). We shall also investigate the 
special case when the additional fine tuning in 
Eqs.~(\ref{sc Einstein equation: order hbar: tt 3G})--(\ref{alpha Riem2: fine tuning})
is exacted, such that the quantum energy-momentum tensor reduces to 
Eqs.~(\ref{sc Einstein equation: order hbar: tt 3G2})
and~(\ref{sc Einstein equation: order hbar: thetatheta 5}).

\section{Solving semiclassical Einstein's equations}
\label{Solving semiclassical Einstein's equations}

Here we investigate stability of the topological wormhole by solving the corresponding semiclassical Einstein's equation~(\ref{semiclassical Einstein equation: order hbar}) to the leading order in the Planck constant $\hbar$. Based on this analysis we arrive at a quantitative stability criterion for the semiclassical wormhole.
We also study how traversability of the wormhole is affected by the quantum backreaction.

\bigskip
\noindent
{\bf Time dependent Ansatz.} 
In what follows we consider the wormhole spacetime in the presence of 
the one-loop energy-momentum tensor in 
Eqs.~(\ref{sc Einstein equation: order hbar: tt 3})--(\ref{sc Einstein equation: order hbar: thetatheta 3}) generated  by the quantum fluctuations of a massive scalar 
inside the wormhole. We assume that the cosmological constant 
counterterm is chosen as in Eq.~(\ref{Lambda: vanishing backreaction in Mink}), 
which subtracts the vacuum energy of the Minkowski vacuum fluctuations.
This choice ensures that the vacuum fluctuations in the Minkowski space surrounding the 
wormhole do not contribute to the quantum energy-momentum tensor, which is supported by observations. For our considerations the presumed cosmological constant driving the observed accelerated expansion of the Universe can be neglected as long as $aH_0\ll 1$ (here 
$H_0\simeq 78~{\rm km/s/Mpc}$ is the present expansion rate of the Universe), 
which is amply satisfied. If, in addition, 
the Newton constant and the Riemann tensor counterterms are chosen as in 
Eqs.~(\ref{sc Einstein equation: order hbar: tt 3G}) and~(\ref{alpha Riem2: fine tuning}), 
the source in the semiclassical Einstein equation $\frac{\kappa^2}{2}\Big[ T_\mu^{\;\nu} +
    {\big\langle\Delta \hat{T}_\mu^{\,\nu}\big\rangle}_{\tt ren}^{(1)} \Big]$
simplifies to that in 
Eqs.~(\ref{sc Einstein equation: order hbar: tt 3G2})--(\ref{sc Einstein equation: order hbar: thetatheta 5}).
With these remarks in mind, we are now ready to consider the perturbative quantum
backreaction induced by 
Eqs.~(\ref{sc Einstein equation: order hbar: tt 3})--(\ref{sc Einstein equation: order hbar: thetatheta 3})
or by 
Eqs.~(\ref{sc Einstein equation: order hbar: tt 3G2})--(\ref{sc Einstein equation: order hbar: thetatheta 5}).

The source in Eqs.~(\ref{sc Einstein equation: order hbar: tt 3})--(\ref{sc Einstein equation: order hbar: thetatheta 3})
in the semiclassical Einstein's equations~(\ref{semiclassical Einstein equation: order hbar}) suggests the following
{\it time dependent} {\it Ansatz} for the metric,
\begin{equation}
{\rm d}s^2 = -{\rm d}t^2 +a_z^2(t){\rm d}z^2 + a_\perp^2(t){\rm d}\Omega_2^2
\,,\qquad
\label{sc metric ansatz}
\end{equation}
where   $a_z^2(t)$ and $a_\perp^2(t)$ are the scale factors associated with 
the longitudinal ($z$) and angular directions of the wormhole, respectively.
The corresponding nonvanishing Christoffel symbols 
implied by Eq.~(\ref{sc metric ansatz}) are then, 
\begin{eqnarray}
\Gamma^t_{\mu\nu} &\!\!\!=\!\!\!& \frac{1}{2}\partial_t g_{\mu\nu} 
\;\Rightarrow\; 
\Gamma^t_{zz} = a_z^2 H_z
\,,\quad 
\Gamma^t_{\vartheta\vartheta} = a_\perp^2 H_\perp
\,,\quad
\Gamma^t_{\varphi\varphi} = a_\perp^2 H_\perp\sin^2(\vartheta)
\,,\qquad 
\nonumber\\
\Gamma^z_{\mu\nu} &\!\!=\!\!& 
            \frac{1}{2a_z^2}\big(\partial_\mu g_{\nu z}\!+\!\partial_\nu g_{\mu z}\big) 
\;\Rightarrow\; 
\Gamma^z_{tz} =\Gamma^z_{zt}  = H_z
\,,\quad 
\nonumber\\
\Gamma^\vartheta_{\mu\nu} &\!\!=\!\!&  \frac{1}{2a_\perp^2}
 \big(2\partial_{(\mu} g_{\nu)\vartheta}\!-\!\partial_\vartheta g_{\mu\nu}\big) 
\,
\Rightarrow
\, 
\Gamma^\vartheta_{t\vartheta} =\Gamma^\vartheta_{\vartheta t}  = H_\perp
\,,\;\;
\Gamma^\vartheta_{\varphi\varphi} = -\sin(\vartheta)\cos(\vartheta)
\,,
\nonumber\\
\Gamma^\varphi_{\mu\nu} &\!\!=\!\!&  \frac{1}{a_\perp^2\sin^2(\vartheta)}
 \partial_{(\mu} g_{\nu)\varphi}
\;\Rightarrow \;
\Gamma^\varphi_{t\varphi} =\Gamma^\varphi_{\varphi t}  = H_\perp
\,,\quad \!\!
\Gamma^\varphi_{\vartheta\varphi} = \Gamma^\varphi_{\varphi\vartheta}
    =\frac{\cos(\vartheta)}{ \sin(\vartheta)}
\,,\;\,
\label{sc Christoffel symbols}
\end{eqnarray}
from which it follows that,
\begin{equation}
\Gamma^\mu_{\mu\alpha} = \frac{1}{\sqrt{-g}}\partial_\alpha\sqrt{-g}
\;\Rightarrow \;
\Gamma^\mu_{\mu t}   = H_z + 2H_\perp
\,,\quad 
\Gamma^\mu_{\mu\vartheta} =\frac{\cos(\vartheta)}{ \sin(\vartheta)}
\,,\qquad 
\label{sc Christoffel symbols 2}
\end{equation}
where $H_z=\dot a_z/a_z$ and $H_\perp=\dot a_\perp/a_\perp$.
We can now calculate nonvanishing elements of the Ricci tensor and Ricci scalar.
Notice first the Ricci tensor is diagonal, so its nonvanishing components are, 
\begin{eqnarray}
{R_{t}^{\;t}}^{(1)}  &\!\!=\!\!& \big(\dot H_z \!+\! H_z^2\big)\!+\!2\big(\dot H_\perp \!+\!H_\perp^2\big)
\,,\qquad
\nonumber\\
{R_{z}^{\;z}}^{(1)}  &\!\!=\!\!& \big(\dot H_z \!+\! H_z^2\big)\!+\!2 H_z H_\perp
\,,\qquad
\nonumber\\
{R_{\vartheta}^{\;\vartheta}}^{(1)} &\!\!=\!\!&R_{\varphi}^{\;\varphi} =H_z H_\perp\!+\!\big(\dot H_\perp\!+\! 2H_\perp^2\big)\!+\!\frac{1}{a_\perp^2}
\,,\qquad
\nonumber\\
R^{(1)}  &\!\!=\!\!&
g^{\mu\nu}R_{\mu\nu}^{(1)}
= 2\Big[  \big(\dot H_z \!+\! H_z^2\big)\!+\!2 H_z H_\perp\!+\!\big(2\dot H_\perp \!+\! 3H_\perp^2\big)\Big]
\!+\!\frac{2}{a_\perp^2}
\,,\qquad
\label{sc Ricci tensor}
\end{eqnarray}
such that the Einstein tensor and the corresponding semiclassical Einstein equations read, 
\begin{eqnarray}
{G_{t}^{\;t}}^{(1)} &\!\!=\!\!& -\Big[2H_z H_\perp \!+\!H_\perp^2 \Big]\!-\!\frac{1}{a_\perp^2}
     =\frac{\kappa^2}{2}\Big[T_t^{\;t}\!+\!{\big\langle\Delta \hat{T}^{\,t}_t\big\rangle}_{\tt ren}^{(1)}  \Big]
\,,\qquad
\label{sc Einstein tensor: tt}\\
{G_{z}^{\;z}}^{(1)} &\!\!=\!\!& -\Big[2\dot H_\perp \!+\!3H_\perp^2 \Big]\!-\!\frac{1}{a_\perp^2}
     =\frac{\kappa^2}{2}\Big[T_t^{\;t}\!+\!{\big\langle\Delta \hat{T}^{\,t}_t\big\rangle}_{\tt ren}^{(1)}  \Big]
\,,\qquad
\label{sc Einstein tensor: zz}\\
{G_{\vartheta}^{\;\vartheta}}^{(1)} &\!\!=\!\!& {G_{\varphi}^{\;\varphi}}^{(1)} 
      =-\Big[H_z H_\perp\!+\!\big(\dot H_z \!+\!H_z^2\big)\!+\!\big(\dot H_\perp \!+\!H_\perp^2\big)\Big]
     =\frac{\kappa^2}{2}{\big\langle\Delta \hat{T}^{\,\vartheta}_\vartheta\big\rangle}_{\tt ren}^{(1)}  
\,,\qquad
\label{sc Einstein tensor: thetatheta}
\end{eqnarray}
such that the ${G_{\varphi}^{\;\varphi}}^{(1)}$ equation contains no new information. 
The structure of these equations suggests to subtract the second equation from the first to obtain,
\begin{eqnarray}
\big(\dot H_\perp\!+\!H_\perp^2\big)\!-\! H_z H_\perp &\!\!=\!\!& 0
\,,\qquad
\label{sc Einstein tensor 2a}
\end{eqnarray}
from where we conclude, 
\begin{equation}
H_z =  \frac{\dot H_\perp}{H_\perp}\!+\!H_\perp
\,.\qquad
\label{sc Einstein tensor 3a}
\end{equation}
This can be used in Eq.~(\ref{sc Einstein tensor: tt}) to obtain,
\begin{equation}
\big(2\dot H_\perp \!+\!3H_\perp^2 \big)\!+\!\frac{1}{a_\perp^2}
     + \frac{\kappa^2}{2}\Big[T_t^{\;t}\!+\!{\big\langle\Delta \hat{T}^{\,t}_t\big\rangle}_{\tt ren}^{(1)}  \Big]
     = 0
\,.\qquad
\label{sc Einstein tensor: tt 2}
\end{equation}
This is a closed equation for $a_\perp(t)$ and it can be solved as follows. First notice that, for the quantum source in Eqs.~(\ref{sc Einstein tensor: tt}) and~(\ref{sc Einstein equation: order hbar: tt 3}), 
there exists a time independent solution,
\begin{eqnarray}
\frac{1}{a_\perp^2}  &\!\!=\!\!& \frac{1}{a^2}\Big(1-\frac{\kappa^2}{\bar\kappa^2}\Big)
 \!-\! \frac{\hbar \kappa^2m^2}{192\pi^2a^2}
\bigg\{\bigg(1\!-\!\frac{1}{5(am)^2}\bigg)
  \bigg[\ln\bigg(\frac{m^2}{4\pi\mu^2}\bigg)
             \!+\!\gamma_E\bigg]
\nonumber\\
&&\hskip 0.cm
{
-\,\Big(1\!-\!\frac{4}{105(am)^4}\Big)
}
\!+\!\mathcal{O}\Big(a^{-8}\,,{\rm e}^{-2\pi {am}}\Big)
\bigg\}-2\frac{\kappa^2}{a^4}\alpha_{\tt Riem^2}
\,,\qquad
\nonumber\\
 &\!\!=\!\!&  \frac{1}{a^2} +  \frac{\hbar \kappa^2}{320\pi^2a^4}
  \bigg[\ln\bigg(\frac{m^2}{4\pi\mu^2}\bigg)
             \!+\!\gamma_E { \!-\!\frac{4}{63 (am)^2}
                                  } 
                                  \bigg]
\!+\!\mathcal{O}\Big(a^{-8}\,,{\rm e}^{-2\pi {am}}\Big)
\,,\qquad
\label{sc Einstein tensor: tt 2b}
\end{eqnarray}
where the second equality holds when both $\bar\kappa$ 
and $\alpha_{\tt Riem^2}$ are fine-tuned to the values in
Eqs.~(\ref{sc Einstein equation: order hbar: tt 3G}) and~(\ref{alpha Riem2: fine tuning}),
see Eq.~(\ref{sc Einstein equation: order hbar: tt 3G2}).
A simple consideration shows that iterating Eq.~(\ref{sc Einstein tensor: tt 2b}) would result in a self-consistent solution for semiclassical gravity,
\begin{eqnarray}
\frac{1}{a_\perp^2}  &\!\!=\!\!& \frac{1}{a^2}\Big(1-\frac{\kappa^2}{\bar\kappa^2}\Big)
 \!-\! \frac{\hbar \kappa^2m^2}{192\pi^2a_\perp^2}
\bigg\{\bigg(1\!-\!\frac{1}{5(a_\perp m)^2}\bigg)
  \bigg[\ln\bigg(\frac{m^2}{4\pi\mu^2}\bigg)
             \!+\!\gamma_E\bigg]
\nonumber\\
&&\hskip 0.0cm
{
-\,\Big(1\!-\!\frac{4}{105(a_\perp m)^4}\Big)
}
\bigg\}\!+\!\mathcal{O}\Big(a_\perp^{-8}\,,{\rm e}^{-2\pi {a_\perp m}}\Big)
 - 2\frac{\kappa^2}{a_\perp^4}\alpha_{\tt Riem^2}
\nonumber\\
 &\!\!=\!\!&  \frac{1}{a^2} +  \frac{\hbar \kappa^2}{320\pi^2a_\perp^4}
  \bigg[\ln\bigg(\frac{m^2}{4\pi\mu^2}\bigg)
             \!+\!\gamma_E
              { \!-\!\frac{4}{63 (a_\perp m)^2}
                                  } 
                        \bigg]
\!+\!\mathcal{O}\Big(\frac{1}{a_\perp^8},{\rm e}^{-2\pi {a_\perp m}}\Big)
\,,\qquad\;
\label{sc Einstein tensor: tt 2b}
\end{eqnarray}
which, when solved, gives $a_\perp=a_\perp(a)$.
As before, the second equality holds for the fine tunings in 
Eqs.~(\ref{sc Einstein equation: order hbar: tt 3G}) and~(\ref{alpha Riem2: fine tuning}).
The conjecture in Eqs.~(\ref{sc Einstein tensor: tt 2b}) ought to be 
checked by constructing a self-consistent solution of semiclassical gravity.

Notice that Eq.~(\ref{sc Einstein tensor: tt 2}) allows for a time dependent solutions for $a_\perp(t)$.
To better understand the form of these solutions, it is useful to  introduce a time variable, 
$\sqrt{a_\perp(t(\tau))}{\rm d}\tau = {\rm d}t$, in terms of which Eq.~(\ref{sc Einstein tensor: tt 2}) becomes,
\begin{equation}
a_\perp^{''}(\tau) +\frac{1}{2}
     + a_\perp^2\frac{\kappa^2}{4}\Big[T_t^{\;t}\!+\!{\big\langle\Delta \hat{T}^{\,t}_t\big\rangle}_{\tt ren}^{(1)}  \Big]
     = 0
\,,\qquad
\label{sc Einstein tensor: tt 3}
\end{equation}
which is solved by the Weierstrass elliptic function $\mathcal{P}$ (related to the Jacobi elliptic function), 
\begin{equation}
a_\perp(\tau)  = \frac{1}{\alpha^{1/3}}\!\times\!\mathcal{P}\Big(\alpha^{1/3}\tau,\Big\{\frac{1}{\alpha^{1/3}},C\Big\}\Big)
\,,\quad \alpha=  -\frac{3\kappa^2}{2}\Big[T_t^{\;t}\!+\!{\big\langle\Delta \hat{T}^{\,t}_t\big\rangle}_{\tt ren}^{(1)}  \Big]
\,,\;
\label{ Weierstrass P-function}
\end{equation}
where $C$ is an integration constant (the other integration constant is absorbed in the definition of time).

Next, Eq.~(\ref{sc Einstein tensor: thetatheta}) can be recast as, 
\begin{eqnarray}
\big(\dot H_z \!+\!H_z^2\big)\!+\!2\big(\dot H_\perp \!+\!H_\perp^2\big)
     +\frac{\kappa^2}{2}{\big\langle\Delta \hat{T}^{\,\vartheta}_\vartheta\big\rangle}_{\tt ren}^{(1)}  
=0
\,,\qquad
\label{sc Einstein tensor: thetatheta 3}
\end{eqnarray}
which for $a_\perp = {\rm const.}$ simplifies to, 
\begin{equation}
\big(\dot H_z \!+\!H_z^2\big) = \frac{\ddot a_z}{a_z}
     = -\frac{\kappa^2}{2}{\big\langle\Delta \hat{T}^{\,\vartheta}_\vartheta\big\rangle}_{\tt ren}^{(1)}  
\,,\qquad
\label{sc Einstein tensor: thetatheta 4}
\end{equation}
which is solved by,
\begin{eqnarray}
 a_z(t) =   a_{z(0)} {\cosh}(H_z t)
\,,\quad \;
H_z^2 &=& -\frac{\kappa^2}{2}{\big\langle\Delta \hat{T}^{\,\vartheta}_\vartheta\big\rangle}_{\tt ren}^{(1)}  
  \simeq  
  {
  \frac{2\hbar G}{315\pi a^6m^2} > 0
}
\,,\qquad
\label{sc Einstein tensor: thetatheta 5}
\end{eqnarray}
where $a_{z(0)}= a_z(0)$, where the value chosen here is that of the fine-tuned
solution in Eq.~(\ref{sc Einstein equation: order hbar: thetatheta 5}).~\footnote{  
Had we not included the higher dimensional Riemann tensor squared counterterm
(corresponding to the choice $\alpha_{\rm Riem^2}=0$), then  the pressure in 
Eq.~(\ref{sc Einstein tensor: thetatheta 5}) would be that implied by 
Eq.~(\ref{sc Einstein equation: order hbar: thetatheta 3}), 
which is positive,
\begin{equation}
p_\perp  \approx \frac{\hbar}{480\pi^2a^4}
\bigg\{\ln\bigg(\frac{m^2}{4\pi\mu^2}\bigg)
             \!+\!\gamma_E \!-\!\frac{8}{21(am)^2}\Big)
\bigg\} 
\,.\qquad
\label{sc Einstein equation: order hbar: thetatheta 3 bis} 
\end{equation}
In this case the metric would be static, and the corresponding quantum backreaction could be studied 
in the context of the static metric {\it Ansatz} considered in the following subsection.
}
The solution in Eq.~(\ref{sc Einstein tensor: thetatheta 5})
is chosen such that it is symmetric in $t\rightarrow -t$.
At late times, the exponentially decaying solution in Eq.~(\ref{sc Einstein tensor: thetatheta 5}) 
can be neglected, and the scale factor can be approximated by, 
$a_z(t) =   [a_{z(0)}/2] {\rm e}^{H_z t}$. 
This means that, due to the negative angular pressure { 
$p_\perp\simeq - \hbar/(1250\pi^2 a^6m^2)<0$ 
}
generated by the one-loop vacuum fluctuations inside the wormhole, the wormhole's length (along the $z$-direction)  will tend to expand exponentially.  
The topological wormhole harbors a cosmological horizon along the $z$-direction, $R_z = 1/H_z$, such that for distances larger than 
$R_z$ the wormhole will not be be traversable, {\it i.e.} it will expand faster than the speed of light, as its physical size will grow as, 
\begin{equation}
                L(t)   = a_z(t) L_0  = L_0 {\rm e}^{H_zt}
\,,\qquad
\label{wormhole's physical size}
\end{equation}
which means that, even a relatively short wormhole, whose initial length $L_0\ll R_z$, will grow after some time $\Delta t\simeq R_z \ln(R_z/L_0)$ into 
a Hubble-sized wormhole, thus becoming non-traversable. 
One ought to keep in mind that this conclusion is reached based on 
a perturbative solution of the semiclassical equations. Furthermore, 
as we argue below, 
the exponentially expanding solution is credible only if the initial length of the 
wormhole is sufficiently large, {\it i.e.} if $L_0>1/H_z$, which will typically not be satisfied for the tiny quantum backreaction in Eq.~(\ref{sc Einstein equation: order hbar: thetatheta 5}).

To complete the analysis we need to show that the solutions given in Eqs.~(\ref{sc Einstein tensor: tt 2}) and~(\ref{sc Einstein tensor: thetatheta 5})
are the physical solutions. Recall that the equation for $a_\perp$ also harbours a time-dependent solution given in Eq.~(\ref{ Weierstrass P-function}).
We will now argue that that is not a physical solution. Note first that Eq.~(\ref{sc Einstein tensor: thetatheta}) can be combined with 
Eq.~(\ref{sc Einstein tensor 3a}) (and its derivative) to yield,
\begin{equation}
\frac{\ddot H_\perp}{H_\perp} + 5\dot H_\perp + 3H_\perp^2 
   +\frac{\kappa^2}{2}{\big\langle\Delta \hat{T}^{\,\vartheta}_\vartheta\big\rangle}_{\tt ren}^{(1)}  =0
\,.\qquad
\label{sc Einstein tensor: thetatheta 6}
\end{equation}
One can subtract from this a time derivative of Eq.~(\ref{sc Einstein tensor: tt 2}) (divided by $2H_\perp$) to obtain,
\begin{equation}
2\dot H_\perp + 3H_\perp^2 \!+\!\frac{1}{a_\perp^2}
   +\frac{\kappa^2}{2}{\big\langle\Delta \hat{T}^{\,\vartheta}_\vartheta\big\rangle}_{\tt ren}^{(1)}  =0
\,,\qquad
\label{sc Einstein tensor: thetatheta 6}
\end{equation}
which is consistent with Eq.~(\ref{sc Einstein tensor: tt 2}) only in the isotropic case, in which  
${\big\langle\Delta \hat{T}^{\,\vartheta}_\vartheta\big\rangle}_{\tt ren}^{(1)} = T_t^{\;t}+{\big\langle\Delta \hat{T}^{\,t}_t\big\rangle}_{\tt ren}^{(1)}$,
which is not the case for the topological wormhole. This therefore completes the proof that the unique solution satisfying all
three Einstein's equations~(\ref{sc Einstein tensor: tt})--(\ref{sc Einstein tensor: thetatheta}) is given in Eqs.~(\ref{sc Einstein tensor: tt 2}) and~(\ref{sc Einstein tensor: thetatheta 5}).
One should keep in mind that these solutions are perturbative, and therefore they are not reliable at late times. To get solutions that are reliable at late times,
one would have to solve the self-consistent semiclassical Einstein equations, in which the renormalized energy-momentum tensor is computed with 
the metric in Eq.~(\ref{sc metric ansatz}), which is beyond the scope of the present work.


\bigskip
\noindent
{\bf Static Ansatz.} Let us now consider the following metric {\it Ansatz} for the
semiclassical wormhole,
\begin{equation}
{\rm d}s^2 = -{\rm e}^{2\alpha(z)}{\rm d}t^2 +{\rm e}^{2\beta(z)}{\rm d}z^2 + a_\perp^2{\rm d}\Omega_2^2
\,.\qquad
\label{sc metric ansatz 2}
\end{equation}
Nonvanishing Christoffel symbols are,
\begin{eqnarray}
\Gamma^t_{zt} &\!\!=\!\!& \Gamma^t_{tz} = \alpha'(z) 
\,,\quad 
\Gamma^z_{tt} ={\rm e}^{2(\alpha-\beta)} \alpha' 
\,,\quad 
\Gamma^z_{zz} = \beta' 
\,,\quad 
\nonumber\\
\Gamma^\vartheta_{\varphi\varphi} &\!\!=\!\!& -\sin(\vartheta)\cos(\vartheta)
\,,\quad 
\Gamma^\varphi_{\vartheta\varphi} = \Gamma^\varphi_{\varphi\vartheta}
    =\frac{\cos(\vartheta)}{ \sin(\vartheta)}
\,,\qquad
\label{sc Christoffel symbols: z-dep ansatz}
\end{eqnarray}
and those of the Ricci tensor and Ricci scalar are, 
\begin{eqnarray}
R_{t}^{\;t} &\!\!=\!\!& R_{z}^{\;z} = -{\rm e}^{-2\beta}\left[ \alpha''+{\alpha'}^2-\alpha'\beta' \right] 
\,,\qquad
\nonumber\\
R_{\vartheta}^{\;\vartheta}&\!\!=\!\!&R_{\varphi}^{\;\varphi} =\frac{1}{a_\perp^2}
\,,\qquad
\nonumber\\
R &\!\!=\!\!&-2{\rm e}^{-2\beta}\left[ \alpha''+{\alpha'}^2-\alpha'\beta' \right] 
\!+\!\frac{2}{a_\perp^2}
\,,\qquad
\label{sc Ricci tensor: z-dep Ansatz}
\end{eqnarray}
such that the Einstein tensor and the corresponding semiclassical Einstein equations read, 
\begin{eqnarray}
G_{t}^{\;t} &\!\!=\!\!& G_{z}^{\;z}  = -\frac{1}{a_\perp^2}
     =\frac{\kappa^2}{2}\Big[T_t^{\;t}\!+\!{\big\langle\Delta \hat{T}^{\,t}_t\big\rangle}_{\tt ren}^{(1)}  \Big]
\,,\qquad
\label{sc Einstein tensor: zz: z-dep Ansatz}\\
G_{\vartheta}^{\;\vartheta}&\!\!=\!\!& G_{\varphi}^{\;\varphi} 
      ={\rm e}^{-2\beta}\left[ \alpha''+{\alpha'}^2-\alpha'\beta' \right] 
     =\frac{\kappa^2}{2}{\big\langle\Delta \hat{T}^{\,\vartheta}_\vartheta\big\rangle}_{\tt ren}^{(1)}  
\,.\qquad
\label{sc Einstein tensor: thetatheta: z-dep Ansatz}
\end{eqnarray}
Just as in the former {\it Ansatz} in Eq.~(\ref{sc Einstein equation: order hbar: tt 3}),
Eq.~(\ref{sc Einstein tensor: zz: z-dep Ansatz}) is  
solved by Eq.~(\ref{sc Einstein tensor: tt 2}), 
\begin{eqnarray}
\frac{1}{a_\perp^2}  &\!\!=\!\!& \frac{1}{a^2}\Big(1-\frac{\kappa^2}{\bar\kappa^2}\Big)
 \!-\! \frac{\hbar \kappa^2m^2}{192\pi^2a^2}
\bigg\{\bigg(1\!-\!\frac{1}{5(am)^2}\bigg)
  \bigg[\ln\bigg(\frac{m^2}{4\pi\mu^2}\bigg)
             \!+\!\gamma_E\bigg]
\nonumber\\
&&\hskip 2.cm
{
-\,\Big(1\!-\!\frac{4}{105(am)^4}\Big)
}
\!+\!\mathcal{O}\Big(a_\perp^{-6}\,,{\rm e}^{-2\pi {am}}\Big)
\bigg\}
- 2\frac{\kappa^2}{a^4}\alpha_{\tt Riem^2}
\,.\qquad
\label{sc Einstein tensor: tt 2b: z-dep Ansatz}
\end{eqnarray}
Notice that the second equation~(\ref{sc Einstein tensor: thetatheta: z-dep Ansatz}) can be solved by $\beta=-\alpha$, in which case 
Eq.~(\ref{sc Einstein tensor: thetatheta: z-dep Ansatz}) becomes ({\it cf.} Eq.~(\ref{sc Einstein tensor: thetatheta 5})),
\begin{equation}
{\rm e}^{2\alpha}\left[ \alpha''+2{\alpha'}^2\right] = \frac12\big({\rm e}^{2\alpha}\big)''  = -H_z^2
\,,\qquad 
     H_z^2 \equiv - \frac{\kappa^2}{2}{\big\langle\Delta \hat{T}^{\,\vartheta}_\vartheta\big\rangle}_{\tt ren}^{(1)}  ={\rm const.}
\,,\qquad
\label{sc Einstein tensor: thetatheta: z-dep Ansatz 2}
\end{equation}
whose general solution is,
\begin{equation}
{\rm e}^{2\alpha(z)} = -H_z^2z^2 + \omega_1 z+\omega_0
\,.\qquad 
\label{sc Einstein tensor: thetatheta: z-dep Ansatz 3}
\end{equation}
The two integration constants, $\omega_1$ and $\omega_0$, can be fixed by matching the wormhole metric onto the Minkowski metric outside the wormhole. 
Assuming that the wormhole length is $L_0$ and that the wormhole center is located at 
$z=0$, we have,
\begin{equation}
{\rm e}^{2\alpha(L_0/2)} = 1 =  {\rm e}^{2\alpha(-L_0/2)} 
\,,\qquad 
\label{sc Einstein tensor: thetatheta: z-dep Ansatz 4}
\end{equation}
from which it follows, $\omega_1=0$, $\omega_0 = 1+(H_z L_0/2)^2$, such that the metric in Eq.~(\ref{sc metric ansatz 2}) 
inside the wormhole 
becomes,
\begin{equation}
{\rm d}s^2 = -\Big[1+H_z^2\Big(\tfrac{L_0^2}{4}-z^2\Big)\Big]{\rm d}t^2 
  +\frac{{\rm d}z^2 }{1\!+\!H_z^2\big(\frac{L_0^2}{4}-z^2\big)} + a_\perp^2{\rm d}\Omega_2^2
\,,\quad\!\!\!\! \big(|z|\leq \tfrac{L_0}{2}\big) 
\,,\;
\label{sc metric ansatz 2: z-dep Ansatz}
\end{equation}
which at $r=|z|=L_0/2$ continuously matches onto the outside Minkowski metrics in 
Eq.~(\ref{metrics: flat space}). However, first derivatives of the metric are discontinuous.
Notice that, had we attached only one wormhole-like spacetime onto the Minkowski space, both the metric and 
its first derivative could have been matched continuously. 
Notice also that with a different renormalization choice, namely $\Lambda$ and $\bar\kappa^2$ as in 
Eqs.~(\ref{Lambda: vanishing backreaction in Mink}) and~(\ref{sc Einstein equation: order hbar: tt 3G}),
but $\alpha_{\tt Riem^2}=0$ instead of that in Eq.~(\ref{alpha Riem2: fine tuning}), one would have obtained 
({\it cf.} Eq.~(\ref{sc Einstein equation: order hbar: thetatheta 3})),
\begin{eqnarray}
 H_z^2 &\!\!=\!\!&  -  \frac{\kappa^2}{2}
    {\big\langle\Delta \hat{T}^{\,\vartheta}_\vartheta\big\rangle}_{\tt ren}^{(1)}  
\approx - \frac{\hbar \kappa^2}{960\pi^2a^4}
\bigg\{\ln\bigg(\frac{m^2}{4\pi\mu^2}\bigg)
             \!+\!\gamma_E \!-\!\frac{8}{21(am)^2}\Big)
\bigg\} 
\nonumber\\
 &\!\!\equiv\!\!& - \chi_z^2 < 0
\,,\qquad
\label{sc Hubble rate 2 z: order hbar}
\end{eqnarray}
and the metric in Eq.~(\ref{sc Hubble rate 2 z: order hbar}) would still give a valid solution,
\begin{equation}
{\rm d}s^2 = -\Big[1+\chi_z^2\Big(z^2-\tfrac{L_0^2}{4}\Big)\Big]{\rm d}t^2 
  +\frac{{\rm d}z^2 }{1+\chi_z^2\big(z^2-\tfrac{L_0^2}{4}\big)} + a_\perp^2{\rm d}\Omega_2^2
\,,\quad\!\!\!\! \big(|z|\leq \tfrac{L_0}{2}\big) 
\,.\;
\label{sc metric ansatz 2: z-dep Ansatz: positive p perp}
\end{equation}
Since this metric has no horizon along the $z$ direction, an initially traversable metric will remain
traversable after the quantum backreaction is taken into account. 
The geometry of the metric in Eq.~(\ref{sc metric ansatz 2: z-dep Ansatz: positive p perp}) is that of 
$AdS_2\times S^2$.
A curious fact is that the effect of the quantum backreaction on the geometry depends on the choice of finite gravitational counterterms.

The metric~(\ref{sc metric ansatz 2: z-dep Ansatz}) can be compared with the metric obtained by using the time-dependent {\it Ansatz} in Eq.~(\ref{sc metric ansatz}),
\begin{equation}
{\rm d}s^2 = -{\rm d}t^2 +a^2_{z(0)}\cosh(H_zt){\rm d}z^2 + a_\perp^2(t){\rm d}\Omega_2^2
\,,\qquad
\label{sc metric ansatz: t-dep Ansatz}
\end{equation}
where, by a suitable rescaling of the coordinate $z$, one can set $a_{z(0)}=1$. Below we show that these two 
solutions are physically equivalent.

\bigskip
\noindent
{\bf Quantum energy conditions.} Let us now consider nonvanishing Christoffel symbols 
for the metrics in Eqs.~(\ref{sc metric ansatz 2: z-dep Ansatz})
and~(\ref{metrics: flat space}), whose coordinates are continuously connected at $r_\pm -a_\perp = \pm z - L_0/2$, 
\begin{eqnarray}
\Gamma^t_{zt} &\!\!=\!\!& \Gamma^t_{tz} 
= -\frac{H_z^2z}{1\!+\!H_z^2\big(\frac{L_0^2}{4}-z^2\big)}
\!\times\!\Theta\Big(\frac{L_0}{2}\!-\!|z|\Big)
\,,\qquad 
\nonumber\\
\Gamma^z_{zz} 
&\!\!=\!\!& \frac{H_z^2z}{1\!+\!H_z^2\big(\frac{L_0^2}{4}-z^2\big)}
\!\times\!\Theta\Big(\frac{L_0}{2}\!-\!|z|\Big)
\,,\qquad
\nonumber\\
\Gamma^{r_\pm}_{\vartheta\vartheta} 
&\!\!=\!\!& - \sum_{\pm}r_\pm
            \!\times\!\Theta \Big(r_\pm\!-\!a_\perp \Big)
\,,\quad
\Gamma^{r_\pm}_{\varphi\varphi}
= - \sum_{\pm}r_\pm\sin^2(\vartheta)\!\times\!\Theta \Big(r_\pm\!-\!a_\perp \Big)
\,,
\nonumber\\
\Gamma^\vartheta_{r_\pm\vartheta} &\!\!=\!\!& 
  \sum_{\pm}\frac{1}{r_\pm}\!\times\!\Theta \Big(r_\pm\!-\!a_\perp \Big)
\,,\quad 
\Gamma^\vartheta_{\varphi\varphi} 
            = -\sin(\vartheta)\cos(\vartheta)
\,,\quad 
\nonumber\\
\Gamma^\varphi_{\vartheta\varphi} &\!\!=\!\!&  \Gamma^\varphi_{\varphi\vartheta}
    =\frac{\cos(\vartheta)}{ \sin(\vartheta)}
\,.\;\,
\label{sc Christoffel symbols 2}
\end{eqnarray}
A comparison of these Christoffel symbols with those in  
Eq.~(\ref{classical Christoffel symbols})
reveals that, apart from the
singular components listed in 
Eqs.~(\ref{Riemann tensor: singular 1})--(\ref{Riemann tensor: singular 2}), additional singular Riemann tensor components
arise,
\begin{eqnarray}
R_{tztz} 
&\!\!=\!\!& R_{ztzt} 
= - R_{tzzt} 
= - R_{zttz} 
= - \frac{H_z^2L_0}{2}\sum_\pm\delta\Big(z\!\mp\!\frac{L_0}{2}\Big)
\nonumber\\
&&\hskip 4.6cm
 = - \frac{H_z^2L_0}{2}\sum_\pm\delta(r_\pm\!-\!a_\perp)
\,.\qquad 
\label{Riemann tensor: singular 0}
\end{eqnarray}
When these singular contributions are added to the regular contributions one obtains for nonvanishing components of the Ricci tensor and Ricci scalar ({\it cf.} 
Eq.~(\ref{Ricci tensor and Ricci scalar})),
\begin{eqnarray}
R_{t}^{\;t} &\!\!=\!\!& \frac{H_z^2L_0}{2}
      \sum_\pm\delta(r_\pm\!-\!a_\perp )
      +H_z^2\Theta\Big(\frac{L_0}{2}-|z|\Big)
\,,\qquad
\nonumber\\
R_{z}^{\;z} &\!\!=\!\!& 
\Big(\frac{H_z^2L_0}{2}-\frac{2}{a_\perp }\Big)
      \sum_\pm\delta(r_\pm\!-\!a_\perp )
      +H_z^2\Theta\Big(\frac{L_0}{2}-|z|\Big)
\,,\qquad
\nonumber\\
R_{\vartheta}^{\;\vartheta} &\!\!=\!\!&
R_{\varphi}^{\;\varphi}=
- \frac{1}{a_\perp} \sum_\pm\delta(r_\pm\!-\!a_\perp )
   + \frac{1}{a_\perp^2} \Theta\Big(\frac{L_0}{2}-|z|\Big)
\,,\qquad
\nonumber\\
R &\!\!=\!\!& \Big(H_z^2L_0-\frac{4}{a_\perp }\,\Big)
            \sum_\pm\delta(r_\pm\!-\!a_\perp )
+ \Big(H_z^2+\frac{2}{a_\perp^2}\Big)
  \Theta\Big(\frac{L_0}{2}-|z|\Big)
\,,\qquad
\label{Ricci tensor and Ricci scalar sc}
\end{eqnarray}
from where one arrives at nontrivial components of the Einstein tensor ({\it cf.} Eq.~(\ref{Einstein tensor}),
\begin{eqnarray}
 G_{t}^{\;t} &\!\!=\!\!& \frac{2}{a_\perp }
  \sum_\pm\delta(r_\pm\!-\!a_\perp )
 -\frac{1}{a_\perp ^2}\Theta\Big(\frac{L_0}{2}-|z|\Big)
\,,\qquad
\nonumber\\
  G_{z}^{\;z} &\!\!=\!\!& 
  - \frac{1}{a_\perp ^2}\Theta\Big(\frac{L_0}{2}-|z|\Big)
\,,\qquad
\nonumber\\
  G_{\vartheta}^{\;\vartheta} &\!\!=\!\!& 
  G_{\varphi}^{\;\varphi} =
  \Big(\!-\frac{H_z^2L_0}{2}+\frac{1}{a_\perp }\,\Big)
       \sum_\pm\delta(r_\pm\!-\!a_\perp )
    - H_z^2 \Theta\Big(\frac{L_0}{2}-|z|\Big)
\,.\qquad
\label{Einstein tensor: sc}
\end{eqnarray}
The semiclassical Einstein equation~(\ref{semiclassical Einstein equation: order hbar}) then implies that 
non-vanishing components of the energy-momentum tensor are
({\it cf.} Eq.~(\ref{energy-momentum tensor: topological wormhole})),
\begin{eqnarray}
T_{t}^{\;t} +\langle \Delta \hat{T}_{t}^{\;t} \rangle^{(1)} 
&\!\!=\!\!&
    \frac{1}{4\pi G a_\perp}\!\times\!\sum_\pm\delta(r_\pm - a_\perp)
 -\frac{1}{8\pi G a_\perp^2}
  \!\times\!\Theta\Big(\frac{L_0}{2}-|z|\Big)
    \,,\qquad
\nonumber\\
 T_{z}^{\;z}+\langle \Delta \hat{T}_{z}^{\;z} \rangle^{(1)} 
 &\!\!=\!\!&
  - \frac{1}{8\pi G a_\perp^2}
  \!\times\!\Theta\Big(\frac{L_0}{2}-|z|\Big)
\nonumber\\
 T_{\vartheta}^{\;\vartheta}
   +\langle \Delta \hat{T}_{\vartheta}^{\;\vartheta} \rangle^{(1)} 
 &\!\!=\!\!& 
 T_{\varphi}^{\;\varphi}  
    +\langle \Delta \hat{T}_{\varphi}^{\;\varphi} \rangle^{(1)} 
\nonumber\\
   &&  \hskip -2.9cm
=\, 
\frac{1}{8\pi G}\Big(\!-\frac{H_z^2L_0}{2}+\frac{1}{a_\perp}\,\Big)
    \!\times\!\sum_\pm\delta(r_\pm \!-\! a_\perp) 
   - \frac{H_z^2}{8\pi G} \!\times\!\Theta\Big(\frac{L_0}{2}-|z|\Big)
\,.\qquad
\label{energy-momentum tensor: topological wormhole sc}
\end{eqnarray}
The corresponding results for the metric in Eq.~(\ref{sc metric ansatz 2: z-dep Ansatz: positive p perp})
are obtained by exacting the change $H_z^2\rightarrow - \chi_z^2$ in 
Eqs.~(\ref{sc Christoffel symbols 2})--(\ref{energy-momentum tensor: topological wormhole sc}),
as well as in the energy conditions discussed in Eqs.~(\ref{NEC 1: quantum})--(\ref{AWEC: quantum}) below.
Comparing this with the classical energy-momentum tensor in Eqs.~(\ref{energy-momentum tensor: topological wormhole}) we see that inside the wormhole null energy condition reads
({\it cf.} Eqs.~(\ref{NEC 1})--(\ref{NEC 2})),
 \begin{eqnarray}
    T_{\mu\nu}\ell_1^\mu\ell_1^\nu 
 &\!\! =\!\!& - T_{t}^{\;t} - \langle \Delta \hat{T}_{t}^{\;t} \rangle^{(1)}
    + T_{z}^{\;z} + \langle \Delta \hat{T}_{z}^{\;z} \rangle^{(1)}  = 0
 \,,\qquad
 \label{NEC 1: quantum}\\
     T_{\mu\nu}\ell_2^\mu\ell_2^\nu 
  &\!\! =\!\!&  - T_{t}^{\;t} - \langle \Delta \hat{T}_{t}^{\;t} \rangle^{(1)}
   + T_{\vartheta}^{\;\vartheta} + \langle \Delta \hat{T}_{\vartheta}^{\;\vartheta} \rangle^{(1)} 
  = \frac{1}{8\pi G}\left(\frac{1}{a_\perp^2} - H_z^2\right)
  \nonumber\\
  &\!\! =\!\!& \frac{1}{8\pi G}\left(\frac{1}{a_\perp^2} 
  -  {
  \frac{2\hbar G}{315\pi a^6m^2} 
    }
\right) > 0
 \,,\qquad
 \label{NEC 2: quantum}
 \end{eqnarray}
 where we made use of Eq.~(\ref{sc Einstein tensor: thetatheta 5}).
The last inequality in Eq.~(\ref{NEC 2: quantum}) follows from $a_\perp\simeq a$, $G\ll a^2$
and $am\gg 1$.
Note that the null vectors in Eqs.~(\ref{NEC 1: quantum})--(\ref{NEC 2: quantum})
are constructed for the semiclassical metric in Eq.~(\ref{sc metric ansatz 2: z-dep Ansatz}), 
\begin{eqnarray}
\ell_1^\mu &\!\!=\!\!& \Big(\Big[1+H_z^2\big(\tfrac{L_0^2}{4}-z^2\big)\Big]^{-1/2},
             \Big[1+H_z^2\big(\tfrac{L_0^2}{4}-z^2\big)\Big]^{1/2},0,0\Big)
\,,\quad
\nonumber\\
\ell_2^\mu &\!\!=\!\!& \Big(\Big[1+H_z^2\big(\tfrac{L_0^2}{4}-z^2\big)\Big]^{-1/2},0,a_\perp^{-1},0\Big)
\,,\qquad
\end{eqnarray}
and that they indeed satisfy $\ell_1^2=0$, $\ell_2^2=0$ and they are normalized as usually,
$(\ell_1)^t(\ell_1)_t=-1=(\ell_2)^t(\ell_2)_t$.
A simple consideration shows that quantum average null energy condition (ANEC) 
for the metric in Eq.~(\ref{sc metric ansatz 2: z-dep Ansatz}) reads,
 \begin{eqnarray}
    \bigl\langle T_{\mu\nu}+\langle\Delta \hat{T}_{\mu\nu}\rangle\bigr\rangle_V  \ell_1^\mu\ell_1^\nu 
&\!\!=\!\!& -\bigl\langle T_{t}^{\;t}\!+\!\langle\Delta \hat{T}_{t}^{\;t}\rangle\bigr\rangle_{V}  
   + \bigl\langle T_{z}^{\;z}\!+\!\langle\Delta \hat{T}_{z}^{\;z}\rangle\bigr\rangle_V   
\nonumber\\
   &\!\!=\!\!& -  \frac{1}{2\pi G a_\perp L_0} <0
 \,,\qquad
 \label{ANEC 1: quantum}\\
 \!   \bigl\langle T_{\mu\nu}\!+\!\langle\Delta \hat{T}_{\mu\nu}\rangle\bigr\rangle_V \ell_2^\mu\ell_2^\nu 
&\!\!=\!\!&  -\bigl\langle T_{t}^{\;t}\!+\!\langle\Delta \hat{T}_{t}^{\;t}\rangle\bigr\rangle_{V} 
   + \bigl\langle T_{\vartheta}^{\;\vartheta}
                  +\langle\Delta \hat{T}_{\vartheta}^{\;\vartheta}\rangle\bigr\rangle_V 
 \nonumber\\
   &\!\!=\!\!&\!                 
  \frac{1}{8\pi G a_\perp^2}\Big(1-\frac{2a_\perp}{L_0}-2a_\perp^2 H_z^2\Big)  
  >0 \quad\!\!\! (L_0\gg a_\perp)
 \,,\qquad\;\;
 \label{ANEC 2: quantum}
 \end{eqnarray}
which is to be compared with classical ANEC in Eqs.~(\ref{ANEC 1})--(\ref{ANEC 2}).
As explained above, the last contribution in Eq.~(\ref{ANEC 2: quantum}) is expected to be small,
and therefore, for sufficiently long wormholes, $L_0\gg a_\perp$,
Eq.~(\ref{ANEC 2: quantum}) is satisfied. However,
as in the classical case, quantum ANEC is violated by Eq.~(\ref{ANEC 1: quantum}),
for the null vector $\ell_1^\mu$ pointing in the $z$-direction along the wormhole.
The only difference between the classical and quantum cases is in a slight change in the radius of 
the wormhole, $a\rightarrow a_\perp$, which is quantitatively insignificant. An interesting question 
worth investigating is whether this conclusion holds true if one solves the semiclassical gravity equations~(\ref{semiclassical Einstein equation}) self-consistently. 

\medskip
Finally we give a couple of remarks on weak energy condition (WEC). 
For a time-like vector $t^\mu$ ($t^2=-1$) we have for quantum WEC and quantum 
average weak energy condition (AWEC), 
 \begin{eqnarray}
    T_{\mu\nu}t^\mu t^\nu 
 &\!\! =\!\!& - T_{t}^{\;t} - \langle \Delta \hat{T}_{t}^{\;t} \rangle^{(1)}
     = \frac{1}{8\pi a_\perp^2} >0
 \,,\qquad
 \label{WEC: quantum}\\
    \bigl\langle T_{\mu\nu}+\langle\Delta \hat{T}_{\mu\nu}\rangle\bigr\rangle_V  t^\mu t^\nu 
&\!\!=\!\!& -\bigl\langle T_{t}^{\;t}\!+\!\langle\Delta \hat{T}_{t}^{\;t}\rangle\bigr\rangle_{V}  
\nonumber\\
&\!\!=\!\!& \frac{1}{8\pi a_\perp^2}\left(1-\frac{4a_\perp}{L_0}\right) > 0
\,,\quad \big(L_0>4a_\perp\big)
\,,\quad
 \label{AWEC: quantum}
 \end{eqnarray}
which is to be compared with Eqs.~(\ref{WEC})--(\ref{AWEC}). The only difference between 
the classical and semiclassical weak energy conditions is in that the classical radius $a$
ought to be replaced by the semiclassical one, $a\rightarrow a_\perp$, 
so that -- provided $L_0>4a_\perp$ is satisfied -- the semiclassical wormhole satisfies both 
WEC and AWEC.

\bigskip
\noindent
{\bf Coordinates on de Sitter.}
To shed light on the question whether the metric is expanding or static, 
let us consider de Sitter space in two spacetime dimensions $(t,z)$,
which can be represented by the following 
three-dimensional flat Minkowski space embedding,
\begin{equation}
{\rm d}S^2 = -\,{\rm d}T^2 + {\rm d}X^2+ {\rm d}Y^2
\,,\qquad
- T^2 + (X^2+Y^2) = \frac{1}{H^2}
\,,\qquad\;
\label{3 dim de Sitter embedding space}
\end{equation}
where $T$ is a time coordinate, $X$ and $Y$ are spatial coordinates
and $1/H$ is the throat radius of the de Sitter space hyperboloid, which in de Sitter 
plays the role of the constant expansion rate.
When the following coordinate transformations,
\begin{equation}
T = \frac{1}{H}\sinh(Ht)
\,,\quad\!\!
X =  \frac{1}{H}\cosh(Ht)\cos(Hz)
\,,\quad\!\!
Y =  \frac{1}{H}\cosh(Ht)\sin(Hz)
\,,
\label{2 dim de Sitter: expanding coordinates}
\end{equation}
are inserted into Eq.~(\ref{3 dim de Sitter embedding space}) one obtains the following metric, 
\begin{equation}
{\rm d}s^2 = -\,{\rm d}t^2 + \cosh^2(Ht){\rm d}z^2
\,.\qquad
\label{2 dim de Sitter: expanding coordinates 2}
\end{equation}
This is the global metric of de Sitter space, in which a contracting phase is followed by an expanding phase.
On the other hand, the coordinate transformations,
\begin{equation}
\!\!
T = \frac{1}{H}\sqrt{1\!-\!(Hz)^2}\sinh(Ht)
\,,\quad\;
X =  \frac{1}{H}\sqrt{1\!-\!(Hz)^2}\cosh(Ht)
\,,\quad\;
Y =  z
\,,\quad\!\!
\label{2 dim de Sitter: static coordinates}
\end{equation}
yield the following metric,
\begin{equation}
{\rm d}s^2 = -\,\big(1-H^2z^2\big){\rm d}t^2 + \frac{{\rm d}z^2}{1-H^2z^2}
\,\,,\qquad
\label{2 dim de Sitter: static coordinates 2}
\end{equation}
such that the two metrics in Eqs.~(\ref{2 dim de Sitter: expanding coordinates 2}) and~(\ref{2 dim de Sitter: static coordinates 2})
describe the same 2-dimensional de Sitter space. With this insight in mind, we see that 
the two semiclassical metrics in Eqs.~(\ref{sc metric ansatz: t-dep Ansatz}) and~(\ref{sc metric ansatz 2: z-dep Ansatz}) describe 
the same four dimensional spacetime: $dS_2\times S^2$, and therefore they are physically equivalent.
The metric in Eq.~(\ref{2 dim de Sitter: static coordinates 2}) can be transformed to the $(t,z)$ part of the metric in
Eq.~(\ref{sc metric ansatz 2: z-dep Ansatz}) by a linear coordinate transformation $z\rightarrow z/\sqrt{1+(H_z L_0/2)^2}$ and 
$t\rightarrow t\sqrt{1+(H_z L_0/2)^2}$,
and therefore these two-dimensional metric sections are physically equivalent. 
We choose the form in Eq.~(\ref{sc metric ansatz 2: z-dep Ansatz}) is to match it continuously to
the external Minkowski metrics in Eq.~(\ref{metrics: flat space}).

The metric in static coordinates in Eq.~(\ref{sc metric ansatz 2: z-dep Ansatz}) possesses a cosmological horizon where the vector orthogonal to $z={\rm const.}$ becomes null,
$\|\partial_\mu z\|^2 = g^{zz} = 0$. When this is solved it gives, 
\begin{equation}
  z_H = \pm\sqrt{\frac{1}{H_z^2}+\frac{L_0^2}{4}} 
\,,\qquad
\label{horizon in z-direction}
\end{equation}
which is always outside the wormhole ($|z|\leq L_0/2<|z_H|$)
and therefore it is not part of the spacetime manifold.~\footnote{ We thank Farshid Jafarpour for pointing this to us.}  Below we show that consequently a traversable wormhole will remain traversable after the quantum backreaction is taken into account.
Note that typically $1/H_z = 6\sqrt{35}\pi a^3m M_P \gg  L_0/2$, where 
we used 
Eq.~(\ref{sc Einstein tensor: thetatheta 5}) and we defined a reduced Planck mass,
$M_P=1/\sqrt{8\pi G \hbar}\simeq 2.5\times 10^{18}~{\rm GeV}$.
To see that let us compare $R_z\equiv 1/H_z$
with the cosmological Hubble radius today, $R_{H_0}\simeq 1.3\times 10^{26}~{\rm m}$. From
\begin{equation}
\frac{R_z}{R_{H_0}} \simeq 6\sqrt{35}\pi \frac{a}{R_{H_0}} (a^2m M_P) \simeq 1\times 10^{25} 
                                 \left(\frac{m}{{\rm MeV}}\right) \big(a[{\rm m}]\big)^3
\,,\qquad\;
\label{condition for traversability}
\end{equation}
we see that $R_z$ is indeed very large. 
For example, for a typical particle whose mass is equal to that of the electron 
($m_e\simeq 0.5~{\rm MeV}$), one finds
$R_z\sim 10^{25}R_{H_0}$, which is huge. This means that the quantum backreaction on the wormhole generated by the quantum vacuum fluctuations of a massive scalar field is tiny. We expect analogous results to hold for
massive particles with spin.

We now know that the expanding and 
static wormhole metrics in 
Eqs.~(\ref{2 dim de Sitter: expanding coordinates 2})
and~(\ref{2 dim de Sitter: static coordinates 2}) 
are physically equivalent. Nevertheless, a legitimate question to ask is which of the two metrics
gives the correct physical picture. 
A subsystem of test particles, whose energy is too small to escape from the wormhole, will
move on virialized trajectories whose average distance from 
the wormhole center will be less than $L_0/2$ and independent on time. On the other hand, energetic test particles, whose energy is larger than the escape energy,  
will move on trajectories whose average distance from the wormhole center will grow in time. These particles will perceive a wormhole that is expanding in the $z$-direction.

\bigskip
\noindent
{\bf Traversing the wormhole.} Here we shall estimate the time it takes to traverse the wormhole. Firstly, a traversable wormhole 
satisfies Eq.~(\ref{condition for traversability}), and its metric is given by Eq.~(\ref{sc metric ansatz 2: z-dep Ansatz}). 
Since the wormhole possesses spherical symmetry, it suffices 
to consider motion in the equatorial plane of the sphere ($\vartheta=\pi/2$). The two Killing vectors we need is 
$K=\partial_t$ and $R=\partial_\varphi$, which imply 
a conserved energy $E=-mK_\mu \frac{{\rm d}x^\mu}{{\rm d}\tau}$ 
and a conserved momentum $L=mR_\mu \frac{{\rm d}x^\mu}{{\rm d}\tau}$,
which for the metric in Eq.~(\ref{sc metric ansatz 2: z-dep Ansatz}) give,
\begin{equation}
    E = m\Big[1+H_z^2\Big(\tfrac{L_0^2}{4}-z^2\Big)\Big]
    \frac{{\rm d}t}{{\rm d}\tau}
    \,,\qquad
     L = ma_\perp^2 \frac{{\rm d}\varphi}{{\rm d}\tau}
\,,\qquad
\label{conserved energy and momentum}
\end{equation}
where $\tau$ is proper time (${\rm d}\tau^2 = -{\rm d}s^2$).
When these are used in Eq.~(\ref{sc metric ansatz 2: z-dep Ansatz}) one obtains,
\begin{equation}
  \frac{\frac{E^2}{m^2}-\Big(\frac{{\rm d}z}{{\rm d}\tau}\Big)^2}{1\!+\!H_z^2\big(\frac{L_0^2}{4}-z^2\big)}
  =     1+ \frac{L^2}{(a_\perp m)^2} 
\,.\qquad
\label{sc metric ansatz 2: z-dep Ansatz 2}
\end{equation}
Next, one can perform a separation of variables to obtain,
\begin{equation}
\int \frac{{\rm d}z}
{\sqrt{\frac{E^2}{m^2} - \Big[1+\frac{L^2}{(a_\perp m)^2}\Big]\Big[1+H_z^2\Big(\frac{L_0^2}{4}-z^2\Big)\Big]}}  
  = \tau
\,.\qquad
\label{sc metric ansatz 2: z-dep Ansatz 3}
\end{equation}
This integral is solved by $\sim\ln[H_zz+\sqrt{(H_zz)^2+\alpha^2}]$
which, when inverted, results in,
\begin{equation}
z(\tau) 
=  \frac{1}{2H_z}
 \left\{{\rm e}^{\omega\tau}-\left[\frac{(E/m)^2}{1+\Big(\frac{L}{a_\perp m}\Big)^2} - \Big[1+\Big(\frac{H_zL_0}{2}\Big)^2\right]{\rm e}^{-\omega\tau}
 \right\} 
\,,\qquad
\label{sc metric ansatz 2: z-dep Ansatz 4}
\end{equation}
with $\omega=\sqrt{1+\Big(\frac{L}{a_\perp m}\Big)^2}H_z$.
From this it follows that the proper time interval $\Delta\tau$ 
needed to traverse the wormhole is given by,
\begin{equation}
\Delta\tau 
=  \frac{1}{H_z\sqrt{1+\Big(\frac{L}{a_\perp m}\Big)^2}}
\!\times\!\ln\left\{
\sqrt{\frac{(E/m)^2}{1+\Big(\frac{L}{a_\perp m}\Big)^2}-1\,}\,-\frac{H_zL_0}{2}
\right\}
\,,\qquad
\label{traversal time}
\end{equation}
which is well defined only when the following condition is satisfied, 
\begin{equation}
\frac{(E/m)^2}{1+\Big(\frac{L}{a_\perp m}\Big)^2}
> 1+ \Big(\frac{H_zL_0}{2}\Big)^2 
\,.\qquad
\label{sc metric ansatz 2: z-dep Ansatz 6}
\end{equation}
The condition in Eq.~(\ref{sc metric ansatz 2: z-dep Ansatz 6})
implies a minimum energy
$E_{\rm min}$ (in units of $c^2$) a particle must have in order to be able to traverse the wormhole,
\begin{equation}
E> E_{\rm min} = m \sqrt{1+\Big(\frac{L}{a_\perp m}\Big)^2}
\sqrt{1+ \Big(\frac{H_zL_0}{2}\Big)^2} 
\,.\qquad
\label{sc metric ansatz 2: z-dep Ansatz 6B}
\end{equation}
As expected, the condition in Eq.~(\ref{sc metric ansatz 2: z-dep Ansatz 6B}) becomes less restrictive for a particle
whose angular momentum vanishes \Big(for $L=0$,
$E_{\rm min} = m \sqrt{1\!+\! \big(H_zL_0/2\big)^2}\,$\Big), 
as in this case 
the particle does not lose time circling around the internal 
sphere.
This means that, as long as a particle enters the wormhole with a large enough energy satisfying 
Eq.~(\ref{sc metric ansatz 2: z-dep Ansatz 6B}), it will traverse the wormhole in a finite proper time given in Eq.~(\ref{traversal time}).

In what follows we show that the finite proper time interval
in Eq.~(\ref{traversal time}) will remain finite when measured 
in the coordinate time of outside Minkowski space observers, which is the the last step
needed to establish traversability of the wormhole.
We can get the coordinate time interval $\Delta t$ by solving
Eq.~(\ref{conserved energy and momentum}).

First note that the solution for $t(\tau)$ can be written as, 
\begin{equation}
 t= 
 \frac{E/(2m)}{\sqrt{1+\Big(\frac{H_zL_0}{2}\Big)^{\!2}}}
\left[ \int\!\frac{{\rm d}\tau}
 {\!\sqrt{1\!+\!\Big(\frac{H_zL_0}{2}\Big)^{\!2}}\!+\!H_zz(\tau)}
+\!\int\!\frac{{\rm d}\tau}{\!\sqrt{1\!+\!\Big(\frac{H_zL_0}{2}\Big)^{\!2}}\!-\!H_zz(\tau)}
\right]
.
\label{t of tau}
\end{equation}
Upon inserting Eq.~(\ref{sc metric ansatz 2: z-dep Ansatz 4}) 
into Eq.~(\ref{t of tau}) and integrating
over $\tau$ one obtains,
\begin{eqnarray}
 t(\tau)= 
 \frac{\sqrt{1+\Big(\frac{L}{a_\perp m}\Big)^{\!2}}}{\sqrt{1+\Big(\frac{H_zL_0}{2}\Big)^{\!2}}}\ln
\left\{\frac{{\rm e}^{2\omega\tau}
   \!-\!\left[\sqrt{1\!+\!\big(H_zL_0/2\big)^{\!2}}
     -\frac{E/m}{\sqrt{1+\big(L/(a_\perp m)\big)^{\!2}}}\right]^2}
 {{\rm e}^{2\omega\tau}
   \!-\!\left[\sqrt{1\!+\!\big(H_zL_0/2\big)^{\!2}}
     +\frac{E/m}{\sqrt{1+\big(L/(a_\perp m)\big)^{\!2}}}\right]^2}
\right\}
\,,
\nonumber\\
\label{t of tau 2}
\end{eqnarray}
where $t_0\equiv t(0)$ is not zero. The desired time interval $\Delta t$
is obtained by subtracting $t(0)$ from $t(\tau)$ in Eq.~(\ref{t of tau 2})
and inserting $\Delta \tau$ from Eq.~(\ref{traversal time}),
\begin{eqnarray}
 \Delta t = 
 \frac{\ell}{\ell_0}\ln
\left[\frac{\Big(\sqrt{\mathcal{E}^2-1}\!-\!\frac{H_zL_0}{2}\Big)^2
   \!-\!\big(\mathcal{E}-\ell_0\big)^2}
 {\Big(\sqrt{\mathcal{E}^2-1}\!-\!\frac{H_zL_0}{2}\Big)^2
   \!-\!\big(\mathcal{E}+\ell_0\big)^2}
   \!\times\!\frac{1\!-\!\big(\mathcal{E}+\ell_0\big)^2}
         {1\!-\!\big(\mathcal{E}-\ell_0\big)^2}
\right]
\,,\qquad
\label{t of tau 3}
\end{eqnarray}
where we introduced a shorthand notation,
\begin{equation}
\ell_0 = \sqrt{1\!+\!\Big(\frac{H_zL_0}{2}\Big)^{\!2}}
\,,\quad \!
\ell = \sqrt{1\!+\!\Big(\frac{L}{a_\perp m}\Big)^{\!2}}
\,,\quad \!
\mathcal{E} = \frac{E/m}{\sqrt{1\!+\!\big(L/(a_\perp m)\big)^{\!2}}}
\,.
\label{shorthand notation}    
\end{equation}
These calculations give traversal times for the case when the quantum backreaction is 
given by Eq.~(\ref{sc Einstein equation: order hbar: thetatheta 5}), and
the effective angular pressure is negative, $p_\perp <0$.
In this case $H_z^2 > 0$, such that the quantum backreaction tends to stretch the wormhole 
in the $z$ direction, thereby destabilizing it. 
On the other hand, when the Riemann tensor squared coupling in 
Eq.~(\ref{sc Einstein equation: order hbar: thetatheta 3}) is chosen to vanish,
$\alpha_{\tt Riem^2}\rightarrow 0$, the corresponding 
metric is shown in Eq.~(\ref{sc metric ansatz 2: z-dep Ansatz: positive p perp}), which 
is a direct product of the two-dimensional anti-de Sitter space and the sphere,
$AdS_2\times S^2$. In that case the quantum backreaction tends to stabilize the wormhole, and 
a classically traversable wormhole will remain traversable. We leave to the reader to repeat the above calculation of the minimum energy and traversal times in this case.

\section{Discussion and outlook}

In this work we consider the quantum backreaction in
a timelike wormhole of topology $M_2\times S^2$, 
where $M_2$ denotes two-dimensional Minkowski space. The wormhole metric in 
Eq.~(\ref{wormhole metric})
is supported by the classical energy-momentum tensor of an anisotropic fluid
in Eq.~(\ref{energy-momentum tensor: topological wormhole}),  which marginally satisfies null energy condition~(\ref{NEC 1})--(\ref{NEC 2}), however it breaks 
average null energy condition~(\ref{ANEC 1})--(\ref{ANEC 2}). 

In a comprehensive appendix we calculate the quantum one-loop backreaction
of a minimally coupled, massive scalar field induced by the corresponding 
quantum vacuum fluctuations, 
in which we use dimensional regularization 
scheme to  renormalize the energy-momentum tensor.
The two counterterms needed are the cosmological constant counterterm and the Ricci
scalar counterterm. The finite parts of these counterterms in 
Eqs.~(\ref{Lambda: vanishing backreaction in Mink}) 
and~(\ref{sc Einstein equation: order hbar: tt 3G}) are chosen such to
yield a vanishing observed cosmological constant and a continuous Newton constant in the entire spacetime. 
To simplify the results even further we have used a finite Riemann tensor squared 
counterterm in Eq.~(\ref{alpha Riem2: fine tuning}).
 We neglected the tiny cosmological constant giving rise to dark energy, which is justified as long as the size of the wormhole is much smaller than the cosmological horizon today $H_0$, {\it i.e.} $aH_0 \ll 1$, which is amply satisfied. It would be easy to include this  contribution, but that would not change our results in any significant way.
The renormalized quantum energy-momentum tensor 
in Eqs.~(\ref{sc Einstein equation: order hbar: tt 3})--(\ref{sc Einstein equation: order hbar: thetatheta 3}), and its simpler cousin in 
Eqs.~(\ref{sc Einstein equation: order hbar: tt 3G2})--(\ref{sc Einstein equation: order hbar: thetatheta 5}), are then used as the source 
in the equations of motion of semiclassical gravity~(\ref{semiclassical Einstein equation}). We solve these equations 
perturbatively  and find a static metric solution in 
Eq.~(\ref{sc metric ansatz 2: z-dep Ansatz}),
with an inverse expansion rate $1/H_z$
in the $z$ direction
(see Eqs.~(\ref{sc Einstein tensor: thetatheta 5}) and~(\ref{condition for traversability})),
\begin{equation}
R_{H_z}\equiv \frac{1}{H_z} 
   \approx 6\sqrt{35}\pi a^3m M_P
\,,\qquad
\label{condition on L0}
\end{equation}
where, to get Eq.~(\ref{condition on L0}), we used the fine-tuned
energy-momentum tensor in Eq.~(\ref{sc Einstein equation: order hbar: thetatheta 5}).
The length scale in Eq.~(\ref{condition on L0}) is typically very large,
much larger than the wormhole length, $R_{H_z}\gg L_0$,
which means that the quantum backreaction is tiny and does not significantly
affect the wormhole geometry.

We also study traversability~\cite{Shatskiy:2008us} of the semiclassicaly-corrected wormhole by sending a test-particle through the wormhole. We find that large wormholes
remain generally traversable, provided the particle has a sufficient 
energy~(\ref{sc metric ansatz 2: z-dep Ansatz 6}) at the entrance to the wormhole.
The proper and external observer time intervals needed to traverse the wormhole are given in 
Eqs.~(\ref{traversal time}) and~(\ref{t of tau 3}), respectively.
From these results one sees that quantum fluctuations do not significantly affect
traversability of the wormhole: 
\begin{quote}
{\it If a classical wormhole is traversable, it will remain traversable when 
the quantum backreaction is taken into account.} 
\end{quote}

Even though we solve here the semiclassical Einstein equations perturbatively (to linear order in $\hbar$),
based on the fact that the quantum backreaction is very small for large wormholes, we expect that the 
self-consistent solution of the semiclassical gravity will maintain the principal features of the perturbative solution. This statement becomes more quantitatively accurate for larger wormholes.  

In this paper we do not address the question how to create wormholes. 
Classical wormholes are characterized by a nontrivial topology~\cite{Visser:1995cc}, 
which is protected by the classical evolution. Therefore, to create a wormhole, one probably needs 
quantum fluctuations~\cite{Morris:1988cz}.~\footnote{See, however, Ref.~\cite{Marunovic:2014hla}
for an example of how the topology of spatial slices of an expanding 
Universe can be changed by global monopoles.} 
The Appendix of this paper contains exact expressions~(\ref{T tt 4 A})--(\ref{T zz 4 A})
for the one-loop quantum backreaction of a massive scalar field.
Here we are primarily interested in the quantum backreaction of large wormholes,  
and these are obtained from asymptotic expansions of the exact results. On the other hand,
small argument expansions of Eqs.~(\ref{T tt 4 A})--(\ref{T zz 4 A})
contain information on the quantum backreaction of small wormholes,
and therefore they would be used to get an insight into the formation of wormholes {\it via} quantum fluctuations.

A natural extension of this work would be to consider a self-consistent quantum backreaction.
That can be achieved by firstly constructing the scalar field propagator
in a presumed wormhole spacetime, and then use that propagator to compute the quantum backreaction and solve the corresponding semiclassical equations for the metric tensor.
Furthermore, one could study the quantum backreaction in the wormhole background induced by massive
fermions and gauge fields, which constitute particle content of the standard model (we do not expect
gravitons or photons to yield any significant backreaction, since they are both massless particles).
Finally, one can modify the shape and topology of the classical wormhole studied 
in this work from $M_2\times S^2$ to some other shape/topology and investigate how that would change 
the quantum backreaction.

\section*{Acknowledgements}

We are grateful to Farshid Jafarpour, who 
suggested to us to study topological wormholes, 
for his careful reading of the manuscript and for useful suggestions
based on which we improved the paper. We thank Daniel Frolovsky for useful suggestions.
T.P. acknowledges support by the Delta ITP consortium, a program of the
Netherlands' Organisation for Scientific Research (NWO), which is funded by the Dutch Ministry of
Education, Culture and Science (OCW) -- NWO project number 24.001.027; by the
NWA ORC 2023 consortium grant: {\it Cosmic emergence: from abstract simplicity to complex
diversity}, 
and by {\it The Magnetic Universe} grant} -- NWO project number OCENW.XL.23.147.


\section*{Appendix}

\section*{One-loop energy-momentum tensor}

The goal of this appendix is to compute the leading order 
quantum energy-momentum tensor in 
Eq.~(\ref{one loop energy-momentum tensor: T*2}), where 
the propagator obeys the equation of motion in flat space given in 
Eq.~(\ref{Feynman propagator: eom: order hbar}).
In this work we adopt a $T^*$ ordering, as mentioned in the main text. However, in our
calculations we shall use the more standard $T$-ordering as the difference 
in the ordering is immaterial for the calculation of the one-loop energy-momentum tensor.

From Eq.~(\ref{field decomposition}) we have,
\begin{eqnarray}
\partial_t\hat\phi(t,\mathbf{z},\vartheta,\varphi) 
&\!\!=\!\!& \sum_{\ell m}\int \frac{{\rm d}^{D\!-\!3}k}{(2\pi)^{D\!-\!3}}
\Big(-i\omega_\ell{\rm e}^{i \mathbf{k}\cdot\mathbf{z}}
   u_{\ell m}(t,\mathbf{k}) \hat{a}_{\ell m}(\mathbf{k})
   Y_{\ell m}(\vartheta,\varphi)
\nonumber\\
&&\hskip 3.cm
+\,i\omega_\ell{\rm e}^{- i \mathbf{k}\cdot\mathbf{z}}
     u^*_{\ell m}(t,\mathbf{k}) 
              \hat{a}^\dagger_{\ell m}(\mathbf{k}) 
              Y^*_{\ell m}(\vartheta,\varphi)
 \Big)
 \,,\qquad\;
\label{field decomposition: time derivative}\\
\partial_{\mathbf{z}}\hat\phi(t,\mathbf{z},\vartheta,\varphi) 
&\!\!=\!\!& \sum_{\ell m}\int \frac{{\rm d}^{D\!-\!3}k}{(2\pi)^{D\!-\!3}}
\Big(i\mathbf{k}{\rm e}^{i \mathbf{k}\cdot\mathbf{z}}
   u_{\ell m}(t,\mathbf{k}) \hat{a}_{\ell m}(\mathbf{k})
   Y_{\ell m}(\vartheta,\varphi)
\nonumber\\
&&\hskip 3.cm
-\,i\mathbf{k}{\rm e}^{- i \mathbf{k}\cdot\mathbf{z}}
     u^*_{\ell m}(t,\mathbf{k}) 
              \hat{a}^\dagger_{\ell m}(\mathbf{k}) 
              Y^*_{\ell m}(\vartheta,\varphi)
 \Big)
 \,,\qquad\;
\label{field decomposition: z derivative}\\
\partial_{\vartheta}\hat\phi(t,\mathbf{z},\vartheta,\varphi) 
&\!\!=\!\!& \sum_{\ell m}\int \frac{{\rm d}^{D\!-\!3}k}{(2\pi)^{D\!-\!3}}
\Big({\rm e}^{i \mathbf{k}\cdot\mathbf{z}}
   u_{\ell m}(t,\mathbf{k}) \hat{a}_{\ell m}(\mathbf{k})
   \partial_{\vartheta}Y_{\ell m}(\vartheta,\varphi)
\nonumber\\
&&\hskip 3.cm
+\,{\rm e}^{- i \mathbf{k}\cdot\mathbf{z}}
     u^*_{\ell m}(t,\mathbf{k}) 
              \hat{a}^\dagger_{\ell m}(\mathbf{k}) 
              \partial_{\vartheta}Y^*_{\ell m}(\vartheta,\varphi)
 \Big)
 \,,\qquad\;
\label{field decomposition: theta derivative}\\
\partial_{\varphi}\hat\phi(t,\mathbf{z},\vartheta,\varphi) 
&\!\!=\!\!& \sum_{\ell m}\int \frac{{\rm d}^{D\!-\!3}k}{(2\pi)^{D\!-\!3}}
\Big({\rm e}^{i \mathbf{k}\cdot\mathbf{z}}
   u_{\ell m}(t,\mathbf{k}) \hat{a}_{\ell m}(\mathbf{k})
   \partial_{\varphi}Y_{\ell m}(\vartheta,\varphi)
\nonumber\\
&&\hskip 3.cm
+\,{\rm e}^{- i \mathbf{k}\cdot\mathbf{z}}
     u^*_{\ell m}(t,\mathbf{k}) 
              \hat{a}^\dagger_{\ell m}(\mathbf{k}) 
              \partial_{\varphi}Y^*_{\ell m}(\vartheta,\varphi)
 \Big)
 \,,\qquad\;
\label{field decomposition: phi derivative}
\end{eqnarray}
from which we conclude that only the diagonal terms $\mu=\nu$ contribute 
to $\big\langle\partial_\mu\hat\phi(x)\partial_\nu\hat\phi(x)\big\rangle$.
We therefore have,
\begin{eqnarray}
\Big\langle\big(\partial_t\hat\phi(x)\big)^2 \Big\rangle 
&\!\!=\!\!& \sum_{\ell m}\int \frac{{\rm d}^{D\!-\!3}k}{(2\pi)^{D\!-\!3}}
\Big(\omega_\ell^2\,
   |u_{\ell m}(t,\mathbf{k})|^2  
   \,Y_{\ell m}(\vartheta,\varphi) Y^*_{\ell m}(\vartheta,\varphi)
 \Big)
 \,,\qquad\;
\label{field decomposition: time derivative 2}\\
\Big\langle\big(\partial_{\mathbf{z}}\hat\phi(x)\big)^2 \Big\rangle 
&\!\!=\!\!& \sum_{\ell m}\int \frac{{\rm d}^{D\!-\!3}k}{(2\pi)^{D\!-\!3}}
\Big(\|\mathbf{k}\|^2\,
   |u_{\ell m}(t,\mathbf{k})|^2  
   \,Y_{\ell m}(\vartheta,\varphi) Y^*_{\ell m}(\vartheta,\varphi)
 \Big)
 \,,\qquad\;
\label{field decomposition: z derivative 2}\\
\Big\langle\big(\partial_{\vartheta}\hat\phi(x)\big)^2 \Big\rangle
&\!\!=\!\!& \sum_{\ell m}\int \frac{{\rm d}^{D\!-\!3}k}{(2\pi)^{D\!-\!3}}
\Big(|u_{\ell m}(t,\mathbf{k})|^2 
   \partial_{\vartheta}Y_{\ell m}(\vartheta,\varphi)
              \partial_{\vartheta}Y^*_{\ell m}(\vartheta,\varphi)
 \Big)
 \,,\qquad\;
\label{field decomposition: theta derivative 2}\\
\Big\langle\big(\partial_{\varphi}\hat\phi(x)\big)^2 \Big\rangle
&\!\!=\!\!& \sum_{\ell m}\int \frac{{\rm d}^{D\!-\!3}k}{(2\pi)^{D\!-\!3}}
\Big(|u_{\ell m}(t,\mathbf{k})|^2 
   \partial_{\varphi}Y_{\ell m}(\vartheta,\varphi)
              \partial_{\varphi}Y^*_{\ell m}(\vartheta,\varphi)
 \Big)
 \,,\qquad\;
\label{field decomposition: phi derivative 2}
\end{eqnarray}
where we used the commutation relations
in Eq.~(\ref{commutation relation a and a+})
and $\hat a_{\ell m}(\mathbf{k})|\Omega\rangle=0$, and no time ordering is yet imposed.
The next question we address is how the $T$-ordering would affect the results in 
Eqs.~(\ref{field decomposition: time derivative 2})--(\ref{field decomposition: phi derivative 2}).
Comparing the definition of time ordering, $T\big[\hat A\hat B\big] = \Theta(\Delta t)\big[\hat A\hat B\big] 
+ \Theta(-\Delta t)\big[\hat B\hat A\big]$, with the operator structure 
of Eqs.~(\ref{field decomposition: time derivative 2})--(\ref{field decomposition: phi derivative 2}),
we see that in these equations $\hat A=\hat B$. Since $T\big[\hat A\hat A\big] = (\hat A)^2$,
operator ordering in Eqs.~(\ref{field decomposition: time derivative 2})--(\ref{field decomposition: phi derivative 2}) is immaterial. In other words, calculating 
Eqs.~(\ref{field decomposition: time derivative 2})--(\ref{field decomposition: phi derivative 2})
is equivalent to calculating the time-ordered product of the same quantities. With insight this in mind we  
shall drop time ordering in the rest of the appendix. 
Next, making use of Eqs.~(\ref{spherical harmonics: completeness relation 2})
and~(\ref{canonically normalized mode functions}), 
Eqs.~(\ref{field decomposition: time derivative 2}) 
and~(\ref{field decomposition: z derivative 2}) can be simplified to, 
\begin{eqnarray}
\Big\langle\big(\partial_t\hat\phi(x)\big)^2 \Big\rangle 
&\!\!=\!\!&\frac{\hbar}{8\pi a^2}   \int \frac{{\rm d}^{D\!-\!3}k}{(2\pi)^{D\!-\!3}} \, \sum_{\ell}(2\ell+1)\omega_\ell
 \,,\qquad\;
\label{field decomposition: time derivative 3}\\
\Big\langle\big(\partial_{\mathbf{z}}\hat\phi(x)\big)^2 \Big\rangle 
&\!\!=\!\!& \frac{\hbar}{8\pi a^2}\int \frac{{\rm d}^{D\!-\!3}k}{(2\pi)^{D\!-\!3}}\,
 \sum_{\ell}(2\ell+1)\frac{\|\mathbf{k}\|^2}{\omega_\ell}
 \,.\qquad\;
\label{field decomposition: z derivative 3}
\end{eqnarray}
To evaluate 
the angular contributions~(\ref{field decomposition: theta derivative 2})--(\ref{field decomposition: phi derivative 2}), the following spherical harmonics addition theorem is useful,
\begin{equation}
\sum_{m=-\ell}^{\ell} Y_{\ell m}(\vartheta,\varphi)
\, Y_{\ell m}^*(\vartheta',\varphi')
= \frac{2\ell+1}{4\pi} P_\ell(\mu)
\,,\qquad
\label{useful formula for spherical harmonics: 1}
\end{equation}
where
\begin{equation}
\mu = \hat{r}\cdot \hat{r}' = \cos(\vartheta)\cos(\vartheta')
+ \sin(\vartheta) \sin(\vartheta') \cos(\varphi-\varphi')
\,,\qquad
\label{useful formula for spherical harmonics: 1b}
\end{equation}
from where it is straightforward to obtain,
\begin{eqnarray}
\sum_m \partial_\varphi Y_{\ell m}(\vartheta,\varphi) \partial_\varphi Y_{\ell m}^*(\vartheta,\varphi)
&\!\!=\!\!& \sin^2(\vartheta) \frac{2\ell+1}{4\pi}
\big[P_\ell'(\mu)\big]_{\mu=1} 
\nonumber\\
&\!\!=\!\!& \sin^2(\vartheta) \frac{2\ell+1}{4\pi} \frac{\ell(\ell+1)}{2}
\,,\qquad
\label{useful formula for spherical harmonics: 2}\\
\sum_m \partial_\vartheta Y_{\ell m}(\vartheta,\varphi) \partial_\vartheta Y_{\ell m}^*(\vartheta,\varphi)
&\!\!=\!\!& \frac{2\ell+1}{4\pi}
\big[P_\ell'(\mu)\big]_{\mu=1} 
\nonumber\\
&\!\!=\!\!&  \frac{2\ell+1}{4\pi} \frac{\ell(\ell+1)}{2}
\,,\qquad
\label{useful formula for spherical harmonics: 2b}
\end{eqnarray}
where we used $\big[P_\ell'(\mu)\big]_{\mu=1} =\ell(\ell+1)/2$,
which can be derived from the Rodrigues formula. With these 
results in mind Eqs.~(\ref{field decomposition: theta derivative 2})--(\ref{field decomposition: phi derivative 2}) can be simplified to,
\begin{eqnarray}
\Big\langle\big(\partial_{\vartheta}\hat\phi(x)\big)^2 \Big\rangle
&\!\!=\!\!&\frac{\hbar}{8\pi a^2}   \int \frac{{\rm d}^{D\!-\!3}k}{(2\pi)^{D\!-\!3}} \, \sum_{\ell}(2\ell+1) 
   \frac{\ell(\ell+1)}{2\omega_\ell}
 \,,\qquad\;
\label{field decomposition: theta derivative 3}\\
\Big\langle\big(\partial_{\varphi}\hat\phi(x)\big)^2 \Big\rangle
&\!\!=\!\!& \sin^2(\vartheta)\frac{\hbar}{8\pi a^2}   \int \frac{{\rm d}^{D\!-\!3}k}{(2\pi)^{D\!-\!3}} \, \sum_{\ell}(2\ell+1) \frac{\ell(\ell+1)}{2\omega_\ell}
 \,.\qquad\;
\label{field decomposition: phi derivative 3}
\end{eqnarray}
Finally, from 
Eq.~(\ref{one loop energy-momentum tensor: T*2})
we see that we also need $\big\langle\hat\phi^2(x)\big\rangle$.
Repeating the same steps as above, one obtains,
\begin{eqnarray}
\big\langle\hat\phi^2(x)\big\rangle
&\!\!=\!\!&\frac{\hbar}{8\pi a^2}   \int \frac{{\rm d}^{D\!-\!3}k}{(2\pi)^{D\!-\!3}} \, \sum_{\ell}(2\ell+1) \frac{1}{\omega_\ell}
 \,.\qquad\;
\label{field decomposition: phi2}
\end{eqnarray}
Making use of the wormhole metric~(\ref{flat metric}) one obtains,
\begin{eqnarray}
    \big\langle \hat{\mathcal{L}}\big\rangle
    &\!\equiv\!& -\frac12 g_{(0)}^{\alpha\beta} \,
     \big\langle \partial_\alpha\hat\phi(x)
     \partial_\beta\hat\phi(x) \big\rangle
     -\frac12 m^2\big\langle  \hat\phi^2(x) \big\rangle
\,,\qquad
\nonumber\\
&\!\!=\!\!& -\frac12 \frac{\hbar}{8\pi a^2}   
\int\! \frac{{\rm d}^{D\!-\!3}k}{(2\pi)^{D\!-\!3}}  \sum_{\ell}(2\ell\!+\!1) \left[-\omega_\ell
    \!+\!\frac{\|\mathbf{k}\|^2}{\omega_\ell}
    \!+\!\frac{\ell(\ell\!+\!1)}{a^2\omega_\ell}
    \!+\!\frac{m^2}{\omega_\ell}\right]
\qquad\!\!
\nonumber\\
&\!\!=\!\!& 0
\,,\qquad
\label{expectation value: Lagrangian}
\end{eqnarray}
such that the energy-momentum tensor in 
Eq.~(\ref{one loop energy-momentum tensor: T*2}) simplifies to,
\begin{eqnarray}
    \big\langle\Delta \hat{T}_{\mu}^{\;\nu}(x)\big\rangle^{(1)}
     &\!\!=\!\!& g_{(0)}^{\nu\rho}\,\big\langle \partial_\mu\hat\phi(x)
               \partial_\rho\hat\phi(x) \big\rangle
\,,\qquad
\label{one loop energy-momentum tensor: order hbar 2}
\end{eqnarray}
and, therefore, its nonvanishing components are, 
\begin{eqnarray}
    \big\langle\Delta \hat{T}_{t}^{\;t}(x)\big\rangle^{(1)}
     &\!\!=\!\!& - 
\Big\langle\big(\partial_t\hat\phi(x)\big)^2 \Big\rangle 
= - \frac{\hbar}{8\pi a^2}   \int \frac{{\rm d}^{D\!-\!3}k}{(2\pi)^{D\!-\!3}} \, \sum_{\ell}(2\ell+1)\omega_\ell
 \,,\qquad\;
\label{T tt}\\
    \big\langle\Delta \hat{T}_{z^i}^{\;z^i}(x)\big\rangle^{(1)}
\!\!&\!\!\!=\!\!\!&\!
\Big\langle\big(\partial_{z^i}\hat\phi(x)\big)^2 \Big\rangle 
=\frac{\hbar}{8\pi a^2}\int\! \frac{{\rm d}^{D\!-\!3}k}{(2\pi)^{D\!-\!3}}
 \sum_{\ell}(2\ell\!+\!1)\frac{\|\mathbf{k}\|^2}{(D\!-\!3)\omega_\ell}
\, ,\qquad\;\,
\label{T zz}\\
    \big\langle\Delta \hat{T}_{\vartheta}^{\;\vartheta}(x)\big\rangle^{(1)}
&\!\!\!=\!\!\!&
\big\langle\Delta \hat{T}_{\varphi}^{\;\varphi}(x)\big\rangle^{(1)}
= \frac{\hbar}{8\pi a^2}   \int \frac{{\rm d}^{D\!-\!3}k}{(2\pi)^{D\!-\!3}} \, \sum_{\ell}(2\ell\!+\!1) 
      \frac{\ell(\ell\!+\!1)}{2a^2\omega_\ell}
 \,.\qquad
\label{T thetatheta}
\end{eqnarray}

\bigskip
\noindent
{\bf Renormalization.} The primitive one-loop energy-momentum tensor 
in Eqs.~(\ref{T tt})--(\ref{T thetatheta}) is expressed in terms of 
composite field operators, and therefore it is divergent.

Let us now first consider vacuum fluctuations in
an infinite wormhole, $a\rightarrow \infty$
({\it i.e.} the flat Minkowski space limit), for which
the propagator obeys the equation, 
\begin{equation}
\big(\partial^2 \!-\! m^2\big)i\Delta_M(x;x') = i\hbar \delta^D(x\!-\!x' )
\,,\qquad
\big(\partial^2 = \eta^{\mu\nu} \partial_\mu\partial_\nu\big)
\,.\qquad
\label{propagator: Minkowski: eom}
\end{equation}
The Minkowski space propagator is given by, 
\begin{eqnarray}
i\Delta_M(x;x' )  &\!\!=\!\!& \frac{\hbar}{2}\!  \int \!\frac{{\rm d}^{D-1}k}{(2\pi)^{D-1}} 
     \frac{{\rm e}^{i\mathbf{k}\cdot\mathbf{x}}}{\omega(\mathbf{k})}
\left[\Theta(t\!-\!t'){\rm e}^{-i\omega(\mathbf{k})(t-t')}
  \!+\!\Theta(t'\!-\!t){\rm e}^{i\omega(\mathbf{k})(t-t')}\right]
\nonumber\\
 &\!\!=\!\!& \frac{\hbar m^{D-2}}{(2\pi)^{D/2}}\frac{K_{\frac{D-2}{2}}\big(m\sqrt{\Delta x^2}\big)}
   {\big(m\sqrt{\Delta x^2}\big)^{\frac{D-2}{2}}}
\,,\qquad
\label{propagator: Minkowski}
\end{eqnarray}
where $K_\nu(z)$ is the Macdonald function (the Bessel function of the second kind),
\begin{equation}
 \Delta x^2(x;x') = -\big(|t\!-\!t' |\!-\!i\varepsilon\big)^2 \!+\! \|\mathbf{x}\!-\!\mathbf{x}'\|^2
\,,\qquad
\label{Delta x2: Feynman propagator}
\end{equation}
is the (Lorentz) invariant distance squared between $x$ and $x'$,
 and the $i\varepsilon$ piece 
($\varepsilon>0$ and $\varepsilon\rightarrow 0$) 
generates the imaginary part of the propagator that 
corresponds that of the vacuum Feynman (time ordered) propagator.
The coincident limit of the Feynman propagator in Eq.~(\ref{propagator: Minkowski}) reads,
\begin{equation}
i\Delta_M(x;x ) = \big\langle T\big[\hat\phi^2(x)\big]\big\rangle  
 = \frac{\hbar}{2}\!  \int\!\! \frac{{\rm d}^{D-1}k}{(2\pi)^{D-1}} 
     \frac{1}{\omega(\mathbf{k})}
    = \frac{\hbar m^{D-2}}{(4\pi)^{\frac{D}{2}}}
\Gamma\Big(1\!-\frac{D}{2}\Big)
\,,
\label{coincident propagator: Minkowski}
\end{equation}
where to get this result we made use of the following master integral~\cite{Peskin:1995ev},
\begin{equation}
\int\! \frac{d^{d}k}{(2\pi)^{d}}
\frac{1}{(k^{2}+M^{2})^{\beta}}
= \frac{1}{(4\pi)^{d/2}}
\frac{\Gamma\!\left(\beta-\frac{d}{2}\right)}{\Gamma(\beta)}
\left(M^{2}\right)^{\frac{d}{2}-\beta}
\,,\qquad
\label{master integral}
\end{equation}
where $\omega(\mathbf{k})=\sqrt{\|\mathbf{k}\|^2+m^2}$.
It is now not hard to show that, 
\begin{eqnarray}
\big\langle \big(\partial_t\hat\phi(x)\big)\big(\partial^t\hat\phi(x)\big)
\big\rangle^{(1)}_M 
&\!\!\!=\!\!\!&  - \frac{\hbar}{2}\!\int\! \frac{{\rm d}^{D-1}k}{(2\pi)^{D-1}} 
     \omega(\mathbf{k})
  = \frac{\hbar}{2}\frac{m^D}{(4\pi)^{\frac{D}{2}}}\Gamma\Big(\!\!-\!\frac{D}{2}\Big)
\,,\qquad
\label{dt dt: Mink}\\
\big\langle \big(\partial_i\hat\phi(x)\big)\big(\partial^j\hat\phi\big)
           \big\rangle^{(1)}_M
&\!\!=\!\!& 
\frac{\hbar}{2}\frac{\delta_i^{\;j}}{D\!-\!1}\int \frac{{\rm d}^{D-1}k}{(2\pi)^{D-1}} 
     \frac{\|\mathbf{k}\|^2}{\omega(\mathbf{k})}
\nonumber\\
 &&\hskip -3.8cm
=\,  \frac{\hbar}{2}\frac{\delta_i^{\;j}}{D\!-\!1}\int\! \frac{{\rm d}^{D-1}k}{(2\pi)^{D-1}} 
    \left(\!\omega(\mathbf{k}) \!-\! \frac{m^2}{\omega(\mathbf{k})}
    \right)
= \frac{\hbar}{2}\delta_i^{\;j}\frac{m^D}{(4\pi)^{\frac{D}{2}}}\Gamma\Big(\!\!-\!\frac{D}{2}\Big)
\,,\quad\;\;\;
\label{T ii: Mink}
\end{eqnarray}
such that,
\begin{eqnarray}
\big\langle  \mathcal{\hat{L}}\big\rangle^{(1)}_M
\equiv -\frac12\eta^{\mu\nu}\big\langle
   \big(\partial_\mu\hat\phi(x)\big)\big(\partial_\nu\hat\phi(x)\big)
             \big\rangle^{(1)}_M 
\!-\! \frac{ m^2}{2}\big\langle \big(\hat\phi(x)\big)^2\big\rangle^{(1)}_M
    = 0
\,,
\nonumber\\
&&
\label{L: Mink}
\end{eqnarray}
where we made use of the fact that operator ordering in the above calculations is immaterial.
From here it then follows that
the nonvanishing elements of 
the one-loop energy-momentum tensor in Minkowski vacuum are,
\begin{eqnarray}
\big\langle \Delta\hat T_{t}^{\;t}\big\rangle^{(1)}_M 
&\!\!=\!\!& 
\big\langle \big(\partial_t\hat\phi(x)\big)\big(\partial^t\hat\phi(x)\big)
\big\rangle^{(1)}_M  
= \frac{\hbar}{2}\frac{m^D}{(4\pi)^{\frac{D}{2}}}\Gamma\Big(\!-\frac{D}{2}\Big)
\,,\qquad
\label{T tt: Mink}\\
\big\langle \Delta\hat T_{i}^{\;j}\big\rangle^{(1)}_M 
&\!\!=\!\!&\big\langle \big(\partial_i\hat\phi(x)\big)\big(\partial^j\hat\phi(x)\big)
\big\rangle^{(1)}_M  
= \frac{\hbar}{2}\delta_i^{\;j}\frac{m^D}{(4\pi)^{\frac{D}{2}}}\Gamma\Big(\!-\frac{D}{2}\Big)
\,,\qquad
\label{T ij: Mink}
\end{eqnarray}
such that,
\begin{eqnarray}
 \big\langle \Delta\hat T_{\mu}^{\;\nu}\big\rangle^{(1)}_M 
&\!\!=\!\!&  \frac{\hbar}{2}\!\times\! \frac{m^D}{(4\pi)^{D/2}}
     \Gamma\Big(\!-\frac{D}{2}\Big) \delta_{\mu}^{\;\nu}
\nonumber\\
&\!\!=\!\!& - \frac{\hbar}{2}\!\times\! \frac{m^4}{(4\pi)^{2}}
     \frac{\mu^{D-4}}{D\!-\!4}
     \left[1+\frac{D\!-\!4}{2}\left(
     \ln\Big(\frac{m^2}{4\pi\mu^2}\Big)
     \!+\!\gamma_E\!-\!\frac32\right)\right]\delta_{\mu}^{\;\nu}
\,,\qquad\;\;
\label{T mn: Mink 3}
\end{eqnarray}
where to get the last equality we used,
\begin{equation}
    \Gamma\Big(\!-\frac{D}{2}\Big) 
    = -\frac{1}{D\!-\!4}\left[1+\frac{D\!-\!4}{2}\left(\gamma_E-\frac32\right)
      \right]
\,,\qquad
\label{Gamma -D/2}
\end{equation}
where $\gamma_E = -\psi(1) = 0.577\cdots$ is the Euler-Mascheroni constant.
In Eq.~(\ref{T mn: Mink 3}) we have introduced a fiducial scale $\mu$, which will later turn out to be useful for renormalization.

In what follows we use the method of Minkowski vacuum subtraction,
which is often used in Casimir effect studies~\cite{Bordag:2009zz}.
It is convenient to subtract the one-loop contribution from 
the Minkowski space vacuum fluctuations, in which fluctuations on
the sphere $S^2$ are replaced by fluctuations on the plane $\mathbb{R}^2$.

Our Minkowski space coordinates are denoted as $t\in \mathbb{R}$,
$\mathbf{z}\in \mathbb{R}^{D-3}$ and $\mathbf{x}\in \mathbb{R}^2$,
and the corresponding momenta are denoted as 
$\mathbf{k}\in \mathbb{R}^{D-3}$ and $\mathbf{p}\in \mathbb{R}^{2}$.
With these remarks in mind, subtracting the Minkowski vacuum contributions 
in Eqs.~(\ref{T tt: Mink})--(\ref{T ij: Mink}) 
from the energy-momentum tensors in 
Eqs.~(\ref{T tt})--(\ref{T thetatheta}) results in,   
\begin{eqnarray}
&&\hskip -0.7cm    
\big\langle\Delta \hat{T}_{t}^{\;t}(x)\big\rangle^{(1)}
    -\big\langle\Delta \hat{T}_{t}^{\;t}(x)\big\rangle^{(1)}_M
\nonumber\\
&& \hskip 0cm
=\,  - \frac{\hbar}{2}   \int \frac{{\rm d}^{D-3}k}{(2\pi)^{D-3}} 
\left[\frac{1}{4\pi a^2}  \sum_{\ell}(2\ell+1)\omega_\ell
   \!-\!\int\! \frac{{\rm d}^2p}{(2\pi)^2} 
     \omega(\mathbf{k},\mathbf{p})
     \right]
 \,,\qquad\;
\label{T tt 2}\\
 &&\hskip -0.7cm  
 \big\langle\Delta \hat{T}_{z^i}^{\;z^i}(x)\big\rangle^{(1)}
    -\big\langle\Delta \hat{T}_{z^i}^{\;z^i}(x)\big\rangle^{(1)}_M
\nonumber\\
&& \hskip 0.cm
=\,
\frac{\hbar}{2}\int \frac{{\rm d}^{D-3}k}{(2\pi)^{D-3}}
\frac{\|\mathbf{k}\|^2}{(D\!-\!3)}
\left[
\frac{1}{4\pi a^2}
 \sum_{\ell}(2\ell\!+\!1)\frac{1}{\omega_\ell}
 \!-\!\!\int\! \frac{{\rm d}^2p}{(2\pi)^2} 
     \frac{1}{\omega(\mathbf{k},\mathbf{p})}
     \right]
 \,,\qquad
\label{T zz 2}\\
 &&\hskip -0.7cm 
    \big\langle\Delta \hat{T}_{\vartheta}^{\;\vartheta}(x)\big\rangle^{(1)}
    -\big\langle\Delta \hat{T}_{\vartheta}^{\;\vartheta}(x)\big\rangle^{(1)}_M
    = \big\langle\Delta \hat{T}_{\varphi}^{\;\varphi}(x)\big\rangle^{(1)}
    -\big\langle\Delta \hat{T}_{\varphi}^{\;\varphi}(x)\big\rangle^{(1)}_M
\nonumber\\
&& \hskip 0.cm
=\,
\frac{\hbar}{2}   \int \frac{{\rm d}^{D-3}k}{(2\pi)^{D-3}} \left[\frac{1}{4\pi a^2}  \sum_{\ell}(2\ell+1) \frac{\ell(\ell+1)}{2a^2\omega_\ell}
\!-\!\!\int\! \frac{{\rm d}^2p}{(2\pi)^2}
     \frac{\|\mathbf{p}\|^2}{2\omega(\mathbf{k},\mathbf{p})}
   \right]
 \,,\qquad\;
\label{T thetatheta 2}
\end{eqnarray}
where $\omega(\mathbf{k},\mathbf{p})=\sqrt{\|\mathbf{k}\|^2
 +\|\mathbf{p}\|^2+m^2}$, with no summation over $z^i$ in Eq.~(\ref{T zz 2}).
 
The expressions in square brackets can be evaluated by 
using Abel-Plana formulas~\cite{Saharian:2018}. 
Let us begin our analysis by introducing the following substitutions,
\begin{equation}
\nu \rightarrow \ell + \tfrac{1}{2}
\,, \qquad 
  \ell(\ell+1) \rightarrow \nu^2 - \tfrac{1}{4}
\,, \qquad
\|\mathbf{p}\|^2 \rightarrow \frac{1}{a^2}\left(\nu^2 - \frac14\right)
\,,\qquad 
\end{equation}
after which the expressions in square brackets in 
Eqs.~(\ref{T tt 2}) and~(\ref{T zz 2}) become,
\begin{eqnarray}
B_\omega
&\!\!\equiv\!\!& \frac{1}{4\pi a^2}  \sum_{\ell}(2\ell+1)\omega_\ell
   \!-\!\int\! \frac{{\rm d}^2p}{(2\pi)^2} 
     \omega(\mathbf{k},\mathbf{p})
 \,,\qquad\;
 \nonumber\\
 &\!\!=\!\!& \frac{1}{2\pi a^3} 
 \left[\sum_{\nu=\frac12,\frac32,\dots}\nu\sqrt{u^2-\frac14+\nu^2}
   \!-\!\int_\frac12^\infty {\rm d}\nu \nu
                    \sqrt{u^2-\frac14+\nu^2}
     \right]
     \,,\qquad
\label{T tt 3}\\
B_{\frac{1}{\omega}}
&\!\!=\!\!&
\frac{1}{2\pi a} 
 \left[\sum_{\nu=\frac12,\frac32,\dots}
 \frac{\nu}{\sqrt{u^2-\frac14+\nu^2}}
   \!-\!\int_\frac12^\infty\! {\rm d}\nu \,
   \frac{\nu}{\sqrt{u^2-\frac14+\nu^2}}
   \right]
 \,,
\label{T zz 3}
\end{eqnarray}
where $u^2 = a^2(m^2+\|\mathbf{k}\|^2)$.
Noting that $\int_\frac12^\infty {\rm d}\nu
= \int_0^\infty {\rm d}\nu - \int_0^\frac12{\rm d}\nu$,
we can use the half-integer Abel-Plana formula~\cite{Saharian:2018},
\begin{equation}
\sum_{n=0}^\infty f(n+\tfrac{1}{2}) - \int_0^\infty f(x)\,{\rm d}x
= -i \int_0^\infty \frac{f(it)-f(-it)}{e^{2\pi t}+1}\, {\rm d}t
\,,\qquad
\label{half-integer Abel-Plana formula}
\end{equation}
which for 
\begin{equation}
f(\nu) \rightarrow \nu \sqrt{u^2 + \nu^2 - \tfrac{1}{4}}
\quad {\rm and}\quad
f(\nu) \rightarrow \frac{\nu}{\sqrt{u^2 + \nu^2 - \tfrac{1}{4}}}
\,,\qquad
\end{equation}
gives,
\begin{eqnarray}
B_\omega
 &\!\!=\!\!& \frac{1}{\pi a^3} 
 \left[
 \int_0^{\sqrt{u^2-\frac14}} \!
\frac{t\sqrt{u^2-\frac14 - t^2}}{e^{2\pi t} + 1}\,{\rm d}t 
   \!+\!\frac12\int_0^\frac12 {\rm d}\nu \nu
                    \sqrt{u^2-\frac14+\nu^2}
     \right]
 \,,\qquad
\label{T tt 4}\\
B_{\frac{1}{\omega}}
&\!\!=\!\!&
\frac{1}{\pi a} 
 \left[
 \int_{0}^{\sqrt{u^2-\frac14}}\!\! \frac{t\,{\rm d}t}{\sqrt{u^2-\frac14-t^2}\,({e^{2\pi t}+1})}
   \!+\!\frac12\int_0^\frac12
   \frac{\nu\,{\rm d}\nu}{\sqrt{u^2-\frac14+\nu^2}}
   \right]
 \,,\qquad\;
\label{T zz 4}
\end{eqnarray}
where we assumed that $t<\sqrt{u^2-\frac14}$, which is a reasonable assumption for large wormholes, for which $u\geq am\gg 1$.
For light particles, for which $am\leq 1/2$, there is no cut contribution from the
Abel-Plana formula~(\ref{half-integer Abel-Plana formula}) to $B_\omega$ and $B_\frac{1}{\omega}$ in 
Eqs.~(\ref{T tt 4})--(\ref{T zz 4}), and we have,
\begin{equation}
    B_\omega = 0 = B_\frac{1}{\omega} \,,\qquad \Big(am\leq \frac12\Big)
\,,\qquad
\label{Bomega and B1/omega for light particles}
\end{equation}
implying that light and massless scalar particles satisfying $am\leq 1/2$ 
do not produce any quantum backreaction at the one-loop level inside the wormhole.
It would be of interest to investigate whether analogous results holds for particles with spin.

The latter integrals in Eqs.~(\ref{T tt 4})--(\ref{T zz 4}) 
are easily evaluated, 
\begin{eqnarray}
\frac12\int_0^\frac12 {\rm d}\nu \,\nu \sqrt{u^2-\frac14+\nu^2} 
    &\!\!=\!\!& \frac16\left[u^3 -\left(u^2-\frac14\right)^\frac32\right]
\,,\qquad
\nonumber\\
\frac12\int_0^\frac12 {\rm d}\nu\,\frac{\nu}{\sqrt{u^2-\frac14+\nu^2}}
   &\!\!=\!\!& \frac12\left[u -\sqrt{u^2-\frac14}\right]
\,.\qquad
\label{2 simple integrals}
\end{eqnarray}
The former integrals in Eqs.~(\ref{T tt 4})--(\ref{T zz 4}) are hard, 
and one has to resort to approximations to evaluate them.
Expanding the integrand in powers of ${\rm e}^{-2\pi t}$
and changing the integration variable, $t\rightarrow u_0\tau$, with $u_0\equiv \sqrt{u^2-\frac14}$,
one obtains,
\begin{eqnarray}
B_\omega
 &\!\!=\!\!& \frac{1}{\pi a^3} 
 \left[\!-u_0^3\sum_{n=1}^\infty(-1)^n
 \int_0^1 \tau\sqrt{1- \tau^2}{\rm e}^{-2\pi n \tau}\,{\rm d}\tau 
   \!+\!\frac16\big( u^3 \!- \! u_0^3\big)
     \right]
\,,\qquad
\label{T tt 4 A}\\
B_{\frac{1}{\omega}}
&\!\!=\!\!&
\frac{1}{\pi a} 
 \left[\!-u_0\sum_{n=1}^\infty(-1)^n
 \int_{0}^1 \frac{\tau\,{\rm d}\tau}{\sqrt{1-\tau^2}} {\rm e}^{-2\pi u_0\tau}
   \!+\!\frac12\big(u \!-\! u_0\big)
   \right]
 \,.\qquad
\label{T zz 4 A}
\end{eqnarray}
These integrals can be evaluated,
\begin{eqnarray}
B_\omega
 &\!\!=\!\!& \frac{1}{\pi a^3} 
 \biggl\{\!-u_0^3\sum_{n=1}^\infty(-1)^n
  \biggl[\frac13\!+\!\frac{(2\pi n u_0)^2}{15}\!\times\!{}_1F_2\Big(1;\frac12,\frac32;(\pi n u_0)^2\Big)
 \nonumber\\
 &&\hskip 3cm
 -\,\frac{1}{4n u_0}\!\times\!I_2\big(2\pi n u_0\big)\biggr]
    \!+\!\frac16\big( u^3 \!- \! u_0^3\big)\biggr\}
\,,\qquad
\label{T tt 4 B}\\
B_{\frac{1}{\omega}}
\!&\!\!=\!\!&\!
\frac{1}{\pi a} 
 \biggl\{\!-u_0\sum_{n=1}^\infty(-1)^n
 \Bigl[ {}_1F_2\Big(1;\frac12,\frac32;(\pi n u_0)^2\Big)\!-\frac{\pi}{2}I_1\big(2\pi n u_0\big)\Bigr]
     \Bigr]
 \nonumber\\
 &&\hskip 3cm
   +\,\frac12\big(u \!-\! u_0\big)\Bigr]
   \biggr\}
 ,\qquad
\label{T zz 4 B}
\end{eqnarray}
where ${}_1F_2$  and $I_\nu(z)$ denote a generalized hypergeometric function and a modified Bessel function
of the first kind, respectively.
We are primarily interested in large wormholes, which corresponds to the large argument
limit (asymptotic expansion) of special functions in Eqs.~(\ref{T tt 4 B})--(\ref{T zz 4 B}), which can 
be found, for example, in section 16.11 of Ref.~\cite{NIST:2026}. Since both functions have 
 branch cuts at large argument values, to get the correct expression for the integrals, one should take  average value above and below the cuts, to obtain,

\begin{eqnarray}
B_\omega
 &\!\!=\!\!& \frac{1}{\pi a^3} 
 \biggl\{\!-u_0^3\sum_{n=1}^\infty(-1)^n
  \biggl[
     \frac{1}{(2\pi nu_0)^2} \!-\! \frac{3}{(2\pi nu_0)^4}
     \!-\! \frac{15}{(2\pi nu_0)^6} \!-\! \frac{315}{(2\pi nu_0)^8}
 \nonumber\\
&&\hskip 2.8cm
    -\, \frac{14175}{(2\pi nu_0)^{10}}
   \!+\!\mathcal{O}\big(u_0^{-12}\big)
 \biggr]
    \!+\!\frac16\big( u^3 \!- \! u_0^3\big)\biggr\}
\,,
\label{T tt 4 C}\\
B_{\frac{1}{\omega}}
\!&\!\!=\!\!&\!
\frac{1}{\pi a} 
 \biggl\{\!-u_0\sum_{n=1}^\infty(-1)^n
 \biggl[\frac{1}{(2\pi nu_0)^2} \!+\! \frac{3}{(2\pi nu_0)^4}
     \!+\! \frac{45}{(2\pi nu_0)^6} \!+\! \frac{1575}{(2\pi nu_0)^8}
\nonumber\\
&&\hskip 2.8cm
 +\, \frac{99225}{(2\pi nu_0)^{10}}
  \!+\!\mathcal{O}\big(u_0^{-12}\big)
 \biggr]
   \!+\!\frac12\big(u \!-\! u_0\big)\Bigr]
   \biggr\}
 \,.\qquad
\label{T zz 4 C}
\end{eqnarray}

Next, one can perform the sums to obtain,
\begin{eqnarray}
B_\omega
 &\!\!=\!\!& \frac{1}{\pi a^3} 
 \biggl\{u_0^3
  \biggl[\frac{1}{48 u_0^2} \!-\! \frac{7}{3840 u_0^4} \!-\! \frac{31}{129024 u_0^6}
     \!-\! \frac{127}{983040 u_0^8} 
 \nonumber\\
&&\hskip 2.8cm
    -\, \frac{2555}{17301504 u_0^{10}}
   \!+\!\mathcal{O}\big(u_0^{-12}\big)
 \biggr]
    \!+\!\frac16\big( u^3 \!- \! u_0^3\big)\biggr\}
\,,\qquad
\label{T tt 4 D}\\
B_{\frac{1}{\omega}}
\!&\!\!=\!\!&\!
\frac{1}{\pi a} 
 \biggl\{u_0
 \biggl[\frac{1}{48 u_0^2} \!+\! \frac{7}{3840 u_0^4} \!+\! \frac{31}{43008 u_0^6}
  \!+\! \frac{127}{196608 u_0^8}
\nonumber\\
&&\hskip 2.8cm
 +\, \frac{17885}{17301504 u_0^{10}}
  \!+\!\mathcal{O}\big(u_0^{-12}\big)
 \biggr]
   \!+\!\frac12\big(u \!-\! u_0\big)\Bigr]
   \biggr\}
\, ,\qquad\;
\label{T zz 4 D}
\end{eqnarray}
where to get these results we made use of, 
\begin{equation}
\sum_{n=1}^\infty \frac{(-1)^{n}}{{n}^z}
= -\left(1-\frac{1}{2^{z-1}}\right)\zeta(z)
\,.\qquad
\label{useful sum}
\end{equation}
Here $\zeta(z)=\sum_{n=1}^\infty \frac{1}{{n}^z}$ is the Riemann zeta function, which when evaluated at even integers takes simple values: $\zeta(2)=\pi^2/6$,  $\zeta(4)=\pi^4/90$,
$\zeta(6)=\pi^6/945$,  $\zeta(8)=\pi^8/9450$,  $\zeta(10)=\pi^{10}/93555$, {\it etc.}
In general, one can obtain $\zeta(2n)\, (n\in \mathbb{N})$ from its generating  function,
\begin{equation}
\sum_{n=0}^\infty \zeta(2n) x^{2n} = -\frac{\pi x}{2}\coth(\pi x) 
\,,\qquad
\label{generating function zeta 2n}
\end{equation}
and from here one can show that $\zeta(2n)$ can be expressed in terms of Bernoulli numbers $B_{2n}$ as,
$\zeta(2n)=(-1)^{n+1}B_{2n}(2\pi)^{2n}/[2(2n)!]$. 

An inspection of the coefficients of the asymptotic series in the square brackets in 
Eqs.~(\ref{T tt 4 B})--(\ref{T zz 4 B}) shows that they can be written with a help of the Pochhammer symbols, 
$(a)_{n'} \equiv \Gamma(a+n')/\Gamma(a)$, as
$2^{2(n'-1)}\big(\!-\frac12\big)_{n'-1}\big(\frac32\big)_{n'-1}$ ${}/(2\pi n u_0)^{2n'}$ and 
$-2^{2n'}\big(\!-\frac12\big)_{n'}\big(\frac12\big)_{n'}/(2\pi n u_0)^{2n'}$ ($n'=1,2,\dots$), 
respectively. This is however hard to prove from the complicated form of the asymptotic series as given in Ref.~\cite{NIST:2026}.
In what follows, we prove that by construction, {\it i.e} by evaluating the integrals in 
Eqs.~(\ref{T tt 4})--(\ref{T zz 4}) by a different approximate method. We begin by
expanding the square roots in Eqs.~(\ref{T tt 4})--(\ref{T zz 4}),
\begin{eqnarray}
  \sqrt{u^2-\tfrac14 - t^2} 
 &\!\!=\!\!&u_0\sum_{n=0}^\infty \frac{\big(\!-\frac12\big)_n}{n!}
    \left(\frac{t}{u_0}\right)^{2n}
\,,\quad
\nonumber\\
  \frac{1}{ \sqrt{u^2-\tfrac14 - t^2} } 
  &\!\!=\!\!& \frac{1}{u_0}\sum_{n=0}^\infty \frac{\big(\frac12\big)_n}{n!}
    \left(\frac{t}{u_0}\right)^{2n}
,\quad
\label{expansion: square roots}
\end{eqnarray}
where $(a)_n=\Gamma(a+n)/\Gamma(a)$ denotes the Pochammer symbol and, 
\begin{equation}
  u_0=\sqrt{u^2-\tfrac14} = \sqrt{(am)^2 + \big(a\|\mathbf{k}\|\big)^2-\tfrac14} 
  \,\,.\qquad
\label{u0: def}
\end{equation}
Both series in Eq.~(\ref{expansion: square roots})
converge for $t<u_0$, which suffices for our purposes.
Inserting Eqs.~(\ref{2 simple integrals}) 
and~(\ref{expansion: square roots}) into Eqs.~(\ref{T tt 4})--(\ref{T zz 4})
gives,
\begin{eqnarray}
\!B_\omega
 &\!\!\!=\!\!\!& \frac{1}{\pi a^3} 
\Bigg\{
\sum_{n=0}^\infty \frac{\big(\!-\frac12\big)_n}{n!}u_0^{1-2n}\!
    \int_{0}^{u_0} \frac{t^{1+2n}\,{\rm d}t}{{e^{2\pi t}\!+\!1}}
    \!+\!\frac16\left(u^3 \!-\!u_0^3\right)
\Bigg\}
,\!\!
\nonumber\\
\label{T tt 5}\\
\!B_{\frac{1}{\omega}}
&\!\!\!=\!\!\!&
\frac{1}{\pi a} 
 \Bigg\{\sum_{n=0}^\infty \frac{\big(\frac12\big)_n}{n!}
\frac{1}{u_0^{1+2n}}
   \!\! \int_{0}^{u_0} \frac{t^{1+2n}\,{\rm d}t}{{e^{2\pi t}\!+\!1}}
    \!+\!\frac12\left(u \!-\!u_0\right)
   \Bigg\}
\,.\quad\;\;\;
\label{T zz 5}
\end{eqnarray}
The integral can be evaluated by firstly rewriting it as,  
\begin{equation}
\int_0^{u_0}\frac{t^{1+2n}\,{\rm d}t}{{e^{2\pi t}\!+\!1}}
= \int_0^\infty  \frac{t^{1+2n}\,{\rm d}t}{{e^{2\pi t}\!+\!1}}
-\int_{u_0}^\infty  \frac{t^{1+2n}\,{\rm d}t}{{e^{2\pi t}\!+\!1}}
\,.\qquad
\label{integral up to u: up to inf min}
\end{equation}
The second integral is exponentially suppressed in $u_0$ as,
\begin{equation}
\int_{u_0}^\infty  \frac{t^{1+2n}\,{\rm d}t}{{e^{2\pi t}\!+\!1}}
   = \frac{{u_0}^{1+2n}}{2\pi}\,{\rm e}^{-2\pi u_0} 
 \left[1+\mathcal{O}\left(\frac{1}{u_0},{\rm e}^{-2\pi {u_0}}\right)\right]
\,,\qquad
\label{integral up to u: up to inf min: 2}
\end{equation}
and the first integral in 
Eq.~(\ref{integral up to u: up to inf min}) can be evaluated
exactly,
\begin{eqnarray}
\int_0^\infty  \frac{t^{1+2n}\,{\rm d}t}{{e^{2\pi t}\!+\!1}}
&\!\!=\!\!&
-\sum_{n'=1}^\infty \frac{(-1)^{n'}}{(2\pi n')^{2+2n}}\int_0^\infty v^{1+2n}
   {\rm e}^{-v} {\rm d}v
\nonumber\\
  &\!\!=\!\!&\frac{1}{(2\pi)^{2+2n}}
  \left(1\!-\!\frac{1}{2^{1+2n}}\right)\Gamma(2\!+\!2n)\zeta(2\!+\!2n)
\,,\qquad
\label{exact integral}
\end{eqnarray}
where we made use of Eq.~(\ref{useful sum}).
Inserting Eqs.~(\ref{integral up to u: up to inf min: 2})
and~(\ref{exact integral}) into 
Eqs.~(\ref{T tt 5})--(\ref{T zz 5}) results in,
\begin{eqnarray}
B_\omega
 &\!\!=\!\!& \frac{1}{\pi a^3} 
\Bigg\{
\sum_{n=0}^\infty \frac{\big(\!-\frac12\big)_n}{n!}
    \bigg[\frac{{u_0}^{1-2n}}{(2\pi)^{2+2n}}
  \left(1\!-\!\frac{1}{2^{1+2n}}\right)\Gamma(2\!+\!2n)\zeta(2\!+\!2n)
\nonumber\\
&& \hskip 3.4cm
 +\,\mathcal{O}\big({\rm e}^{-2\pi {u_0}}\big)
 \bigg]
    \!+\!\frac16\left(u^3 \!-\!u_0^3\right)
\Bigg\}
\,,\qquad\;
\label{T tt 6}\\
B_{\frac{1}{\omega}}
&\!\!=\!\!&
\frac{1}{\pi a} 
 \Bigg\{\sum_{n=0}^\infty \frac{\big(\frac12\big)_n}{n!}
   \bigg[ \frac{1}{(2\pi)^{2+2n}}\frac{1}{{u_0}^{1+2n}}
  \left(1\!-\!\frac{1}{2^{1+2n}}\right)\Gamma(2\!+\!2n)\zeta(2\!+\!2n)
\nonumber\\
&& \hskip 2.8cm
   +\,\mathcal{O}\big({\rm e}^{-2\pi {u_0}}\big)
  \bigg]
    \!+\!\frac12\left(u \!-\!u_0\right)
   \Bigg\}
 \,,\qquad\;
\label{T zz 6}
\end{eqnarray}
where, instead of inserting the results from Eq.~(\ref{integral up to u: up to inf min: 2}),
we have inserted just an estimate $\mathcal{O}\big({\rm e}^{-2\pi u_0}\big)$, as we are not sure of the precise form of  the asymptotic expansion that includes exponential suppression. This is because the results in 
Eq.~(\ref{integral up to u: up to inf min: 2}) multiply a series that, when summed, either gives zero or diverges. The reason for that ambiguous result can be traced back to the fact  
that the original integral in Eq.~(\ref{integral up to u: up to inf min}) contains a cut at $t\geq u_0$.

With these results we have achieved the intended goal, namely we have obtained an explicit form for the asymptotic series coefficients in Eqs.~(\ref{T tt 4 D})--(\ref{T zz 4 D}), which in light of Eq.~(\ref{useful sum}), can be used to reconstruct the coefficients of the asymptotic series in Eqs.~(\ref{T tt 4 C})--(\ref{T zz 4 C}).

Indeed, upon inserting Eqs.~(\ref{T tt 6})--(\ref{T zz 6}) into 
Eqs.~(\ref{T tt 2})--(\ref{T thetatheta 2}) one obtains,
\begin{eqnarray}
&&\hskip -0.7cm    
\big\langle\Delta \hat{T}_{t}^{\;t}(x)\big\rangle^{(1)}
    -\big\langle\Delta \hat{T}_{t}^{\;t}(x)\big\rangle^{(1)}_M
= -\frac{\hbar}{2}   \int \!\frac{{\rm d}^{D\!-\!3}k}{(2\pi)^{D\!-\!3}} 
B_\omega
\nonumber\\
&& \hskip -0.7cm
\approx\,
 \frac{\hbar}{2\pi a}
\Bigg\{\frac{16M^{D-2}}{(4\pi)^\frac{D+3}{2}a}\sum_{n=0}^\infty
   \frac{\Gamma\big(n+\frac32\big)\Gamma\big(n+1-\frac{D}{2}\big)}{(\pi aM)^{2n}}
  \left(1\!-\!\frac{1}{2^{1+2n}}\right)
    \zeta(2\!+\!2n)
\nonumber\\
 &&\hskip 1.cm
     - \,\,\frac{a}{4(4\pi)^\frac{D-2}{2}} \big(m^D\!-\!M^D\big)\Gamma\Big(\!-\frac{D}{2}\Big)
    + \mathcal{O}\big({\rm e}^{-2\pi aM}\big) \Bigg\}
 \,,\qquad\;
\label{T tt: exact}
\end{eqnarray}
\begin{eqnarray}
&&\hskip -0.7cm  
 \big\langle\Delta \hat{T}_{z^i}^{\;z^i}(x)\big\rangle^{(1)}
    -\big\langle\Delta \hat{T}_{z^i}^{\;z^i}(x)\big\rangle^{(1)}_M
=
\frac{\hbar}{2}\int \frac{{\rm d}^{D\!-\!3}k}{(2\pi)^{D\!-\!3}}
\frac{\|\mathbf{k}\|^2}{(D\!-\!3)}
B_{\frac{1}{\omega}}
\nonumber\\
&& \hskip -0.4cm
\approx\,\frac{\hbar}{2\pi a} 
 \Bigg\{\frac{16M^{D-2}}{(4\pi)^{\frac{D+3}{2}}a}
 \sum_{n=0}^\infty \frac{\Gamma\big(n\!+\!\frac32\big)\Gamma\big(n+1-\frac{D}{2}\big)}{(\pi aM)^{2n}}
  \left(1\!-\!\frac{1}{2^{1+2n}}\right)\zeta(2\!+\!2n)
\nonumber\\
&&\hskip 1.3cm
    -\,\frac{a}{4(4\pi)^{\frac{D-2}{2}}}\Gamma\Big(\!-\!\frac{D}{2}\Big)
   \big(m^{D}\!-\!M^{D}\big)
   \!+\!\mathcal{O}\big({\rm e}^{-2\pi aM}\big)
   \Bigg\}
 ,\quad
\label{T zz: exact}\\
 &&\hskip -0.7cm 
    \big\langle\Delta \hat{T}_{\vartheta}^{\;\vartheta}(x)\big\rangle^{(1)}
    -\big\langle\Delta \hat{T}_{\vartheta}^{\;\vartheta}(x)\big\rangle^{(1)}_M
    = \big\langle\Delta \hat{T}_{\varphi}^{\;\varphi}(x)\big\rangle^{(1)}
    -\big\langle\Delta \hat{T}_{\varphi}^{\;\varphi}(x)\big\rangle^{(1)}_M
\nonumber\\
&& \hskip 0.cm
=\,
\frac{\hbar}{4} \int\! \frac{{\rm d}^{D\!-\!3}k}{(2\pi)^{D\!-\!3}} \left[B_\omega 
  - \big(\|\mathbf{k}\|^2+m^2\big) B_{\frac{1}{\omega}}
   \right]
 \,,\quad
 \nonumber\\
&& \hskip -0.4cm
\approx\,\frac{\hbar}{4\pi a} 
 \Bigg\{\!\!-\frac{16M^{D-2}}{(4\pi)^{\frac{D+3}{2}}a}
 \sum_{n=0}^\infty \frac{\Gamma\big(n\!+\!\frac32\big)\Gamma\big(n+1-\frac{D}{2}\big)}{(\pi aM)^{2n}}
  \left(1\!-\!\frac{1}{2^{1+2n}}\right)
\nonumber\\
&& \hskip 4.4cm
  \times\bigg(2n\!+\!\frac{n\!+\!1\!-\!\frac{D}{2}}{2a^2M^2}\bigg)\zeta(2\!+\!2n)
\nonumber\\
&& \hskip 0.cm 
    -\,\frac{a}{2(4\pi)^{\frac{D-2}{2}}}
   \bigg(m^{D}\!-\!M^{D}\!-\!\frac{D}{8a^2}M^{D-2}\bigg)\Gamma\Big(\!\!-\!\frac{D}{2}\Big)
    \!+\!\mathcal{O}\big({\rm e}^{-2\pi aM}\big)
\,.\qquad
\label{T thetatheta: exact}
\end{eqnarray}
Summing up all of the contributions in Eqs.~(\ref{T tt: exact})--(\ref{T thetatheta: exact}) results in,
\begin{eqnarray}
 &&\hskip -0.7cm 
 \sum_{\mu=0}^{D-1} \Big(\big\langle\Delta \hat{T}_{\mu}^{\;\mu}(x)\big\rangle^{(1)}
    -\big\langle\Delta \hat{T}_{\mu}^{\;\mu}(x)\big\rangle^{(1)}_M\Big)
=\,
-\frac{\hbar}{2} \int\! \frac{{\rm d}^{D\!-\!3}k}{(2\pi)^{D\!-\!3}} \left[
        m^2 B_{\frac{1}{\omega}}
  \right]
\nonumber\\
&& \hskip 0.cm
\approx\, - \frac{\hbar}{2\pi a} 
 \Bigg\{\frac{32M^{D-2}}{(4\pi)^{\frac{D+3}{2}}a}
 \sum_{n=0}^\infty \frac{\Gamma\big(n\!+\!\frac32\big)\Gamma\big(n+2-\frac{D}{2}\big)}{(\pi aM)^{2n}}
  \left(1\!-\!\frac{1}{2^{1+2n}}\right)
\nonumber\\
&& \hskip 3.4cm
  \times\bigg(1\!+\!\frac{1}{4a^2M^2}\bigg)\zeta(2\!+\!2n)
\nonumber\\
&& \hskip 0.cm 
    -\,\frac{a}{2(4\pi)^{\frac{D-2}{2}}}
   \bigg(m^{D}\!-\!M^{D}\!-\!\frac{1}{4a^2}M^{D-2}\bigg)\Gamma\Big(\!1\!-\!\frac{D}{2}\Big)
     \!+\!\mathcal{O}\big({\rm e}^{-2\pi aM}\big)
     \Bigg\}
\nonumber\\
&& \hskip 0.4cm    
    = -m^2\Big(\bigl\langle\hat\phi^2(x)\bigr\rangle^{(1)}\!-\!\bigl\langle\hat\phi^2(x)\bigr\rangle^{(1)}_M\Big)
\,,\qquad\;\;
\label{m2 phi2: exact}
\end{eqnarray}
which is consistent with Eq.~(\ref{expectation value: Lagrangian}),
representing a nontrivial check of our results.

For large wormholes, $u\geq am \gg 1$,
in most applications one can neglect the exponentially small 
correction $\propto {\rm e}^{-2\pi {u_0}}$, such that the 
series expressions in Eqs.~(\ref{T tt 6})--(\ref{T zz 6})
are rapidly converging towards the desired answer
if truncated at some reasonably high 
value of $n$. With this in mind, one can insert 
Eqs.~(\ref{T tt 6})--(\ref{T zz 6}) into 
Eqs.~(\ref{T tt 2})--(\ref{T thetatheta 2}) to obtain,
\begin{eqnarray}
&&\hskip -0.7cm    
\big\langle\Delta \hat{T}_{t}^{\;t}(x)\big\rangle^{(1)}
    -\big\langle\Delta \hat{T}_{t}^{\;t}(x)\big\rangle^{(1)}_M
= -\frac{\hbar}{2}   \int \!\frac{{\rm d}^{D\!-\!3}k}{(2\pi)^{D\!-\!3}} 
B_\omega
\nonumber\\
&& \hskip 0cm
=\,
- \frac{\hbar}{2}\!\int\!\frac{{\rm d}^{D\!-\!3}k}{(2\pi)^{D\!-\!3}} 
\frac{1}{\pi a^3}\bigg[\bigg(\frac{{u_0}}{48}
   - \frac{7}{3840 {u_0}}
    - \frac{31}{129024 {u_0}^3}
    +\mathcal{O}\big({u_0}^{-5},{\rm e}^{-2\pi {u_0}}\big)
    \bigg)
\nonumber\\
 &&\hskip 4.cm    
     +\,\frac16\left(u^3 \!-\!u_0^3\right)\bigg]
 \,,\qquad\;
\label{T tt 7}\\
 &&\hskip -0.7cm  
 \big\langle\Delta \hat{T}_{z^i}^{\;z^i}(x)\big\rangle^{(1)}
    -\big\langle\Delta \hat{T}_{z^i}^{\;z^i}(x)\big\rangle^{(1)}_M
=
\frac{\hbar}{2}\int \frac{{\rm d}^{D\!-\!3}k}{(2\pi)^{D\!-\!3}}
\frac{\|\mathbf{k}\|^2}{(D\!-\!3)}
B_{\frac{1}{\omega}}
\nonumber\\
&& \hskip -0.4cm
=\,\frac{\hbar}{2}\!\int \!\frac{{\rm d}^{D\!-\!3}k}{(2\pi)^{D\!-\!3}}
\frac{\|\mathbf{k}\|^2}{(D\!-\!3)}
\frac{1}{\pi a}\bigg[\!\bigg(\frac{1}{48{u_0}}
   \!+\! \frac{7}{3840 {u_0}^3}
    \!+\! \frac{31}{43008 {u_0}^5}
    \!+\!\mathcal{O}\big({u_0}^{-7}\!,{\rm e}^{-2\pi {u_0}}\big)
    \bigg)
\nonumber\\
 &&\hskip 4.2cm    
     +\,\frac12\left(u \!-\!u_0\right)\bigg]
 ,\quad
\label{T zz 7}\\
 &&\hskip -0.7cm 
    \big\langle\Delta \hat{T}_{\vartheta}^{\;\vartheta}(x)\big\rangle^{(1)}
    -\big\langle\Delta \hat{T}_{\vartheta}^{\;\vartheta}(x)\big\rangle^{(1)}_M
    = \big\langle\Delta \hat{T}_{\varphi}^{\;\varphi}(x)\big\rangle^{(1)}
    -\big\langle\Delta \hat{T}_{\varphi}^{\;\varphi}(x)\big\rangle^{(1)}_M
\nonumber\\
&& \hskip 0.cm
=\,
\frac{\hbar}{4} \int\! \frac{{\rm d}^{D\!-\!3}k}{(2\pi)^{D\!-\!3}} \left[B_\omega 
  - \big(\|\mathbf{k}\|^2+m^2\big) B_{\frac{1}{\omega}}
   \right]
 \,,\quad
\label{T thetatheta 7}
\end{eqnarray}
where $u=a\sqrt{\|\mathbf{k}\|^2\!+\!m^2}$, 
$u_0=a\sqrt{\|\mathbf{k}\|^2\!+\!m^2\!-\!\frac14}\,$.

The next step is to evaluate the integrals in Eqs.~(\ref{T tt 7})--(\ref{T thetatheta 7}).
For this we use the master integral~(\ref{master integral}) with 
$\beta =-\frac{3}{2}, -\frac12,\frac12,\frac32,\frac52$ and $M^2=m^2,m^2-\frac{1}{4a^2},$
 to obtain, 
\begin{eqnarray}
&&\hskip -0.7cm    
\big\langle\Delta \hat{T}_{t}^{\;t}(x)\big\rangle^{(1)}
    -\big\langle\Delta \hat{T}_{t}^{\;t}(x)\big\rangle^{(1)}_M
= \, \frac{\hbar}{(4\pi)^{\frac{D}{2}}}
\bigg[\frac{\Gamma\big(\!\!-\!\frac{D-2}{2}\big)}{24a^2}
    \Big(m^2\!-\!\frac{1}{4a^2}\Big)^{\!\!\frac{D-2}{2}}
\nonumber\\
&&
+\frac{7}{960a^4}\Gamma\big(2\!-\!\tfrac{D}{2}\big)
    \Big(m^2\!-\!\frac{1}{4a^2}\Big)^{\!\frac{D-4}{2}}
+\frac{31}{16128a^6}\Gamma\big(3\!-\!\tfrac{D}{2}\big)
    \Big(m^2\!-\!\frac{1}{4a^2}\Big)^{\!\!\frac{D-6}{2}}
\nonumber\\
&&
-\,\frac{1}{2}\Gamma\big(\!-\tfrac{D}{2}\big) m^D
 + \frac{1}{2}\Gamma\big(\!-\tfrac{D}{2}\big)\Big(m^2\!-\!\frac{1}{4a^2}\Big)^{\!\frac{D}{2}}
    \!+\mathcal{O}\Big(a^{-8}\,,{\rm e}^{-2\pi {am}}\Big)
 \bigg]
\,,\quad
\label{T tt 8}\\
\nonumber\\
 &&\hskip -0.7cm  
 \big\langle\Delta \hat{T}_{z^i}^{\;z^i}(x)\big\rangle^{(1)}
    -\big\langle\Delta \hat{T}_{z^i}^{\;z^i}(x)\big\rangle^{(1)}_M
=\,\frac{\hbar}{(D\!-\!3)}
\frac{1}{(4\pi)^{\frac{D}{2}}}\bigg[\!- \frac{\Gamma\big(1\!-\!\frac{D}{2}\big)}{24a^2}
    \Big(m^2\!-\!\frac{1}{4a^2}\Big)^{\!\!\frac{D-2}{2}}
\nonumber\\
 &&\hskip 1.cm
+\frac{7}{960a^4}\Gamma\big(2\!-\!\tfrac{D}{2}\big)
    \Big(m^2\!-\!\frac{1}{4a^2}\Big)^{\!\!\frac{D-4}{2}}
 \!+\! \frac{31}{5376 a^6}\Gamma\big(3\!-\!\tfrac{D}{2}\big)
    \Big(m^2\!-\!\frac{1}{4a^2}\Big)^{\!\!\frac{D-6}{2}}
\nonumber\\
&&\hskip 1.cm
 +\, \frac{3}{2}\Gamma\big(\!-\tfrac{D}{2}\big)m^D
-\frac{3}{2}\Gamma\big(\!-\tfrac{D}{2}\big)
   \Big(m^2\!-\!\frac{1}{4a^2}\Big)^{\!\!\frac{D}{2}}
 \!+\mathcal{O}\Big(a^{-8}\,,{\rm e}^{-2\pi {am}}\Big)
\bigg]
\nonumber\\
&&\hskip 0cm
-\,\frac{\hbar}{(D\!-\!3)}
\frac{1}{(4\pi)^{\frac{D}{2}}}
\bigg[
\frac{1}{12a^2}\Gamma\big(2\!-\!\tfrac{D}{2}\big)
    \Big(m^2\!-\!\frac{1}{4a^2}\Big)^{\!\!\frac{D-2}{2}}
\nonumber\\
 &&\hskip 1.cm
+\,\frac{7}{480a^4}\Gamma\big(3\!-\!\tfrac{D}{2}\big)
    \Big(m^2\!-\!\frac{1}{4a^2}\Big)^{\!\!\frac{D-4}{2}}
 \!+\! \frac{31}{8064a^6}\Gamma\big(4\!-\!\tfrac{D}{2}\big)
    \Big(m^2\!-\!\frac{1}{4a^2}\Big)^{\!\!\frac{D-6}{2}}
\nonumber\\
&&\hskip 1.cm
 -\,\Gamma\big(1\!-\!\tfrac{D}{2}\big)m^D
\!+\! \Gamma\big(1\!-\!\tfrac{D}{2}\big)
    \Big(m^2\!-\!\frac{1}{4a^2}\Big)^{\!\!\frac{D}{2}}
 \!+\mathcal{O}\Big(a^{-8}\,,{\rm e}^{-2\pi {am}}\Big)
\bigg]
 \,,\qquad
\label{T zz 8}
\end{eqnarray}
which can be simplified to,
\begin{eqnarray}
&&\hskip -0.7cm    
\big\langle\Delta \hat{T}_{t}^{\;t}(x)\big\rangle^{(1)}
\!= \, \frac{\hbar}{(4\pi)^{\frac{D}{2}}}
\bigg[\frac{1}{2}\Gamma\big(\!-\tfrac{D}{2}\big)\Big(m^2\!-\!\frac{1}{4a^2}\Big)^{\!\frac{D}{2}}
\!\!+\!\frac{1}{24a^2}\Gamma\big(1-\!\tfrac{D}{2}\big)
    \Big(m^2\!-\!\frac{1}{4a^2}\Big)^{\!\!\frac{D-2}{2}}
\nonumber\\
&&\hskip 0.9cm
+\,\frac{7}{960a^4}\Gamma\big(2-\tfrac{D}{2}\big)
    \Big(m^2\!-\!\frac{1}{4a^2}\Big)^{\!\frac{D-4}{2}}
\!\!+\!\frac{31}{16128a^6}\Gamma\big(3-\tfrac{D}{2}\big)
    \Big(m^2\!-\!\frac{1}{4a^2}\Big)^{\!\!\frac{D-6}{2}}
\nonumber\\
&&\hskip 0.9cm
 +\, \frac{1}{2}\Gamma\big(\!-\tfrac{D}{2}\big)\Big(m^2\!-\!\frac{1}{4a^2}\Big)^{\!\frac{D}{2}}
    \!+\!\mathcal{O}\Big(a^{-8}\,,{\rm e}^{-2\pi {am}}\Big)
 \bigg]
\,,\quad
\label{T tt 9}\\
\nonumber\\
 &&\hskip -0.7cm  
 \big\langle\Delta \hat{T}_{z^i}^{\;z^i}(x)\big\rangle^{(1)}
 =\big\langle\Delta \hat{T}_{t}^{\;t}(x)\big\rangle^{(1)}
 \,,\qquad
\label{T zz 9}
\end{eqnarray}
where we added the Minkowski vacuum contributions given in 
Eqs.~(\ref{T tt: Mink})--(\ref{T ij: Mink}).
Following the analogous procedure,
from Eq.~(\ref{T thetatheta 7}) we obtain the angular contributions
to the one-loop energy-momentum tensor,
\begin{eqnarray}
 &&\hskip -0.7cm 
    \big\langle\Delta \hat{T}_{\vartheta}^{\;\vartheta}(x)\big\rangle^{(1)}
    = \big\langle\Delta \hat{T}_{\varphi}^{\;\varphi}(x)\big\rangle^{(1)}
\nonumber\\
&& \hskip 0.5cm
=\,
 \frac{\hbar}{(4\pi)^{\frac{D}{2}}}
\bigg[
 \frac{1}{2}\Gamma\big(\!-\tfrac{D}{2}\big)\Big(m^2\!-\!\frac{1}{4a^2}\Big)^{\!\frac{D}{2}}
\!-\!\frac{1}{8a^2}\Gamma\big(1\!-\!\tfrac{D}{2}\big)
    \Big(m^2\!-\!\frac{1}{4a^2}\Big)^{\!\!\frac{D-2}{2}}
\nonumber\\
&&\hskip 1cm
-\,\frac{17}{960a^4}\Gamma\big(2\!-\!\tfrac{D}{2}\big)
    \Big(m^2\!-\!\frac{1}{4a^2}\Big)^{\!\frac{D-4}{2}}
\!-\!\frac{457}{80640a^6}\Gamma\big(3\!-\!\tfrac{D}{2}\big)
    \Big(m^2\!-\!\frac{1}{4a^2}\Big)^{\!\!\frac{D-6}{2}}
\nonumber\\
&&\hskip 1cm
    +\,\mathcal{O}\Big(a^{-8}\,,{\rm e}^{-2\pi {am}}\Big)
 \bigg]
\,.\qquad
\label{T thetatheta 9}
\end{eqnarray}
To complete the calculation, we also evaluate expectation value of the field operator squared,
\begin{eqnarray}
 &&\hskip -0.7cm 
    \big\langle \big(\hat\phi(x)\big)^2 \big\rangle^{(1)}
 = \frac{\hbar}{2} \int\! \frac{{\rm d}^{D\!-\!3}k}{(2\pi)^{D\!-\!3}} B_{\frac{1}{\omega}}
 +  \frac{\hbar}{2} \int\! \frac{{\rm d}^{D\!-\!1}k}{(2\pi)^{D\!-\!1}} 
       \frac{1}{\sqrt{m^2 + \|\vec k\|^2}}
\nonumber\\
&& \hskip 0.5cm
=\,
 \frac{\hbar}{(4\pi)^{\frac{D}{2}}}
\bigg[\Gamma\big(1\!-\!\tfrac{D}{2}\big)\Big(m^2\!-\!\frac{1}{4a^2}\Big)^{\!\frac{D-2}{2}}
\!+\!\frac{1}{12a^2}\Gamma\big(2\!-\!\tfrac{D}{2}\big)
    \Big(m^2\!-\!\frac{1}{4a^2}\Big)^{\!\!\frac{D-4}{2}}
\nonumber\\
&&\hskip 1cm
+\,\frac{7}{480a^4}\Gamma\big(3\!-\!\tfrac{D}{2}\big)
    \Big(m^2\!-\!\frac{1}{4a^2}\Big)^{\!\frac{D-6}{2}}
\!+\!\frac{31}{8064a^6}\Gamma\big(4\!-\!\tfrac{D}{2}\big)
    \Big(m^2\!-\!\frac{1}{4a^2}\Big)^{\!\!\frac{D-8}{2}}
\nonumber\\
&&\hskip 1cm
    +\,\mathcal{O}\Big(a^{-8}\,,{\rm e}^{-2\pi {am}}\Big)
 \bigg]
\,.\qquad
\label{phi2: wormhole}
\end{eqnarray}
As an accuracy check of our calculations, we now evaluate 
expectation value of the quantized Lagrangian,
\begin{eqnarray}
\big\langle \hat{\mathcal{L}} \big\rangle^{(1)}
&\!\! =\!\!& -\frac{1}{2}\left[ \delta_\nu^{\;\mu} \big\langle \Delta\hat T_\mu^{\;\nu}\rangle^{(1)} 
   \!+\!m^2 \big\langle \big(\hat\phi(x)\big)^2 \rangle^{(1)}  \right]
\nonumber\\
 &\!\! =\!\!& 
-\frac{1}{2}\bigg\{(D\!-\!2)\big\langle \Delta\hat T_t^{\;t}\rangle^{(1)}
   \!+\!2 \big\langle \Delta\hat T_\vartheta^{\;\vartheta}\rangle^{(1)}
\nonumber\\
 &&\hskip 1cm 
   \!+\!\bigg[\Big(m^2\!-\!\frac{1}{4a^2}\Big)\!+\!\frac{1}{4a^2}\bigg]
       \big\langle \big(\hat\phi(x)\big)^2 \rangle^{(1)}  
      \bigg\}
\qquad\;\;
\nonumber\\
&\!\! =\!\!& 0
\,,\qquad
\label{expectation value of Lagrangian}
\end{eqnarray}
which is the expected result.

\bigskip

We are now ready to renormalize the result by using dimensional regularization.
To prepare for renormalization, we expand Eqs.~(\ref{T tt 9})--(\ref{T thetatheta 9})  around 
$D\!=\!4$ to obtain,
\begin{eqnarray}
&&\hskip -0.7cm    
\big\langle\Delta \hat{T}_{t}^{\;t}(x)\big\rangle^{(1)}
= \big\langle\Delta \hat{T}_{z^i}^{\;z^i}(x)\big\rangle^{(1)}
\nonumber\\
&&\hskip 0.7cm    
\!\approx \, -\frac{\hbar}{64\pi^2}
\bigg\{\bigg(m^4\!-\!\frac{2m^2}{3a^2}\!+\!\frac{2}{15a^4}\bigg)
  \bigg[\frac{2\mu^{D-4}}{D\!-\!4}+\ln\bigg(\frac{m^2\!-\!\frac{1}{4a^2}}{4\pi\mu^2}\!\bigg)
             \!+\!\gamma_E-\frac32\bigg]
\nonumber\\
&&\hskip 1.25cm
-\,\frac{1}{12a^2}\Big(m^2\!-\!\frac{31}{40a^2}\Big)
           \!-\!\frac{31}{4032a^6}\frac{1}{m^2\!-\!\frac{1}{4a^2}}
\!+\mathcal{O}\Big(a^{-8}\,,{\rm e}^{-2\pi {am}}\Big)
\bigg\}
\,,\qquad\;
\label{T tt 10}
\\
 &&\hskip -0.7cm 
    \big\langle\Delta \hat{T}_{\vartheta}^{\;\vartheta}(x)\big\rangle^{(1)}
    = \big\langle\Delta \hat{T}_{\varphi}^{\;\varphi}(x)\big\rangle^{(1)}
\nonumber\\
&&\hskip 0.7cm    
\!\approx \, -\frac{\hbar}{64\pi^2}
\bigg\{\bigg(m^4\!-\!\frac{2}{15a^4}\bigg)
  \bigg[\frac{2\mu^{D-4}}{D\!-\!4}+\ln\bigg(\frac{m^2\!-\!\frac{1}{4a^2}}{4\pi\mu^2}\!\bigg)
             \!+\!\gamma_E-\frac32\bigg]
\nonumber\\
&&\hskip 1.25cm
+\,\frac{1}{4a^2}\Big(m^2\!-\!\frac{27}{40a^2}\Big)
           \!+\!\frac{457}{20160a^6}\frac{1}{m^2\!-\!\frac{1}{4a^2}}
\!+\mathcal{O}\Big(a^{-8}\,,{\rm e}^{-2\pi {am}}\Big)
\bigg\}
\,,\qquad\;
\label{T thetatheta 10}
\end{eqnarray}
where we dropped the terms $\mathcal{O}(D\!-\!4)$
and we made use of  ({\it cf.} Eq.~(\ref{Gamma -D/2})),
\begin{eqnarray}
\frac{M^{D-2n}}{(4\pi)^{\frac{D}{2}}} &\!\! =\!\!&\frac{M^{4-2n}\mu^{D-4}}{16\pi^2} 
 \bigg[1\!+\!\frac{D-4}{2}\ln\bigg(\frac{M^2}{4\pi\mu^2}\bigg)\bigg]
\!+\!\mathcal{O}\big((D\!-\!4)^2\big)
\,,\qquad\;
\label{expanding M**(D-4)}\\
    \Gamma\Big(\!-\frac{D}{2}\Big) 
   &\!\! =\!\!& -\frac{1}{D\!-\!4}\left[1+\frac{D\!-\!4}{2}\left(\gamma_E-\frac32\right)
      \right]\!+\!\mathcal{O}\big(D\!-\!4\big)
\,,\qquad
\label{Gamma -D/2 2}\\
    \Gamma\Big(1\!-\!\frac{D}{2}\Big) 
    &\!\! =\!\!& \frac{2}{D\!-\!4}\left[1+\frac{D\!-\!4}{2}\left(\gamma_E-1\right)
      \right]\!+\!\mathcal{O}\big(D\!-\!4\big)
\,,\qquad
\label{Gamma 1-D/2 }\\
    \Gamma\Big(2\!-\!\frac{D}{2}\Big) 
   &\!\! =\!\!& -\frac{2}{D\!-\!4}\left[1+\frac{D\!-\!4}{2}\gamma_E
      \right]\!+\!\mathcal{O}\big(D\!-\!4\big)
\,.\qquad
\label{Gamma 2-D/2}
\end{eqnarray}

Observe now that the divergent terms $\propto 1/(D\!-\!4)$ in 
Eqs.~(\ref{T tt 10})--(\ref{T thetatheta 10}) are not identical, such that 
vacuum fluctuations in the topological wormhole background are not isotropic.
That means that apart from the cosmological constant counterterm,
we also need the Hilbert-Einstein counterterm,~\footnote{For the metric at hand, there is a 
degeneracy regarding which counterterms to use to renormalize the
energy-momentum tensor. Instead of the cosmological constant 
and Ricci scalar counterterms, 
one could have used the cosmological constant and any of the higher order counterterms, namely the Ricci tensor squared, the Ricci scalar squared or the Riemann tensor squared
counterterm. In fact, for the problem at hand all three of these higher order counterterms 
are degenerate. 
For the details on how these counterterms contribute to the energy momentum tensor,
see Ref.~\cite{Prokopec:2025jrd}.
}
\begin{eqnarray}
S_{\tt ct}\big[g_{\mu\nu}\big] &\!\!=\!\!& \int {\rm d}^Dx \sqrt{-g}
 \bigg[\delta\Big(\frac{1}{\kappa^2}\Big)R\!-\!2\delta\Big(\frac{\Lambda}{\kappa^2}\Big)
+ \delta\alpha_{\tt Riem^2} R_{\mu\nu\rho\sigma}R^{\mu\nu\rho\sigma}
\bigg]
\,,\qquad 
\label{counterterm action}
\end{eqnarray}
which contribute to the energy-momentum tensor 
$T^{\mu\nu}_{\tt ct}(x) = \frac{2}{\sqrt{-g}}\frac{\delta S_{\tt ct}}{\delta g_{\mu\nu}(x)}$
 as~\cite{Prokopec:2025jrd}, 
\begin{eqnarray}
\big(T_{\tt ct}(x)\big)^{\,\nu}_{\!\mu} &\!\!=\!\!&
-\,2\delta\Big(\frac{1}{\kappa^2}\Big)G_\mu^{\;\nu}
\!-\!2 \delta\Big(\frac{\Lambda}{\kappa^2}\Big)\delta_\mu^{\;\nu} 
\qquad
\nonumber\\
&& \hskip -1cm
+\, \delta \alpha_{\tt Riem^2}
    \Big(\delta_\mu^{\;\nu}\,R_{\alpha\beta\gamma\delta}
                R^{\alpha\beta\gamma\delta}
         \!\!-\! 4R_{\mu\alpha\beta\gamma}R^{\nu\alpha\beta\gamma}
          \!\!+\! 8g_{\mu\rho}\nabla_{\alpha}\nabla_{\beta}R^{\alpha(\rho\nu)\beta}
    \Big)
\,,\qquad\;
\label{energy-momentum tensor: counterterm}
\end{eqnarray}
where we added the Riemann tensor squared counterterm, which we do not need for 
renormalization, but we use it in the main text to discuss the role of 
finite counterterm contributions.
Upon using Eqs.~(\ref{Einstein tensor}) 
and~(\ref{Riemann tensor: singular 1})--(\ref{Riemann tensor: regular}) in
Eq.~(\ref{energy-momentum tensor: counterterm}) one obtains,
\begin{eqnarray}
\big(T_{\tt ct}\big)^{\,\nu}_{\!\mu} &\!\!=\!\!&
 \delta\Big(\frac{1}{\kappa^2}\Big) \frac{2}{a^2}
     \big(\delta_\mu^{\;\nu} \!-\!\tilde{\delta}_\mu^{\;\nu}\big)
 \!-\!2 \delta\Big(\frac{\Lambda}{\kappa^2}\Big)\delta_\mu^{\;\nu}   
\!+\! \delta\alpha_{\tt Riem^2} \frac{2}{a^4}\Big(2\delta_\mu^{\;\nu} \!-\!\tilde{\delta}_\mu^{\;\nu}
\Big)
\,,\qquad
\label{energy-momentum tensor: counterterm 2}
\end{eqnarray}
where $\tilde{\delta}_\mu^{\;\nu}  \equiv {\rm diag}\big(0,0,\cdots,0,1,1\big)$ is the diagonal 
matrix with zeros along the $t$ and $z^i$ directions and one along 
the angular ($\vartheta$ and $\varphi$) directions. 
Note that the result in Eq.~(\ref{energy-momentum tensor: counterterm 2}) 
holds close to $D=4$,
and to work in dimensions that differ by an integer from $D=4$ one would have to add other counterterms.
The counterterms in 
Eq.~(\ref{energy-momentum tensor: counterterm 2}) have the same structure as 
the primitive result in Eqs.~(\ref{T tt 10})--(\ref{T thetatheta 10}) 
\big($(T_{\tt ct})^{\,t}_{\!t}=(T_{\tt ct})^{\,z^i}_{\!z^i}$, 
$(T_{\tt ct})^{\,\vartheta}_{\!\vartheta}=(T_{\tt ct})^{\,\varphi}_{\!\varphi}$\big),
so we need to consider only two linearly independent equations. A simple algebra shows that 
that the following choice of the counterterm coefficients,
\begin{eqnarray}
\delta\Big(\frac{\Lambda}{\kappa^2}\Big)  &\!\!=\!\!& 
   \frac12  \Big[\big\langle\Delta \hat{T}_{\vartheta}^{\;\vartheta}(x)\big\rangle^{(1)}\Big]_{\tt div}
          \!+\! \frac{\Lambda}{\kappa^2}
\nonumber\\
  &\!\!=\!\!& - \frac{\hbar}{64\pi^2}
 \bigg(m^4\!-\!\frac{2}{15a^4}\bigg)\!\times\!\frac{\mu^{D-4}}{D\!-\!4}
   \!+\!\frac{\Lambda}{\kappa^2}
\,,\qquad
\label{counterterm coupling: Lambda}\\
\delta\Big(\frac{1}{\kappa^2}\Big)  &\!\!=\!\!& 
\frac{a^2}2 \bigg(\!- \Big[\big\langle\Delta \hat{T}_{t}^{\;t}(x)\big\rangle^{(1)}\Big]_{\tt div}
    \!+\!  \Big[\big\langle\Delta \hat{T}_{\vartheta}^{\;\vartheta}(x)\big\rangle^{(1)}\Big]_{\tt div}
\bigg)
            \!+\!\frac{1}{\bar{\kappa}^2} 
\nonumber\\
  &\!\!=\!\!& - \frac{\hbar}{96\pi^2}
 \Big(m^2\!-\!\frac{2}{5a^2}\Big)\!\times\!\frac{\mu^{D-4}}{D\!-\!4}
    \!+\!\frac{1}{\bar{\kappa}^2} 
\,,\qquad
\label{counterterm coupling: Newton}
\end{eqnarray}
where $\Lambda/\kappa^2$ and $\bar{G}=\bar{\kappa}^2/(16\pi)$ are the finite 
cosmological constant and Newton constant associated with the counterterms, respectively.
Note that  $\Lambda$ and $\bar{G}$ are the cosmological constant and Newton constant
associated with the counterterms, which are in general not equal to the observed
cosmological constant and Newton constant.
Adding Eq.~(\ref{energy-momentum tensor: counterterm 2}) to 
Eqs.~(\ref{T tt 10})--(\ref{T thetatheta 10}) results in the renormalized energy-momentum
tensor, $\langle\Delta\hat T_\mu^{\;\nu}\rangle_{\tt ren}^{(1)}
\equiv  \langle\Delta\hat T_\mu^{\;\nu}\rangle^{(1)}+(T_{\tt ct})_\mu^{\;\nu}$,
\begin{eqnarray}
&&\hskip -0.7cm    
\big\langle\Delta \hat{T}_{t}^{\;t}(x)\big\rangle^{(1)}_{\tt ren}
= \big\langle\Delta \hat{T}_{z}^{\;z}(x)\big\rangle^{(1)}_{\tt ren}
\nonumber\\
&&\hskip -0.3cm    
= \, -\frac{\hbar}{64\pi^2}
\bigg\{\!\bigg(\!m^4\!-\!\frac{2m^2}{3a^2}\!+\!\frac{2}{15a^4}\!\bigg)
  \bigg[\ln\bigg(\frac{m^2 -\frac{1}{4a^2}}{4\pi\mu^2}\bigg)
             \!+\!\gamma_E-\frac32\bigg]
\!-\!\frac{1}{12a^2}\Big(m^2\!-\!\frac{31}{40a^2}\Big)
\nonumber\\
&&\hskip 0.4cm
           -\,\frac{31}{4032a^6}\frac{1}{m^2\!-\!\frac{1}{4a^2}}
\!+\!\mathcal{O}\Big(a^{-8}\,,{\rm e}^{-2\pi {am}}\Big)
\bigg\}
 \!+\! \frac{2}{a^2}\frac{1}{\bar\kappa^2}   \!-\!2 \frac{\Lambda}{\kappa^2}
\!+\!\frac{4}{a^4} \alpha_{\tt Riem^2}
\,,\qquad
\label{T tt 11}
\\
 &&\hskip -0.7cm 
    \big\langle\Delta \hat{T}_{\vartheta}^{\;\vartheta}(x)\big\rangle^{(1)}_{\tt ren}
    = \big\langle\Delta \hat{T}_{\varphi}^{\;\varphi}(x)\big\rangle^{(1)}_{\tt ren}
\nonumber\\
&&\hskip -0.25cm    
= \, -\frac{\hbar}{64\pi^2}
\bigg\{\bigg(m^4\!-\!\frac{2}{15a^4}\bigg)
  \bigg[\ln\bigg(\frac{m^2 -\frac{1}{4a^2}}{4\pi\mu^2}\bigg)
             \!+\!\gamma_E-\frac32\bigg]
\!+\!\frac{1}{4a^2}\Big(m^2\!-\!\frac{27}{40a^2}\Big)
\nonumber\\
&&\hskip 0.4cm
          +\,\frac{457}{20160a^6}\frac{1}{m^2\!-\!\frac{1}{4a^2}}
\!+\!\mathcal{O}\Big(a^{-8}\,,{\rm e}^{-2\pi {am}}\Big)
\bigg\}
 \!-\!2 \frac{\Lambda}{\kappa^2}
\!+\!\frac{2}{a^4} \alpha_{\tt Riem^2} 
\,,\qquad
\label{T thetatheta 11}
\end{eqnarray}
where we dropped index $i$ on $z^i$ as there is only one $z$ coordinate in $D=4$.
These results are used in 
Eqs.~(\ref{sc Einstein equation: order hbar: tt})--(\ref{sc Einstein equation: order hbar: thetatheta}) in the main text.



\begin{thebibliography}{99}


\bibitem{Einstein:1935tc}
A.~Einstein and N.~Rosen,
``The Particle Problem in the General Theory of Relativity,''
Phys. Rev. \textbf{48} (1935), 73-77
doi:10.1103/PhysRev.48.73

\bibitem{Godel:1949ga}
K.~Godel,
``An Example of a New Type of Cosmological Solutions of Einstein's Field Equations of Gravitation,''
Rev. Mod. Phys. \textbf{21} (1949), 447-450
doi:10.1103/RevModPhys.21.447

\bibitem{MisnerThorneZurek:2009}
Charles W.~Misner, Kip S.~Thorne and Wojciech H. Zurek (2009-04-01). "John Wheeler, relativity, and quantum information". Physics Today. 62 (4) 40–46,  doi:10.1063/1.3120895. ISSN 0031-9228.

\bibitem{Morris:1988cz}
M.~S.~Morris and K.~S.~Thorne,
``Wormholes in space-time and their use for interstellar travel: A tool for teaching general relativity,''
Am. J. Phys. \textbf{56} (1988), 395-412
doi:10.1119/1.15620

\bibitem{Bronnikov:1973fh}
K.~A.~Bronnikov,
``Scalar-tensor theory and scalar charge,''
Acta Phys. Polon. B \textbf{4} (1973), 251-266

\bibitem{Kardashev:2006nj}
N.~S.~Kardashev, I.~D.~Novikov and A.~A.~Shatskiy,
``Astrophysics of Wormholes,''
Int. J. Mod. Phys. D \textbf{16} (2007), 909-926
doi:10.1142/S0218271807010481
[arXiv:astro-ph/0610441 [astro-ph]].

\bibitem{Bronnikov:2013coa}
K.~A.~Bronnikov, L.~N.~Lipatova, I.~D.~Novikov and A.~A.~Shatskiy,
``Example of a stable wormhole in general relativity,''
Grav. Cosmol. \textbf{19} (2013), 269-274
doi:10.1134/S0202289313040038
[arXiv:1312.6929 [gr-qc]].

\bibitem{Bronnikov:2021uta}
K.~A.~Bronnikov and R.~K.~Walia,
``Field sources for Simpson-Visser spacetimes,''
Phys. Rev. D \textbf{105} (2022) no.4, 044039
doi:10.1103/PhysRevD.105.044039
[arXiv:2112.13198 [gr-qc]].

\bibitem{Khabibullin:2005ad}
A.~R.~Khabibullin, N.~R.~Khusnutdinov and S.~V.~Sushkov,
``Casimir effect in a wormhole spacetime,''
Class. Quant. Grav. \textbf{23} (2006), 627-634
doi:10.1088/0264-9381/23/3/006
[arXiv:hep-th/0510232 [hep-th]].

\bibitem{Sushkov:2008zz}
S.~V.~Sushkov, A.~R.~Khabibullin and N.~R.~Khusnutdinov,
``Casimir effect for two spheres in a wormhole spacetime,''
Grav. Cosmol. \textbf{14} (2008), 147-153
doi:10.1134/S0202289308020047


\bibitem{Birrell:1982ix}
N.~D.~Birrell and P.~C.~W.~Davies,
``Quantum Fields in Curved Space,''
Cambridge University Press, 1982,
ISBN 978-0-511-62263-2, 978-0-521-27858-4
doi:10.1017/CBO9780511622632

\bibitem{Butcher:2014lea}
L.~M.~Butcher,
``Casimir Energy of a Long Wormhole Throat,''
Phys. Rev. D \textbf{90} (2014) no.2, 024019
doi:10.1103/PhysRevD.90.024019
[arXiv:1405.1283 [gr-qc]].

\bibitem{Mota:2022qpf}
H.~F.~S.~Mota, C.~R.~Muniz and V.~B.~Bezerra,
``Thermal Casimir Effect in the Einstein Universe with a Spherical Boundary,''
Universe \textbf{8} (2022) no.11, 597
doi:10.3390/universe8110597
[arXiv:2210.06128 [hep-th]].

\bibitem{Garattini:2024uso}
R.~Garattini,
``Effects of additional sources on Casimir Wormholes,''
[arXiv:2411.05522 [gr-qc]].

\bibitem{Garattini:2024jkr}
R.~Garattini and M.~Faizal,
``Hot Casimir wormholes,''
JCAP \textbf{01} (2025) no.081, 081
doi:10.1088/1475-7516/2025/01/081
[arXiv:2403.15174 [gr-qc]].

\bibitem{Hochberg:1996ee}
D.~Hochberg, A.~Popov and S.~V.~Sushkov,
``Selfconsistent wormhole solutions of semiclassical gravity,''
Phys. Rev. Lett. \textbf{78} (1997), 2050-2053
doi:10.1103/PhysRevLett.78.2050
[arXiv:gr-qc/9701064 [gr-qc]].

\bibitem{Avalos:2025hfw}
R.~Avalos, D.~Brito, E.~Fuenmayor and E.~Contreras,
``Hyperbolic Casimir-like wormhole,''
Eur. Phys. J. C \textbf{85} (2025) no.7, 793
doi:10.1140/epjc/s10052-025-14526-x
[arXiv:2507.17906 [gr-qc]].

\bibitem{Agrawal:2024dam}
A.~S.~Agrawal, S.~Tarai, B.~Mishra and S.~K.~Tripathy,
``Matter geometry coupling and Casimir wormhole geometry,''
JHEAp \textbf{47} (2025), 100378
doi:10.1016/j.jheap.2025.100378
[arXiv:2409.12160 [gr-qc]].

\bibitem{Planck:2018vyg}
N.~Aghanim \textit{et al.} [Planck],
Astron. Astrophys. \textbf{641} (2020), A6
[erratum: Astron. Astrophys. \textbf{652} (2021), C4]
doi:10.1051/0004-6361/201833910
[arXiv:1807.06209 [astro-ph.CO]].


\bibitem{Friedman:1993ty}
J.~L.~Friedman, K.~Schleich and D.~M.~Witt,
``Topological censorship,''
Phys. Rev. Lett. \textbf{71} (1993), 1486-1489
[erratum: Phys. Rev. Lett. \textbf{75} (1995), 1872]
doi:10.1103/PhysRevLett.71.1486
[arXiv:gr-qc/9305017 [gr-qc]].

\bibitem{Visser:1995cc}
M.~Visser, ``Lorentzian wormholes: From Einstein to Hawking,''
Springer-Verlag (1996), ISBN 978-1-56396-653-8.

\bibitem{Hochberg:1997wp}
D.~Hochberg and M.~Visser,
``Geometric structure of the generic static traversable wormhole throat,''
Phys. Rev. D \textbf{56} (1997), 4745-4755
doi:10.1103/PhysRevD.56.4745
[arXiv:gr-qc/9704082 [gr-qc]].

\bibitem{Hochberg:1998ii}
D.~Hochberg and M.~Visser,
``The Null energy condition in dynamic wormholes,''
Phys. Rev. Lett. \textbf{81} (1998), 746-749
doi:10.1103/PhysRevLett.81.746
[arXiv:gr-qc/9802048 [gr-qc]].

\bibitem{Krasnikov:2010vw}
S.~Krasnikov,
``No to censorship! Comment on the Friedman-Schleich-Witt theorem,''
Grav. Cosmol. \textbf{19} (2013), 54
doi:10.1134/S0202289313010064
[arXiv:1007.4167 [gr-qc]].


\bibitem{Maldacena:2013xja}
J.~Maldacena and L.~Susskind,
``Cool horizons for entangled black holes,''
Fortsch. Phys. \textbf{61} (2013), 781-811
doi:10.1002/prop.201300020
[arXiv:1306.0533 [hep-th]].

\bibitem{Maldacena:2017axo}
J.~Maldacena, D.~Stanford and Z.~Yang,
``Diving into traversable wormholes,''
Fortsch. Phys. \textbf{65} (2017) no.5, 1700034
doi:10.1002/prop.201700034
[arXiv:1704.05333 [hep-th]].

\bibitem{Maldacena:2018lmt}
J.~Maldacena and X.~L.~Qi,
``Eternal traversable wormhole,''
[arXiv:1804.00491 [hep-th]].

\bibitem{Garcia-Garcia:2019poj}
A.~M.~Garc{\'\i}a-Garc{\'\i}a, T.~Nosaka, D.~Rosa and J.~J.~M.~Verbaarschot,
``Quantum chaos transition in a two-site Sachdev-Ye-Kitaev model dual to an eternal traversable wormhole,''
Phys. Rev. D \textbf{100} (2019) no.2, 026002
doi:10.1103/PhysRevD.100.026002
[arXiv:1901.06031 [hep-th]].
 
\bibitem{Zhou:2020wgh}
T.~G.~Zhou and P.~Zhang,
``Tunneling through an Eternal Traversable Wormhole,''
Phys. Rev. B \textbf{102} (2020), 224305
doi:10.1103/PhysRevB.102.224305
[arXiv:2009.02641 [cond-mat.str-el]].

\bibitem{Maldacena:2018gjk}
J.~Maldacena, A.~Milekhin and F.~Popov,
``Traversable wormholes in four dimensions,''
Class. Quant. Grav. \textbf{40} (2023) no.15, 155016
doi:10.1088/1361-6382/acde30
[arXiv:1807.04726 [hep-th]].

\bibitem{Maldacena:2020sxe}
J.~Maldacena and A.~Milekhin,
``Humanly traversable wormholes,''
Phys. Rev. D \textbf{103} (2021) no.6, 066007
doi:10.1103/PhysRevD.103.066007
[arXiv:2008.06618 [hep-th]].

\bibitem{Allen:1987tz}
B.~Allen and A.~Folacci,
``The Massless Minimally Coupled Scalar Field in De Sitter Space,''
Phys. Rev. D \textbf{35} (1987), 3771
doi:10.1103/PhysRevD.35.3771

\bibitem{Onemli:2002hr}
V.~K.~Onemli and R.~P.~Woodard,
``Superacceleration from massless, minimally coupled phi**4,''
Class. Quant. Grav. \textbf{19} (2002), 4607
doi:10.1088/0264-9381/19/17/311
[arXiv:gr-qc/0204065 [gr-qc]].

\bibitem{Glavan:2022dwb}
D.~Glavan and T.~Prokopec,
``Photon propagator in de Sitter space in the general covariant gauge,''
JHEP \textbf{05} (2023), 126
doi:10.1007/JHEP05(2023)126
[arXiv:2212.13982 [gr-qc]].

\bibitem{Glavan:2022nrd}
D.~Glavan and T.~Prokopec,
``Even the photon propagator must break de Sitter symmetry,''
Phys. Lett. B \textbf{841} (2023), 137928
doi:10.1016/j.physletb.2023.137928
[arXiv:2212.13997 [hep-th]].

\bibitem{Koksma:2011cq}
J.~F.~Koksma and T.~Prokopec,
``The Cosmological Constant and Lorentz Invariance of the Vacuum State,''
[arXiv:1105.6296 [gr-qc]].


\bibitem{Lucat:2018slu}
S.~Lucat, T.~Prokopec and B.~Swiezewska,
``Conformal symmetry and the cosmological constant problem,''
Int. J. Mod. Phys. D \textbf{27} (2018) no.14, 1847014
doi:10.1142/S0218271818470144
[arXiv:1804.00926 [gr-qc]].

\bibitem{Jimu:2024xqm}
D.~Jimu and T.~Prokopec,
``Uniqueness of gravitational constant at low energies from the connection between spin-2 and spin-0 sectors,''
JHEP \textbf{04} (2025), 134
doi:10.1007/JHEP04(2025)134
[arXiv:2410.01449 [gr-qc]].

\bibitem{Appelquist:1974tg}
T.~Appelquist and J.~Carazzone,
``Infrared Singularities and Massive Fields,''
Phys. Rev. D \textbf{11} (1975), 2856
doi:10.1103/PhysRevD.11.2856


\bibitem{Starobinsky:1980te}
A.~A.~Starobinsky,
``A New Type of Isotropic Cosmological Models Without Singularity,''
Phys. Lett. B \textbf{91} (1980), 99-102
doi:10.1016/0370-2693(80)90670-X

\bibitem{Shatskiy:2008us}
A.~Shatskiy, I.~D.~Novikov and N.~S.~Kardashev,
``New analytic models of 'traversable' wormholes,''
Phys. Usp. \textbf{51} (2008), 457-464
doi:10.1070/PU2008v051n05ABEH006581
[arXiv:0810.0468 [gr-qc]].

\bibitem{Marunovic:2014hla}
A.~Marunovi\'c and T.~Prokopec,
``Global monopoles can change Universe's topology,''
Phys. Lett. B \textbf{756} (2016), 268-272
doi:10.1016/j.physletb.2016.03.030
[arXiv:1411.7402 [gr-qc]].

\bibitem{Peskin:1995ev}
M.~E.~Peskin and D.~V.~Schroeder,
``An Introduction to quantum field theory,''
Addison-Wesley, 1995,
ISBN 978-0-201-50397-5, 978-0-429-50355-9, 978-0-429-49417-8
doi:10.1201/9780429503559.

\bibitem{Bordag:2009zz}
M.~Bordag, G.~L.~Klimchitskaya, U.~Mohideen and V.~M.~Mostepanenko,
``Advances in the Casimir effect,''
Int. Ser. Monogr. Phys. \textbf{145} (2009), 1-768
Oxford University Press, 2009,

\bibitem{Saharian:2018}
A. A. ~Saharian,
``The generalized Abel-Plana formula with applications to Bessel functions and Casimir effect,''
[arXiv:0708.1187].

\bibitem{NIST:2026}
https://dlmf.nist.gov/16.11 [accessed on Feb 22, 2026].

\bibitem{Prokopec:2025jrd}
T.~Prokopec,
``Gravitational Noether-Ward identities for scalar field,''
[arXiv:2512.22958 [gr-qc]].





\end{thebibliography}
\end{document}